# The counting house, measuring those who count:
## Presence of Bibliometrics, Scientometrics, Informetrics, Webometrics and Altmetrics in Google Scholar Citations, ResearcherID, ResearchGate, Mendeley, & Twitter

**Alberto Martín-Martín[1], Enrique Orduna-Malea[2], Juan M. Ayllón[1] & Emilio Delgado López-Cózar[1]**

[1] *EC3 Research Group: Evaluación de la Ciencia y de la Comunicación Científica, Universidad de Granada (Spain)*
[2] *EC3 Research Group: Evaluación de la Ciencia y de la Comunicación Científica, Universidad Politécnica de Valencia (Spain)*

## ABSTRACT

Following in the footsteps of the model of scientific communication, which has recently gone through a metamorphosis (from the Gutenberg galaxy to the Web galaxy), a change in the model and methods of scientific evaluation is also taking place. A set of new scientific tools are now providing a variety of indicators which measure all actions and interactions among scientists in the digital space, making new aspects of scientific communication emerge. In this work we present a method for "capturing" the structure of an entire scientific community (the Bibliometrics, Scientometrics, Informetrics, Webometrics, and Altmetrics community) and the main agents that are part of it (scientists, documents, and sources) through the lens of *Google Scholar Citations* (*GSC*).

Additionally, we compare these author "portraits" to the ones offered by other profile or social platforms currently used by academics (*ResearcherID*, *ResearchGate*, *Mendeley*, and *Twitter*), in order to test their degree of use, completeness, reliability, and the validity of the information they provide. A sample of 814 authors (researchers in Bibliometrics with a public profile created in GSC) was subsequently searched in the other platforms, collecting the main indicators computed by each of them. The data collection was carried out on September, 2015. The Spearman correlation (α= 0.05) was applied to these indicators (a total of 31), and a Principal Component Analysis was carried out in order to reveal the relationships among metrics and platforms as well as the possible existence of metric clusters.

We found that it is feasible to depict an accurate representation of the current state of the Bibliometrics community using data from *GSC* (the most influential authors, documents, journals, and publishers). Regarding the number of authors found in each platform, *GSC* takes the first place (814 authors), followed at a distance by *ResearchGate* (543), which is currently growing at a vertiginous speed. The number of *Mendeley* profiles is high, although 17.1% of them are basically empty. *ResearcherID* is also affected by this issue (34.45% of the profiles are empty), as is *Twitter* (47% of the *Twitter* accounts have published less than 100 tweets). Only 11% of our sample (93 authors) have created a profile in all the platforms analyzed in this study. From the PCA, we found two kinds of impact on the Web: first, all metrics related to academic impact. This first group can further be divided into usage metrics (views and downloads) and citation metrics. Second, all metrics related to connectivity and popularity (followers). *ResearchGate* indicators, as well as *Mendeley readers*, present a high correlation to all the indicators from *GSC,* but only a moderate correlation to the indicators in *ResearcherID. Twitter* indicators achieve only low correlations to the rest of the indicators, the highest of these being to *GSC* (0.42-0.46), and to *Mendeley* (0.41-0.46).

Lastly, we present a taxonomy of all the errors that may affect the reliability of the data contained in each of these platforms, with a special emphasis in *GSC*, since it has been our main source of data. These errors alert us to the danger of blindly using any of these platforms for the assessment of individuals, without verifying the veracity and exhaustiveness of the data.

In addition to this working paper, we also have made available a website where all the data obtained for each author and the results of the analysis of the most cited documents can be found: Scholar Mirrors.

**Corresponding author**
**Emilio Delgado López-Cózar.** edelgado@ugr.es

# 1. INTRODUCTION

## 1.1. Disciplines and scientific communities: territories and the tribes of Science

Science, in order to be properly investigated, grasped, and taught, has usually been organized in various areas of knowledge. Over time, each of these areas has been further divided into fields, subfields, disciplines, and specialties, as a result of the ever faster growth of knowledge and the parallel increase in the number of people who form the scientific communities within each of these areas. This process of scientific budding follows the life cycle of a living being (birth, growth, reproduction, and death), and is subject to endless metamorphosis, each discipline displaying its own idiosyncrasies.

Each of these units in which scientific knowledge is structured has its own epistemological properties (its object, its principles, and its methods) that endow them with a characteristic identity as well as boundaries that demarcate their cognitive territory. The inner and outer boundaries are not always clearly defined. There is overlapping between disciplines, gaps, and loops, sometimes quite vague and difficult to trace.

The different areas of knowledge are populated by communities of scientists and professionals, each group using their own tools, methodologies and techniques. These are social groups that share - with more or less consensus - professional practices, forms of work organization, living conditions, social expectations, principles, values, and beliefs.

Whitley (1984) dissected with a precision close to that of a surgeon's scalpel the process by which the academic communities - and their disciplines and specialties - become socially and cognitively institutionalized: how they create organizations that allow them to associate in order to defend their interests, how they erect spaces for the exchange of ideas and social development (conferences, seminars, forums, etc.), how they institute professional (newsletter, discussion list) or scientific (journals) means of communication, how they obtain academic standing by teaching the subject at the university (courses in graduate and postgraduate programs, including Master and PhD degrees), how they create groups, departments, laboratories, and companies dedicated to advance research, how they define research agendas where not only research problems but also ways to address and solve them are addressed, or how to create a common language to establish ideas and principles. Not to mention that the process of social and cognitive institutionalization of disciplines is directly influenced by the geographic location and the different levels of economic and cultural development of the countries where they are based.

As masterfully formulated by Becher and Trowler (2001), there is a close relationship between the disciplines (territories of knowledge) and people who advance them (scientific tribes); between the epistemic properties of the forms of scientific knowledge and the social aspects of academic communities. This is why any analysis of a discipline cannot ignore these two areas: the cognitive (disciplines) and social (community); you cannot understand one without the other.

Therefore, the ultimate aim of this Working Paper is to portray a discipline (Bibliometrics) and those who practice it, because a discipline is what is performed by those who cultivate it. Consequently, identifying the members of the Bibliometric tribe is one of the goals of this work.

## 1.2. A discipline with many names

There are numerous works which address the history of our field of knowledge (Broadus 1987a; Hertzel 1987; Shapiro 1992; Godin, 2006; De Bellis 2009). Its denomination, object of study, and scope have been addressed as well (Lawani, 1981; Bonitz, 1982; Peritz, 1984; Broadus, 1987b; Brookes, 1988; 1990; Sengupta, 1992; Glänzel & Schoepflin 1994; Braun 1994, Gorbea, 1995; Hood & Wilson, 2001; Cronin, 2001; Thelwall, 2008; Larriviere, 2012). There are also several literature reviews about this subject (Narin & Moll, 1977; White & McCain, 1989; Van Raan, 1997; Wilson, 1999; Borgman & Furner, 2002).

Bibliometrics can be synthetically defined as the discipline responsible for measuring communication and, in enlarged form, as the specialty responsible for quantitatively study the production, distribution, dissemination and consumption of information conveyed in any type of document (book, journal, conference, patent, or website) and any intellectual field, but with special attention to scientific information. It is a discipline with peculiar features:

- It is a very young discipline: although rooted in the early twentieth century in the library environment with the idea of measuring the production of knowledge (bibliographic statistics) and to properly manage library collections, it is not after World War II that Bibliometrics really starts to set its foundations. Its epistemic fundamentals are still boiling (they are not fully settled yet).
- It is a discipline best defined by its methods than by the thematic areas covered (the so-called “metrics”: quantitative data analysis applying various statistical techniques).
- It has a strong interdisciplinary character which arises from the incorporation of methods and techniques developed in other fields, and by its application to the study of any subject area. This makes Bibliometrics an open discipline willing to be fertilized by ideas from the most diverse origin and accept scientists from the most diverse disciplinary environments. This is the reason why Bibliometrics resembles a crossroads, a place where different scientific traditions meet.

The young age of the discipline and its interdisciplinary and instrumental character is the reason why this discipline is known by many different names. However, this fact does not mean the subject of study or the borders of the discipline are not clearly defined. Rather, it is a sign of the coexistence of different traditions that have shaped the development of the discipline.

Bibliometrics is the original and most widespread name. It stems from the bibliographic tradition represented by Paul Otlet with his proposal for a "bibliometrie", a Science for measuring all the dimensions of books and other documents, and from the library tradition concerned since ancient times about measuring the growth of knowledge and usage of its holdings.

Scientometrics is oriented towards the quantitative analysis of scientific and technical literature. It comes from the tradition of the science of science (space of confluence of Sociology, History, and Philosophy of science), to which science policy is also linked. It was crucial for this scientometric orientation the creation of the Citation indexes (databases dedicated to the collection of scientific production).

Informetrics is focused on the discovery of mathematical models that explain the properties of information. It is connected with the modern information science. It is a designation so close to Scientometrics that sometimes it is difficult to find differences among them.

Webometrics and Altmetrics are the most recent denominations. They started to gain momentum as the use of the new information and communication technologies began to spread. They are being developed in the tradition of the modern Library and Information Science, a discipline increasingly dedicated to computer science and to computing itself. These new names are strongly influenced by the medium in which information is conveyed rather than by the content itself. They come also to highlight the traditional technological aspect that the different metric specialties have enjoyed since their inception.

An analysis of the terms used in the titles of documents in our field published between 1969 and 2015 and indexed in Google Scholar (Figure 1) shows a clear predominance of the term “Bibliometrics”, followed by “Scientometrics”. However, in the last three years the term “Altmetrics” is being increasingly used, as a result of the novelty of the new social media communication technologies.

**Figure 1. Number of results returned by Google Scholar for the terms Bibliometrics, Scientometrics, Informetrics, Webometrics and Altmetrics contained only within the document titles by year (1969-2015)**

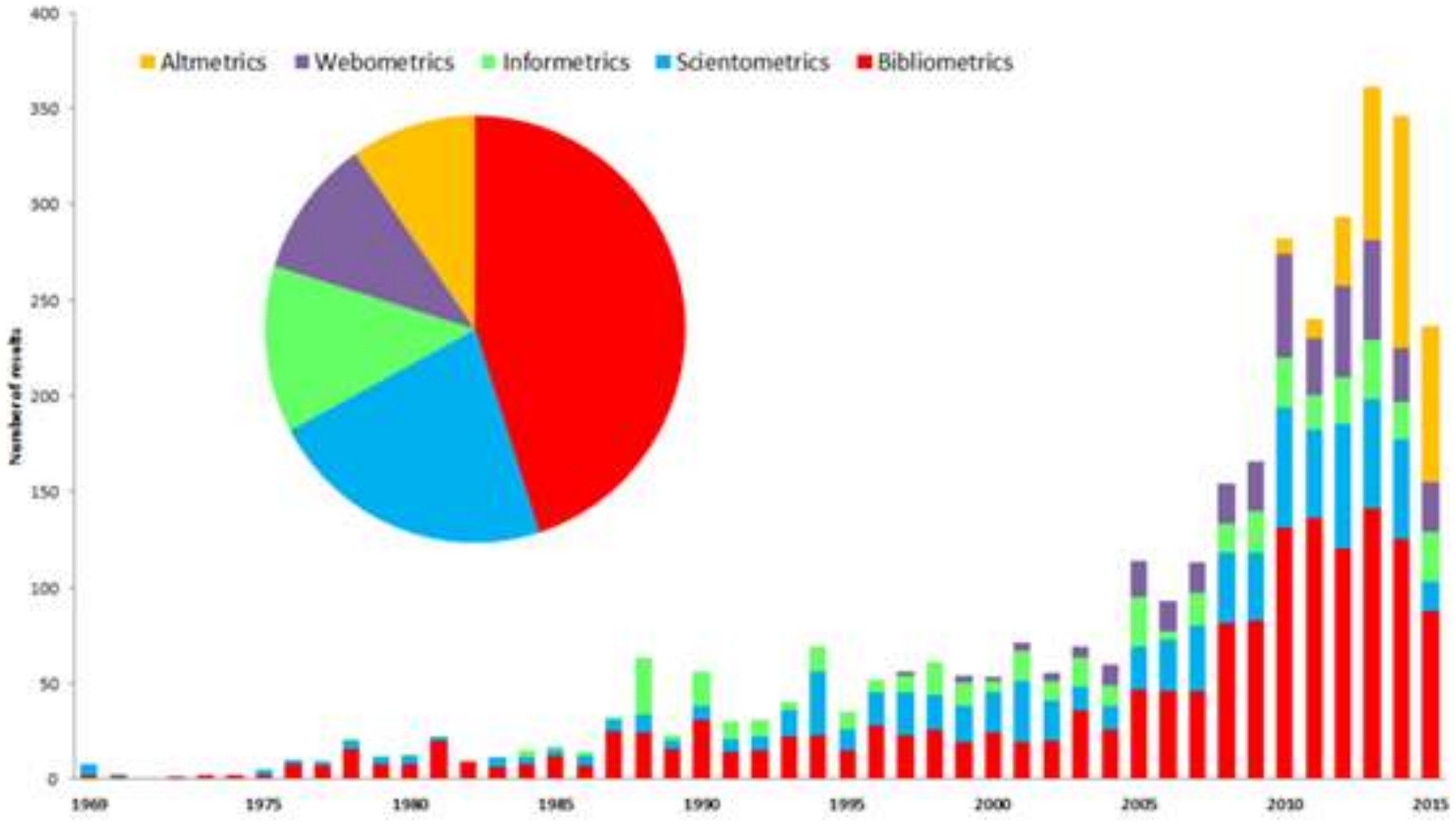

A similar result is obtained when the keywords used by the 814 scientists specialized in Bibliometrics or working sporadically in this field with a public profile on Google Scholar Citations are analyzed (Figure 2). The prevalence of Scientometrics and Bibliometrics is clear, although the weight of the latter would be higher had the terms been properly standardized.

**Figure 2. Word cloud of the keywords used by the researchers with a public Google Scholar Citations profile analyzed in this product (size indicates frequency of use in the sample)**

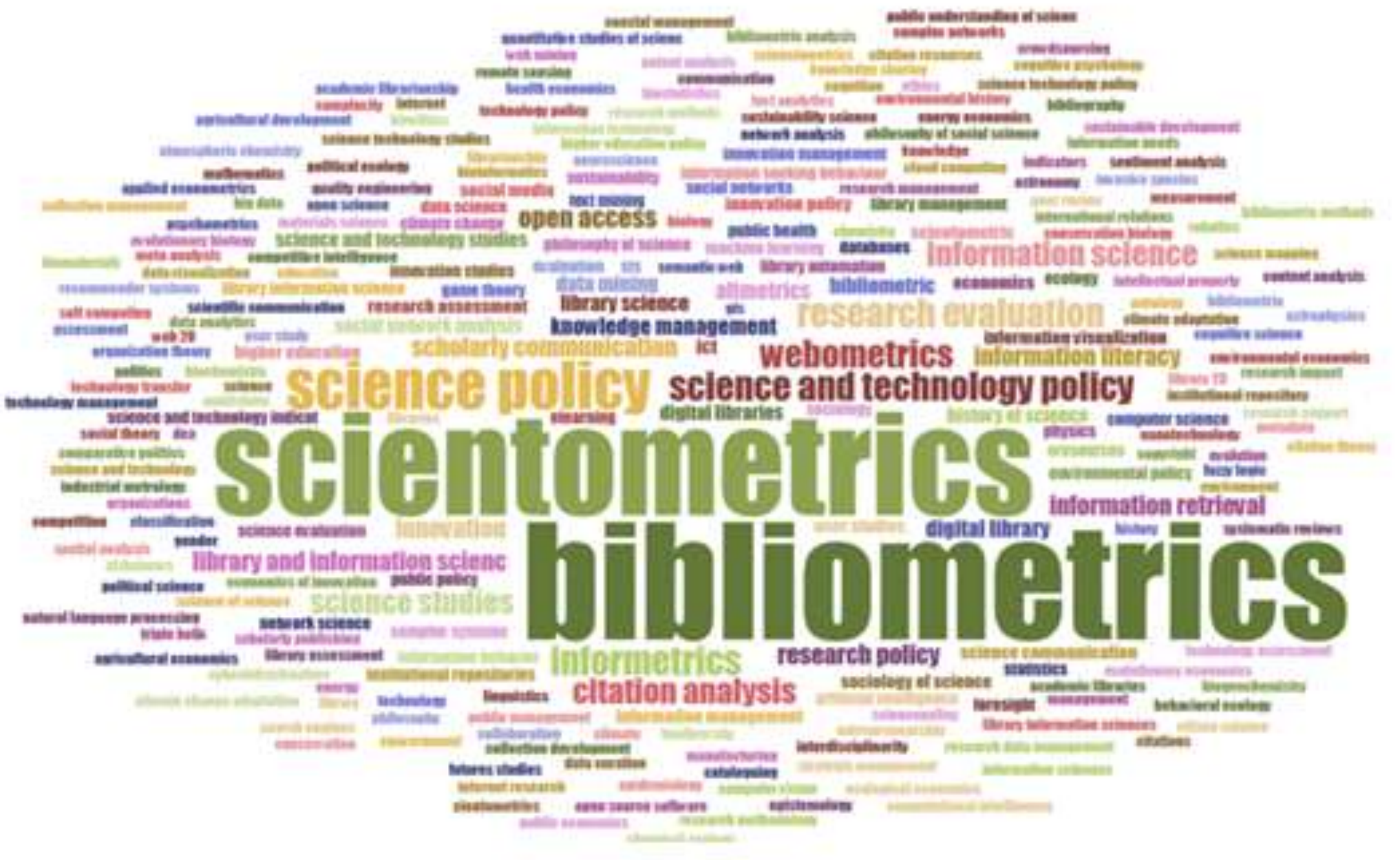

Furthermore, it is of great interest to know which other terms are used by bibliometricians. The list of terms associated with the Library and Information Science are very numerous, which shows how this discipline was the area where Bibliometrics stemmed from. Similarly, the relationship with science and technology studies (and specifically with science policy) is obvious. Lastly, there are also many terms related to research evaluation and citation analysis.

## 1.3. New mirrors and meters of Science: new media and new metrics

There is no better way to learn about a discipline than analysing its scientific literature. The best mirror of a scientific discipline is precisely the intellectual production that its academic community generates. This is the assumption in which Bibliometrics is based when it is used to examine the traits that define other disciplines and specialties.

Knowing the scope of a discipline will not only help characterize and determine its perspective and scientific nature, but it will also indirectly delineate its internal structure, its coherence, its contours, and its location in the overall picture of Sciences. This will enable an understanding of what the research is and has been about in a particular discipline, and how it may evolve in the future.

Today the number of venues in which research results produced in any discipline are published has been remarkably increased. The “Gutenberg paradigm”, which limited research products to the printed world (and more specifically to the journal, the main communication channel), has been challenged since the end of the twentieth century by a plethora of new channels of communication that are created, indexed, searched, located, read, and mentioned in the shared hyperspace (Castells, 2002). All this, of course, made possible by the development and worldwide use of the Internet, and the social web in particular. These are the new mirrors where the disciplines and communities are reflected. Revealing and evaluating the role of these new channels in Bibliometrics is another goal of this paper.

Following in the footsteps of the model of scientific communication, which has recently gone through a metamorphosis (from the Gutenberg galaxy to the web galaxy), a change in the model and methods of scientific evaluation is also taking place. The new media, due to its electronic nature, are supplied with multiple indicators measuring all actions and interactions among scientists in the digital space. In this work we open the door to new platform providers of metric indicators (whose nature is still unknown because of its youth) and snoop inside to see what they tell us about the various facets of scientific communication, complementing in this way some recent works in the topic (Jamali, Nicholas & Herman, 2015; Mikki et al, 2015), where not only the potential of these new mirrors but also their limitations and perceptions are considered.

We intend to bring attention to some of these new metrics and look into their meaning. In this way we position ourselves in the debate about "Altmetrics", but using a different perspective: the perspective of individuals and not just the documents they produce. We observe what these new metrics measure by taking as the object of study precisely those researchers who measure others (bibliometricians). In short, Bibliometrics, and those who measure, are measured.

Following our research line oriented on discovering the inner depths of Google Scholar while testing its suitability as a tool for research evaluation, this time we have turned our efforts to investigate new uses for Google Scholar Citations (sometimes also known as Google Scholar Profiles). We present in this new Working Paper a method to learn about the impact of an entire scientific specialty: a very specific scientific and professional community (the Bibliometrics, Scientometrics, Informetrics, Webometrics, and Altmetrics community), and the main agents that are part of it (scientists, professionals, the documents they produce, and the journals and publishers that publish these documents).

From the scientific output of the members of the metrics and quantitative information science studies community who have made public their profile on *Google Scholar Citations (GSC)*, we can develop a picture of this discipline.

Once we’ve seen the picture of the discipline that can be observed through the data available in GSC, we also want to compare it to its counterparts in other academic web services, like *ResearcherID*, a researcher identification system launched by *Thomson Reuters*, mainly built upon data from *Web of Science* (which has been and still is the go-to source for many researchers in the field of research evaluation), and other profiling services which arose in the wake of the Web 2.0 movement: *ResearchGate*, an academic social network, and *Mendeley*, a social reference manager which also offers profiling features. These are the most widely known tools worldwide for academic profiling.[1,2]

These tools offer researchers the chance to create an academic profile, as well as the chance to upload their publications, which are therefore available for other researchers to access, download, and comment upon. Researchers can also feed these databases with other kinds of data (tagging and following profiles, asking and answering specific questions) which might be useful for the rest of users in the platform.

In addition, we also include the links to the authors' homepages (the first tool researchers used to showcase their scientific activities on the Web), and *Twitter*, the popular microblogging site, in order to learn how much presence bibliometricians have in this platform and the kind of communication activities in which they take part there.

In short, our aim is to present a multifaceted and integral perspective of the discipline, as well as to provide the opportunity for an easy and intuitive comparison of these products and the reflections of scientific activity each of them portrays.

This project can also be considered as an attempt to deconstruct traditional journal, author, and institutional (mainly university) rankings, which are usually built upon data from traditional citation databases (Web of Science, Scopus) and are based exclusively on journal impact indicators. In this product, we are using a bottom-up approach by analyzing the documents that are either published by a group of authors associated with the discipline, those which are published in the main journals of the discipline, or those which use the most common and significant keywords in the discipline.

This is done in keeping with the widespread notion that the impact of the various scientific units (documents, individuals, organizations, subject domains) should be evaluated directly, using appropriate indicators for each unit, and not by using proxies like, for example, the average impact of the journals where a researcher’s or an institution’s documents are published to evaluate that researcher or institution.

---

[1] http://www.nature.com/news/online-collaboration-scientists-and-the-social-network-1.15711
[2] https://101innovations.wordpress.com/2015/06/23/first-1000-responses-most-popular-tools-per-research-activity

In short, the objectives of this study are essentially the following:

1. Applying Google Scholar Citations to radiograph Bibliometrics as a discipline, identifying the core authors, documents, journals, and most influential publishers in the field.
2. Comparing the user metric portraits generated by *Google Scholar Citations* to those offered by new platforms for the management of personal bibliographic profiles (*ResearcherID*, *ResearchGate*, and *Mendeley*) and content dissemination and communication (*Twitter*).
3. Testing the completeness, reliability and validity of the information provided by *Google Scholar Citations* (to generate disciplinary rankings), and by the remaining social platforms (to generate complementary academic mirrors of the scientific community).

# 2. METHODS

## 2.1. Search and identification of relevant authors

The first step was to identify all authors who have published in the areas of Bibliometrics, Scientometrics, Informetrics, Webometrics or Altmetrics, and for whom a *Google Scholar Citations* (GSC) public profile could be found at the time the data was collected (24/07/2015).

In order to locate the set of authors relevant to our study (i.e., those who have published in Bibliometrics and have a public profile in GSC), the following search strategies were used:

a) Keywords

A search was conducted in core selected journals: *Scientometrics*, *Journal of Informetrics*, *Research Evaluation*, *Cybermetrics*, and the *ISSI conferences* (*International Conference on Scientometrics and Informetrics*) with the goal of extracting the most frequently used and representative words in the discipline. The selected keywords were:

- Altmetrics
- Bibliometrics
- Citation Analysis
- Citation Count
- H-Index
- Impact Factor
- Informetrics
- Patent Citation
- Quantitative Studies of Science and Technology
- Research Assessment
- Research Evaluation
- Research Policy
- Science and Technology Policy
- Science Evaluation
- Science Policy
- Science Studies
- Scientometrics
- Webometrics

All public GSC profiles containing any of these keywords as one of the research interests were selected (GSC allows authors to display up to five research interests).

The lack of normalization in the use of keywords sometimes forced us to search alternatives of these keywords. These variants included misspelled words, the same keywords in other languages, etc. As an example, these are all the variants we found of the keyword "bibliometrics": bibliometric; bibliometría; bibliometria; bibliometric analysis; bibliometric methods; bibliometics; bibliometircs; bibliometric analysis in mining sciences;

bibliometric mapping; bibliometric studies; bibliometric visualization; bibliometric.; bibliometrics methodology; bibliometrics of social sciences and…; bibliometrics.; bibliometrics...; bibliométrie; bibliometry.

b) Institutional affiliation

All the profiles associated with research centers working on Bibliometrics were also selected. As an example, the profiles with these verified e-mail domains were selected: cwts.leidenuniv.nl, cwts.nl, science-metrix.com, etc.

c) Additional searches

Since there may be some authors working in the discipline who have created a public GSC profile, but who haven't added significant keywords or appropiately filled the institution field in their profile, we also conducted a topic search on *Google Scholar* (using the same keywords as before) as well as a journal search (all the documents indexed in *Google Scholar* published in the following journals: *Scientometrics*, *Journal of Informetrics*, *Research Evaluation*, *Cybermetrics*, and *ISSI proceedings*), with the aim of finding authors we might have missed with the previous two strategies. These searches returned roughly 15,000 documents. Additionally, these searches allowed us to find documents written by authors with no public GSC profile, but which are nonetheless extremely relevant to the discipline.

All these searches were conducted on the 24th of July, 2015.

## 2.2. Filtering and classification of author profiles

Since *Google Scholar Citations* gives authors complete control over how to set their profile (personal information, institutional affiliation, research interests, as well as their scientific production), a systematic manual revision was carried out in order to:

- Detect false positives: authors whose scientific production doesn't have anything to do with this discipline, even though they labeled themselves with one or more of the keywords associated with it.
- Classify authors in two categories:
  - **Core**: authors whose scientific production substantially falls within the field of Bibliometrics.
  - **Related**: authors who have sporadically published bibliometric studies, or whose field of expertise is closely related to Scientometrics (social, political, and economic studies about science), and therefore they can't be strictly considered bibliometricians.

In order to set the limit between the two categories (core and related authors), we decided to consider as "core authors" those who meet a certain criterion: at least half of the documents which contribute to their h-index should fall within the limits of the field of Bibliometrics.

We considered the titles of the documents, as well as the publishing channel where they appeared, focusing our attention in the journals. Our Bradford-like core of journals about Bibliometrics consisted of six journals (*Scientometrics*, *Journal of Informetrics*, *JASIST*, *Research Evaluation*, *Research Policy*, and *Cybermetrics*), followed by other LIS journals which also publish numerous bibliometric studies (*Journal of Information Science*, *Information Processing & Management*, *Journal of Documentation*, *College Research Libraries*, *Library Trends*, *Online Information Review*, *Revista Española de Documentación Científica*, *Aslib Proceedings*, and *El Profesional de la Información*) and lastly, journals devoted to social and political studies about science (*Social Studies of Science*, *Science and Public Policy*, *Minerva*, *Journal of Health Services Research Policy*, *Technological Forecasting and Social Change*, *Science Technology Human Values*, *Environmental Science Policy*, and *Current Science*).

In the end, we selected a total of 814 GSC profiles. 398 of them have been classified as core authors, and the remaining 416 as related authors.

### 2.3. Expansion to a multi-faceted approach: units of scientific analysis

Once we defined the set of authors, we automatically extracted the top 100 most cited documents for each author from their GSC profile. To this set of documents, we added the documents we found on our previous topic and journal searches (the third strategy we used to find authors who work on Bibliometrics).

After deleting duplicates, a set of roughly 41,000 documents remained. In the cases where various versions of the same document were found with different number of citations, the one with the highest citation count was selected. This list was sorted according to the number of citations.

For each of the top 1,000 most cited documents in this list, both the basic bibliographic information (especially the sources: journals and book publishers) and the number of citations according to WoS (*Web of Science*) were collected. For those documents that were not indexed in WoS Core Collection (mostly books), the number of citations in WoS was calculated by searching the document in WoS's Cited Reference Search. By doing this we're trying to highlight the (until now mostly neglected) potential of this tool, which truly offers a wealth of citation data that could be used for the evaluation of non-WoS documents.

Lastly, in the cases when a book is a collective work, the number of citations is the sum of the citations to each of the chapters, in addition to the citations directed to the book as a whole.

### 2.4. Expansion to a multi-faceted approach: social media mirrors

The original 814 authors selected in the previous step (with a public profile created in Google Scholar Citations) were subsequently searched by name in *ResearcherID*, *ResearchGate*, *Mendeley*, and *Twitter*. In the cases where a profile was found in any of these platforms, the main indicators provided by the platform were collected. The data collection from these new academic mirrors was carried out between the 4th and 10th of September, 2015.

Since the maturity of each platform is an important issue to adequately consider its degree of use, the official release date of each platform can be found below:

- *Google Scholar Citations*: a restricted beta release was made on the 20th of July, 2011. It was opened to the general public on the 16th of November, 2011.
- *ResearcherID*: author identification system developed by Thomson Reuters. Released in January 2008.
- *ResearchGate*: academic social network created in May 2008.
- *Mendeley*: social reference manager created in August 2008.
- *Twitter*: online social networking service that enables users to send and read short 140-character messages. Released on the 15th of July, 2006.

The URLs to personal homepages were searched and collected as well. In this case, this information was retrieved from the field “homepage” included in the *Google Scholar Citations* profiles of the authors considered. Since there is not any restriction about the kind of URL an author may use in this field, some authors choose to save the URL of their profile in other platforms (such as ResearchGate), or the URL of the research group, institution, or company they work for, among other cases. In this case, this information was filtered and only personal or institutional websites managed directly by the authors are analyzed.

### 2.5. Author-level metrics: list and scope

All the metrics collected from each of the social media platforms analyzed, as well as their definition and scope can be found in Table 1.

***Table 1. List and explanation of author-level indicators***

| ***Google Scholar Citations*** | |
|---|---|
| **INDICATOR** | **DEFINITION** |
| Citations | Number of citations to all publications. Computed for citations from all years, and citations since 2010 |
| h-index | The largest number h such that h publications have at least h citations. Computed for citations from all years, and citations since 2010 |
| i10 index | Number of publications with at least 10 citations. Computed for citations from all years, and citations since 2010 |
| ***ResearcherID*** | |
| **INDICATOR** | **DEFINITION** |
| Total Articles in Publication List | The number of items in the publication list |
| Articles with Citation Data | Only articles added from *Web of Science Core Collection* can be used to generate citation metrics. The publication list may contain articles from other sources. This value indicates how many articles from the publication list were used to generate the metrics |
| Sum of the Times Cited | The total number of citations to any of the items in the publication list from *Web of Science Core Collection*. The number of citing articles may be smaller than the sum of the times cited because an article may cite more than one item in the set of search results |
| Average Citations per Item | The average number of citing articles for all items in the publication list from *Web of Science Core Collection*. It is the sum of the times cited divided by the number of articles used to generate the metrics |
| h-index | h is the number of articles greater than h that have at least h citations. For example, an h-index of 20 means that there are 20 items that have 20 citations or more |
| ***ResearchGate*** | |
| **INDICATOR** | **DEFINITION** |
| RG Score | It's a metric that measures scientific reputation based on how an author's research is received by his/her peers. The exact method to calculate this metric has not been made public, but it takes into account how many times the contributions (papers, data, etc.) an author uploads to *ResearchGate* are visited and downloaded, and also by whom (reputation) |
| Publications | Total number of publications an author has added to his/her profile in *ResearchGate* (full-text or no) |
| Views | Total number of times an author's contributions to *ResearchGate* have been visualized. This indicator has recently been combined with the "Downloads" indicator to form the new "Reads" indicator, but the data collection for this product was made before this change came into effect |
| Downloads | Total number of times an author's contributions to *ResearchGate* have been downloaded. This indicator has recently been combined with the "Views" indicator to form the new "Reads" indicator, but the data collection for this product was made before this change came into effect |
| Citations | Total number of citations to the documents uploaded to the profile. *ResearchGate* generates its own citation database, and they warn this number might not be exhaustive |
| Impact Points | Sum of the JCR impact factors of the journals where the author has published articles |
| Profile views | Number of times the author's profile has been visited |
| Following | Number of *ResearchGate* users the author follows (the author will receive notifications when those users upload new material to *ResearchGate*) |
| Followers | Number of *ResearchGate* users who follow the author (those *ResearchGate* will receive notifications when the author uploads new materials to *ResearchGate*) |
| ***Mendeley*** | |
| **INDICATOR** | **DEFINITION** |
| Readers | This number represents the total number of times a *Mendeley* user has added a document by this author to his/her personal library |
| Publications | Number of publications the author has uploaded to *Mendeley* and classified as "My Publications" |
| Followers | Number of *Mendeley* users who follow the author |
| Following | Number of *Mendeley* users the author follows |
| ***Twitter*** | |
| **INDICATOR** | **DEFINITION** |
| Tweets | Total number of tweets an author has published according to his profile |
| Followers | Number of *Twitter* users who follow the tweets published by the author |
| Following | Number of *Twitter* users the author follows |
| Days registered | Number of days since the author created an account on *Twitter* |
| Sum Retweets | Number of Retweets obtained for the author. |
| H Retweets | An author has a h-Retweet of "n" when "n" of its tweets has achieved at least "n" Retweets. |

## 2.6. Limitations

Projects of a bibliographic nature like this one can't ever reach perfection, and it is entirely possible that we may have missed relevant authors. The criteria for selecting the authors were two: first, the existence of a public GSC profile about the author by 24/07/2015 (when the data collection was made), and second, that the author works on the fields of Bibliometrics, Scientometrics, Informetrics, Webometrics, or Altmetrics.

We're completely aware that these lists don't include all the researchers in the area, since some haven't created a profile, or they haven't made it public. We should note that we made an exception with Eugene Garfield, one of the fathers of Bibliometrics. Despite the fact that he doesn't have a public GSC profile, we manually searched his production on *Google Scholar* and computed the same indicators GSC displays. We believe this Working Paper would be incomplete without him.

We strongly encourage researchers without a GSC profile, and especially those who have made important contributions to the development of this field, to bring together the scattered bibliographic information *Google Scholar* has already compiled about their works. Sharing this information would not only greatly benefit their online visibility; it would also be very useful to the rest of the scientific community.

## 2.7. Statistical analysis

Spearman correlation ($\alpha = 0.05$) was applied to all 31 metrics considered in each of the platforms (excluding personal webpages), and finally a Principal Component Analysis (Spearman similarity with varimax rotation of axes and uniform weighting) was applied in order to reveal the relationships among metrics and platforms as well as the possible existence of metric clusters.

# 3. RESULTS

## 3.1. The actors of Bibliometrics as a discipline, according to Google Scholar Citations: authors, documents, journals and publishers

### a) Authors

By analyzing the list of most influential authors of the discipline (Table 2) we noticed that the most prominent positions (top ten) include the founders of the discipline (Price and Garfield) and the most influential bibliometricians, almost all of them holders of the Price medal (all except Chen), a prize that recognizes scientists who have contributed with their work to the development of Bibliometrics.

*Table 2. Top 25 influential core authors in Bibliometrics according to Google Scholar Citations*

| AUTHOR | GS CITATIONS | H-INDEX |
|---|---|---|
| Loet Leydesdorff | **26,484** | 73 |
| Eugene Garfield | **22,622** | 55 |
| Mike Thelwall | **13,840** | 61 |
| Derek J. de Solla Price | **13,263** | 33 |
| Francis Narin | **11,297** | 45 |
| Wolfgang Glänzel | **10,796** | 54 |
| Ronald Rousseau | **9,570** | 42 |
| Chaomei Chen | **9,512** | 43 |
| Anthony (Ton) F.J. van Raan | **9,200** | 53 |
| Ben R Martin | **8,975** | 39 |
| András Schubert | **8,655** | 45 |
| Peter Ingwersen | **8,356** | 35 |
| Henk F. Moed | **8,256** | 46 |
| Blaise Cronin | **7,347** | 43 |
| Henry Small | **7,307** | 32 |
| Tibor Braun | **7,231** | 41 |
| Vasily V. Nalimov | **6,343** | 31 |
| Lutz Bornmann | **6,108** | 40 |
| Belver C. Griffith | **5,695** | 26 |
| Howard D. White | **5,569** | 30 |
| Johan Bollen | **5,394** | 33 |
| Katy Borner | **5,326** | 31 |
| Félix de Moya Anegón | **5,074** | 35 |
| Koenraad Debackere | **4,933** | 32 |
| Jose Maria López Piñero | **4,823** | 31 |

Bibliometrics received a decisive boost from the personality and the work of both Price and Garfield, who can be considered the fathers of this discipline. On the one hand, Price, armed with the theoretical foundations laid by John Desmond Bernal and Robert K. Merton, set out to systematically apply quantitative techniques to the History and social studies of Science, developing the theoretical foundations of Scientometrics, born from the combination of the Sociology of science, History, Philosophy of science, and Information science. This approach is characterized by the analysis of the life and activity of Science and scientists from a quantitative perspective. The numbers were used to characterize the production of knowledge and scientists' lives: what they create and produce, with whom they relate to, the sources they used, and the impact and influence they provide/receive to/from other scientists, etc.

On the other hand, Garfield made possible that Bibliometrics became a reality (Mccain 2010; Bensman, 2007): the creation of the "citation index" made possible the quantification of scientific activity through its main output: the publications and citations they generate. Since then, citation analysis and all its variants have become the most widespread analysis technique of this new specialty (this is evidenced by the significant presence of highly cited documents that deal with this topic). Garfield defined the phenotype of the discipline: technology (the basis for the storage and circulation of information) is at the heart of all its tools. That is, Bibliometrics will evolve at the same rate the technologies of information and communication do.

The map of Bibliometrics can also be discerned by analyzing the rest of the authors in the list: the Hungarian school (both Eastern Europe and Russia, like Nalimov), the Dutch school (with its various branches in Leiden and Amsterdam), the Belgian school (with Egghe and Rousseau), the North American School (Small, Griffith, and White), the Spanish school (with López Piñero, Spanish translator of Price's work, and the one who introduced Bibliometrics in Spain), and the new authors that represent the technological transformation of the discipline (mainly Thelwall).

### b) Documents

An analysis of the list of the 25 most cited documents according to Google Scholar (Table 3) reveals several issues:

- The importance of the documents that first introduce new techniques and citation-based indicators, like the ones by Hirsch (3rd), Garfield (9th and 10th), Small (12th), Egghe (23rd), and Griffith and White (37th). Among them we find the most widely known indicator in Bibliometrics (the impact factor) and the one that has come to replace it while extending its capabilities (h-index).
- The excellence both in the work in which Hirsch proposes the h-index and in the articles about the impact factor highlights the strong orientation of Bibliometrics towards evaluation in general and the assessment of the performance of individuals, journals, and institutions... This reveals a clear link between Bibliometrics and Science policy, and explains the use of bibliometric indicators and other bibliometric tools by policymakers.
- As we would expect, among the most cited documents we find texts that have served as textbooks for the discipline (written by Moed, Van Raan, Eghhe, Rousseau, etc.).
- The anomalous institutionalization process of the discipline. The main "bibliometric laws" which still hold true today where established at the dawn of the discipline, even before it was fully instituted (Lotka, Zipf, Bradford), and were developed by authors working outside the discipline. The same happened with the proposal of the h-index by Hirsch, elaborated by this physician in his “leisure time”. Bibliometrics is often revolutionized from outside Bibliometrics.

- The great relevance of some topics such as the "Triple Helix" by Leydersdorff, or the social networks by Barabási, which make a big impact outside the borders of our discipline (Management and Economy in the first case, and sociometrics and computer science in the second).

***Table 3. Top 25 most influential documents in Bibliometrics according to Google Scholar Citations***

| TITLE | AUTHORS | SOURCE | YEAR | GS CITATIONS |
|---|---|---|---|---|
| Little science, big science | de Solla Price | Columbia University Press | 1963 | **5,410** |
| An index to quantify an individual's scientific research output | Hirsch | PNAS | 2005 | **4,860** |
| The dynamics of innovation: from National Systems and "Mode 2" to a Triple Helix of university-industry-government relations | Etzkowitz & Leydesdorff | Research Policy | 2000 | **4,414** |
| Universities and the global knowledge economy: a triple helix of university-industry-government relations | Etzkowitz & Leydesdorff | Pinter Press | 1997 | **2,585** |
| Handbook of Quantitative Science and Technology Research: The Use of Publication and Patent Statistics in Studies of S&T Systems | Moed; Glänzel & Schmoch (ed.) | Springer | 2005 | **2,261** |
| Citation analysis as a tool in journal evaluation. Journals can be ranked by frequency and impact of citations for science policy studies | Garfield | Science | 1972 | **2,166** |
| Citation indexing: Its theory and application in science, technology, and humanities | Garfield | Wiley | 1979 | **2,130** |
| The frequency distribution of scientific productivity | Lotka | Journal of Washington Academy Sciences | 1926 | **2,090** |
| Co-citation in the scientific literature: A new measure of the relationship between two documents | Small | JASIS | 1973 | **1,988** |
| Links and impacts: The influence of public research on industrial R&D | Cohen; Nelson & Walsh | Management Science | 2002 | **1,881** |
| Evolution of the social network of scientific collaborations | Barabasi; Jeong; Neda; Ravasz; Schubert & Vicsek | Physica A | 2002 | **1,851** |
| Citation indexes for science. A new dimension in documentation through association of ideas | Garfield | Science | 1955 | **1,783** |
| What is research collaboration? | Katz & Martin | Research Policy | 1997 | **1,591** |

| TITLE | AUTHORS | SOURCE | YEAR | GS CITATIONS |
|---|---|---|---|---|
| Handbook of quantitative studies of science and technology | Van Raan (ed.) | North-Holland | 1988 | **1,510** |
| The history and meaning of the journal impact factor | Garfield | JAMA | 2006 | **1,487** |
| The increasing linkage between US technology and public science | Narin; Hamilton & Olivastro | Research Policy | 1997 | **1,211** |
| A general theory of bibliometric and other cumulative advantage processes | de Solla Price | JASIST | 1976 | **1,148** |
| Statistical bibliography or bibliometrics? | Pritchard | Journal of Documentation | 1969 | **1,134** |
| Theory and practise of the g-index | Egghe | Scientometrics | 2006 | **1,113** |
| The Web of knowledge: a Festschrift in honor of Eugene Garfield | Garfield; Cronin & Atkins (ed). | Information Today | 2000 | **1,102** |
| Visualizing a discipline: An author co-citation analysis of information science, 1972-1995 | White & McCain | JASIS | 1998 | **1,100** |
| CiteSpace II: Detecting and visualizing emerging trends and transient patterns in scientific literature | Chen | JASIST | 2006 | **1,083** |
| Citation analysis in research evaluation | Moed | Springer | 2005 | **1,060** |
| Citation frequency and the value of patented inventions | Harhoff; Narin; Scherer & Vopel | Review of Economics and Statistics | 1999 | **1,023** |
| Maps of random walks on complex networks reveal community structure | Rosvall & Bergstrom | PNAS | 2008 | **992** |

If we pay attention to the distribution of documents according to their typology (Figure 3), the journal article stands out overwhelmingly (89% of all 1,069 documents processed), showing that formal papers published in peer reviewed journals stand as the main scientific vehicles in this social science discipline.

***Figure 3. Distribution of highly cited documents in Bibliometrics according to Google Scholar citations (n= 1,069)***

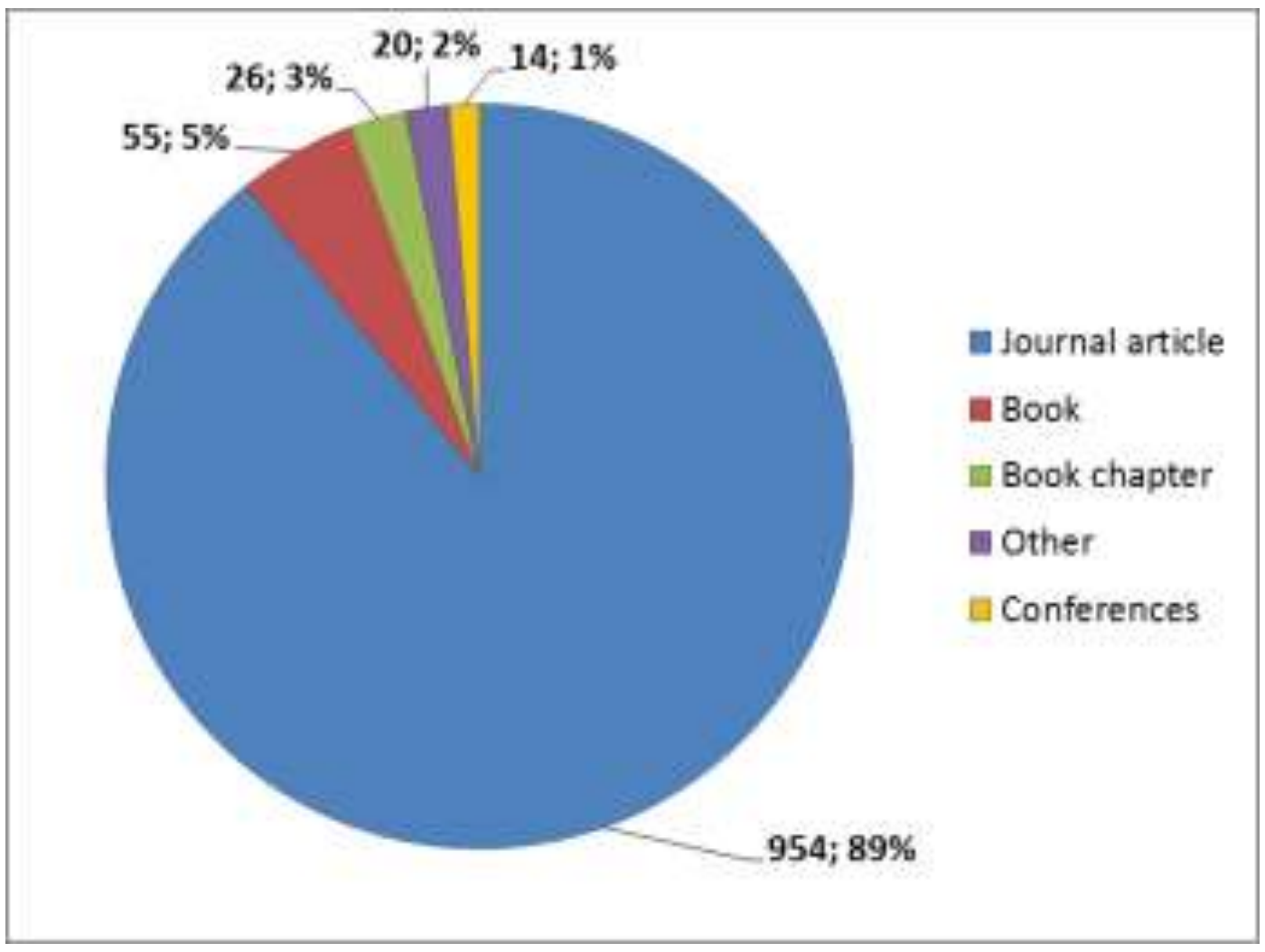

The presence of books (5%) is smaller, but this figure may be misleading. An analysis of the top highly cited documents according to Google Scholar citations shows that within the top 25 documents, 8 of them are books (14 within the top 50). Obviously, the number of published books is lower than the number of articles. The presence of documents from the remaining categories is lower: Book chapters (3%), and other material (including dissertation theses, reports, etc.; 2%). Lastly, the results obtained for conference proceedings (1%) reveal a low impact of this scientific communication channel.

**c) Journals**

The third unit analyzed is the journals in which highly cited documents have been published (i.e., considering only the top 1,000 most cited documents). In Table 4 we provide the top 25 journals according to the number of highly cited documents published. Additionally, we show the total number of citations received by these articles, the percentage of citations per article (C/A), the percentage of highly cited documents in the sample (HCA) and the distribution of citations.

*Scientometrics* is the journal with more articles published within the 1,000 most cited documents (284 articles). It is thus the most influential journal in the discipline. Its birth in 1978 was a milestone in the process of institutionalization of the discipline.

The second place is occupied by *JASIST* (137 articles). This fact shows the important role of this journal in Bibliometrics, although its scope is broader. This journal has maintained since its inception a strong link between Information Science and Bibliometrics, though some authors have noticed a slight specialization towards Bibliometrics over time (Nicolaisen & Frandsen, 2015).

*Journal of informetrics*, focused exclusively on Bibliometrics, Scientometrics, Webometrics, and Altmetrics, appears in the fourth position (36 articles). The young age of this journal (it was created in 2007) explains why there isn’t a greater number of articles published in this journal among the most cited documents in the discipline.

The connection between Library and Information Science and Bibliometrics is noticeable through the presence of other important LIS journals in the list, such as *Journal of Documentation*, *Journal of Information Science*, *Library Trends*, or *Aslib Proceedings*. This connection has been a matter of public record for a long time now (White & McCain 1998; Larivière, Sugimoto, Cronin 2012, Larivière 2012).

Its connections with the field of web technologies from an information science perspective is strongly marked as well (*Cybermetrics*, *Online Information Review*). Additionally, we can see that journals oriented towards the Social Studies of Science (such as *Research Policy*, *Social Studies of Science*, and *Science and Public Policy*) also have strong ties to Bibliometrics.

Lastly, the role of multidisciplinary journals (such as *Nature*, *Science*, *PNAS* or *PLoS One*) should not be forgotten. If we analyze the number of citations instead of the number of articles published, we find the same first three journals occupying the first positions (*Scientometrics*, *JASIST*, and *Research Policy*), but the data also shows a great impact of articles published outside the core journals of the discipline. *Science* gets 9,219 citations from only 8 articles whereas *PNAS* achieves 7,642 citations from 9 articles, and *PLoS One* gets 2,376 citations from 13 articles (the figures for *Nature* are lower, with “only” 1,871 citations from 10 articles).

***Table 4. Top 25 most influential journals in Bibliometrics according to Google Scholar Citations***

| JOURNAL | ARTICLES | CITATIONS | C/A | HCA (%) | CITATIONS (%) |
|---|---|---|---|---|---|
| Scientometrics | **284** | 44,384 | 156 | 29.8 | 22.5 |
| JASIST | **137** | 27,021 | 197 | 14.4 | 13.7 |
| Research Policy | **57** | 18,866 | 330 | 6.0 | 9.6 |
| Journal of Informetrics | **36** | 5,052 | 140 | 3.8 | 2.6 |
| Journal of Documentation | **25** | 5,538 | 221 | 2.6 | 2.8 |
| Information Processing & Management | **24** | 4,404 | 183 | 2.5 | 2.2 |
| Journal of Information Science | **20** | 3,815 | 190 | 2.1 | 1.9 |
| Research Evaluation | **18** | 2,126 | 118 | 1.9 | 1.1 |
| ARIST | **14** | 3,621 | 258 | 1.5 | 1.8 |
| Social Studies of Science | **13** | 3,204 | 246 | 1.4 | 1.6 |
| Science and Public Policy | **13** | 2,875 | 221 | 1.4 | 1.5 |
| Plos One | **13** | 2,376 | 182 | 1.4 | 1.2 |
| Nature | **10** | 1,871 | 187 | 1.0 | 1.0 |
| Current Contents | **10** | 1,696 | 169 | 1.0 | 0.9 |
| PNAS | **9** | 7,642 | 849 | 0.9 | 3.9 |
| Science | **8** | 9,219 | 1,152 | 0.8 | 4.7 |
| Library Trends | **7** | 1,230 | 175 | 0.7 | 0.6 |
| Medicina Clinica | **6** | 958 | 159 | 0.6 | 0.5 |
| Online Information Review | **6** | 806 | 134 | 0.6 | 0.4 |
| Science Technology & Human Values | **5** | 946 | 189 | 0.5 | 0.5 |
| Aslib Proceedings | **5** | 765 | 153 | 0.5 | 0.4 |
| Cybermetrics | **5** | 627 | 125 | 0.5 | 0.3 |
| American Psychologist | **4** | 1,026 | 256 | 0,4 | 0,5 |
| World Patent Information | **4** | 726 | 181 | 0.4 | 0.4 |
| Ethics in Science and Environmental Politics | **4** | 687 | 171 | 0.4 | 0.3 |

## d) Book publishers

The last unit of analysis is the book publishers. Table 5 shows the top 20 publishers according to the percentage of highly cited documents (top 1,000). Additionally, the number of documents, citations (total and percentage of citations respect to the total) and citations per document are offered.

The first position is occupied by Springer, with 10 documents positioned within the set of highly cited books, and receiving 5,766 citations (14.3% of all citations to book publishers). Information Today (10.9%) and Wiley (9.1%) stand on the second and third position respectively.

***Table 5. Top 25 most influential publishers in Bibliometrics according to Google Scholar Citations***

| PUBLISHER | HC | HC (%) | CITATIONS | CITATIONS (%) | C/A |
|---|---|---|---|---|---|
| **Springer** | 10 | **18,2** | 5,766 | 14,3 | 576.60 |
| **Information Today** | 6 | **10,9** | 1,635 | 4,0 | 272.50 |
| **Wiley** | 5 | **9,1** | 3,121 | 7,7 | 624.20 |
| **Lexington** | 4 | **7,3** | 1,627 | 4,0 | 406.75 |
| **Sage** | 4 | **7,3** | 1,324 | 3,3 | 331.00 |
| **UFMG** | 4 | **7,3** | 845 | 2,1 | 211.25 |
| **University of Chicago Press** | 3 | **5,5** | 6,874 | 17,0 | 2,291.33 |
| **Russell Sage Foundation** | 3 | **5,5** | 3,836 | 9,5 | 1,278.67 |
| **North-Holland** | 3 | **5,5** | 2,130 | 5,3 | 710.00 |
| **Blackwell** | 2 | **3,6** | 1,132 | 2,8 | 566.00 |
| **Elsevier** | 2 | **3,6** | 1,071 | 2,7 | 535.50 |
| **Taylor Graham** | 2 | **3,6** | 688 | 1,7 | 344.00 |
| **Scarecrow Press** | 2 | **3,6** | 416 | 1,0 | 208.00 |
| **ISSI** | 2 | **3,6** | 276 | 0,7 | 138.00 |
| **Ablex** | 2 | **3,6** | 193 | 0,5 | 96.50 |
| **FECYT** | 2 | **3,6** | 193 | 0,5 | 96.50 |
| **Columbia University Press** | 1 | **1,8** | 5,410 | 13,4 | 5,410.00 |
| **Pinter Press** | 1 | **1,8** | 2,585 | 6,4 | 2,585.00 |
| **Yale University Press** | 1 | **1,8** | 936 | 2,3 | 936.00 |
| **MIT Press** | 1 | **1,8** | 710 | 1,8 | 710.00 |

We can observe that all publishers achieve high numbers of citations per document. In this case, we should highlight the performance of university presses (such as University of Chicago, Columbia, Yale, or MIT), with a very low presence in terms of productivity but an impressive impact in the number of citations. The ability to attract well-established authors in order to edit specialized books makes a great difference in book publisher rankings.

## 3.2. Online presence of the bibliometric community

Scientists traditionally communicated with their communities both through informal means (letters, meetings, seminars, conferences ...) and formal means (books, journal articles, patents, patents, etc.), and in both of them the scope of these communications was limited by the printed technology in which the contents were transmitted. Today, since the birth of the Web, which brought the chance to create personal pages, and with the emergence of academic social networks, researchers can display their work through a rich variety of channels and electronic formats.

Studies of the level of web presence and impact of scientists' through their personal websites have already been carried out. Barjak, Li & Thelwall (2007) analyzed data from 456 scientists from five scientific disciplines in six European countries, whereas Mas-Bleda, Aguillo, (2013) and Más-Bleda et al (2014) put their focus on 1,498 highly cited researchers working at European institutions, distributed in 22 different countries, using data extracted from the ISIHighlyCited.com database.

In the field of Bibliometrics, the pioneer work by Haustein et al (2014) should also be highlighted. In this study, 1,136 documents authored by the 57 presenters of the 2010 STI conference in Leiden (57 researchers, who together had authored 1,136 papers) were collected using WoS and Scopus. After this, the scholarly and professional social media presence of these authors in several platforms was measured (Google Scholar Citations, LinkedIn, Twitter, Academia.edu, ResearchGate and ORCID).

In this work we intend to expand this sample by considering the social presence of the whole bibliometric community as well as other researchers who are related to the discipline in some way. A total of 814 researchers (398 bibliometricians and 416 researchers who have sporadically published bibliometric studies) have been analyzed.

In Table 6 we find the distribution of authors according to the number of platforms in which they have created a personal profile, regardless of their impact or the degree to which these profiles are updated. We highlight the following points:

- The degree of social presence is high. All 814 authors have at least a personal profile created in one platform; 14.7% of the authors are visible in only one platform.
- Authors with two (19.1%), three (23.5%), or four (21.1%) profiles are the more numerous groups.
- No significant differences between core and related authors are found.
- There is a small group of authors (6.2%) with high media visibility (presence in all social media analyzed), being among them some of the most influential bibliometricians (such as Loet Leydesdorff, Mike Thelwall, Chaomei Chen, Lutz Bornmann, Félix de Moya Anegón, Katy Borner, Judit Bar-Ilan, Nees Jan van Eck, or Isidro F. Aguillo, among others).

*Table 6. Social presence of the bibliometric community*

| NUMBER OF PLATFORMS | AUTHORS | | |
|---|---|---|---|
| | CORE | RELATED | TOTAL |
| **6** | 32 | 19 | 51 |
| **5** | 72 | 51 | 123 |
| **4** | 76 | 96 | 172 |
| **3** | 80 | 112 | 192 |
| **2** | 78 | 78 | 156 |
| **1** | 60 | 60 | 120 |
| **TOTAL** | **398** | **416** | **814** |

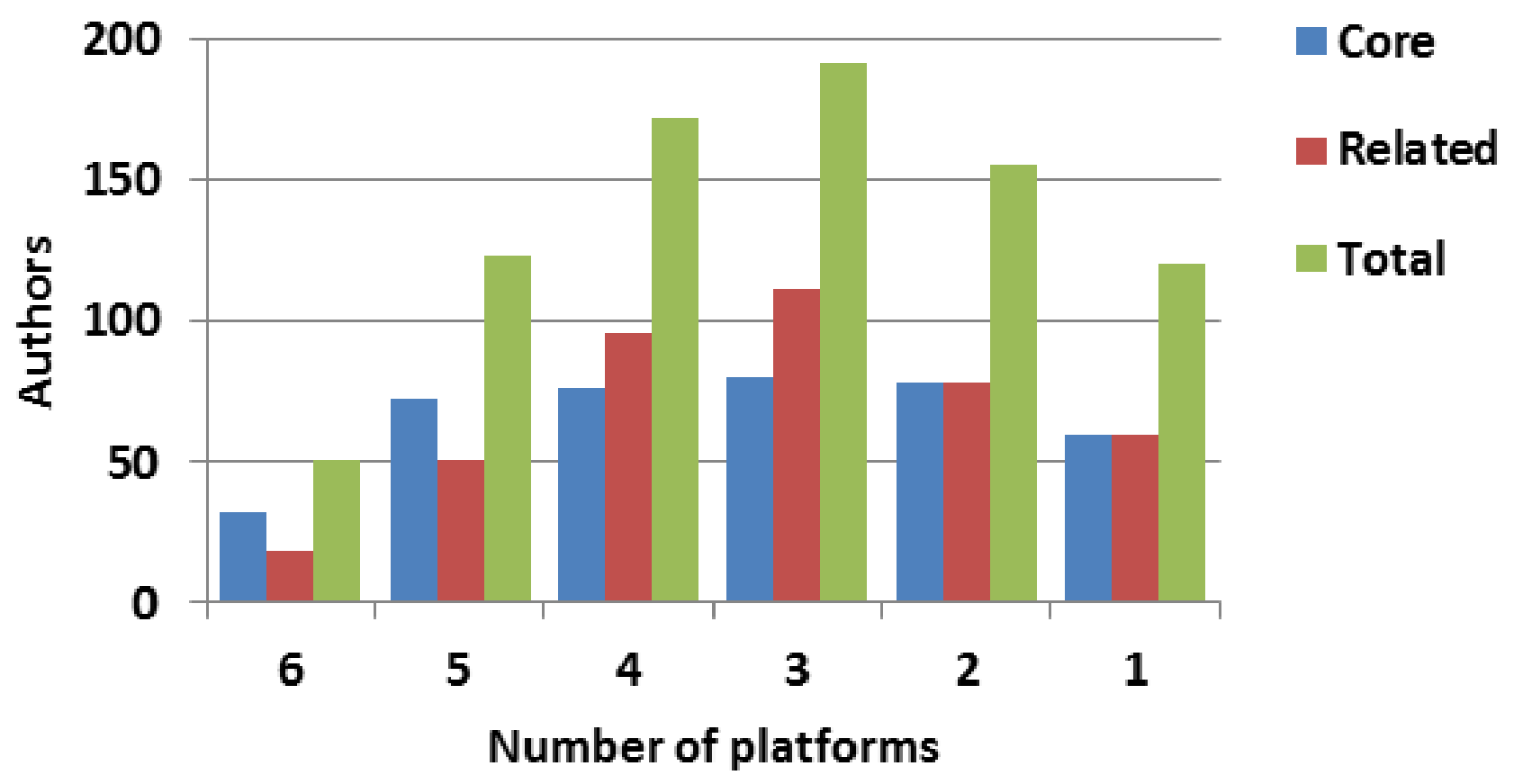

The use of each specific social platform is shown in Table 7. The main results derived from these data are the following:

- *ResearchGate* is (after *Google Scholar Citations*) the second most used platform by these authors (66.7%), followed at some distance by *Mendeley* (41.28%) and homepages (41.15%).
- The number of *Mendeley* profiles is high, although this data by itself is misleading, since 17.1% of the profiles (68 out of 397) are basically empty. *ResearcherID* is also affected by this issue (34.45% of the profiles are empty); as is *Twitter* (47% of the 240 authors with a *Twitter* profile have published less than 100 *tweets*).
- *ResearcherID* presents a wider acceptance among core authors (45.7%) than related authors (35.1%)
- *Twitter* is the least used platform, since only 33.17% of core authors (and 25.96% of related authors) have created a *Twitter* profile.
- Personal homepages are widely used by authors, although this denomination covers a wide range of different website typologies (personal websites outside institutions, institutional websites not managed by authors). The use of social platforms as personal sites is common (22 authors considered their profiles in other academic social sites such as ResearchGate, Academia.edu, Mendeley, and ImpactStory as their personal websites).
- Core and related authors present similar behavior as regards their presence on these social platforms, although there is a slightly higher rate of core authors on *Twitter*, *ResearcherID,* and *Mendeley* than there is of related authors.

***Table 7. Degree of use of social platforms by type of author***

| WEB PLATFORMS | AUTHORS | | | | | |
|---|---|---|---|---|---|---|
| | CORE | % | RELATED | % | TOTAL | % |
| *** Google Scholar Citations** | 398 | 100 | 416 | 100 | **814** | 100 |
| **ResearcherGate** | 260 | 65.33 | 283 | 68.03 | **543** | 66.71 |
| **Mendeley** | 171 | 42.96 | 165 | 39.66 | **336** | 41.28 |
| **** Homepage** | 158 | 39.69 | 177 | 42.54 | **335** | 41.15 |
| **ResearcherID** | 182 | 45.73 | 146 | 35.10 | **328** | 40.29 |
| **Twitter** | 132 | 33.17 | 108 | 25.96 | **240** | 29.48 |

* All authors in the sample have a profile in GSC. ** *ResearchGate* and *Academia.edu* URLs were discarded.

Figure 4 shows the combination of profiles used by the authors (core and related) of the bibliometric community. It should be reminded that all authors in our sample have Google Scholar Citation profiles (this was the main selection criteria).

Personal webpages have been omitted from this analysis since they represent another dimension of web presence, different from those offered by social platforms and academic profiles.

***Figure 4. Combination of profiles used by the bibliometricians in our sample***

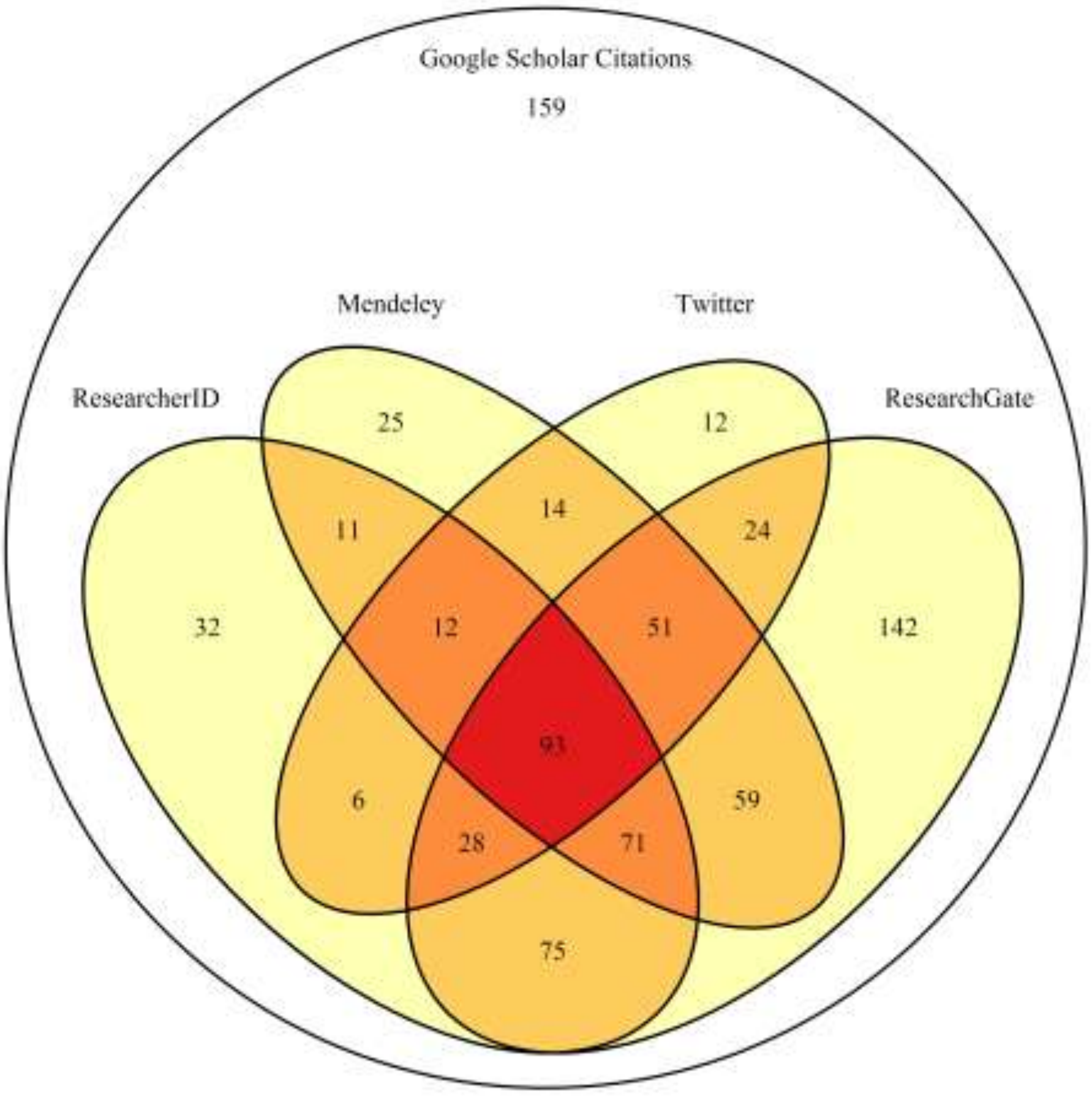

As we can see in Figure 4, there is great number of researchers who only have a profile in *Google Scholar Citations* (159). There are also many authors who only have a profile in GSC and *ResearchGate* (142). The number of researchers who have an account in all the platforms analyzed in this study (GSC, *ResearcherID*, *Mendeley*, *Twitter*, and *ResearchGate*) is 93 (11.4% of our sample).

The remaining combinations seem to be more unusual. For example, there are only 12 authors who use only GSC and *Twitter*, and 14 authors who use only GSC, *Mendeley*, and *Twitter*. In a similar manner, there are only 11 authors who use only GSC, *ResearcherID*, and *Mendeley*.

These results are similar to the ones offered by Van Noorden (2014) about the presence of scientists in social networks and the provisional results by Bosman and Kramer (2015).

Despite the fact that this sample suffers from a bias in favor of *Google Scholar Citations* because of how the data were collected, there is no doubt that GSC is the platform authors currently prefer to display their publications, followed at a distance by *ResearchGate*, but a distance that is increasingly shorter. 66.7% of the authors with a GSC profile have also a *ResearchGate* profile. This is significant enough, although these results must be tempered by the degree of use and update frequency of each platform, aspects which will be discussed later in greater detail.

These results should be especially contextualized within the bibliometric community, which undoubtedly has a certain bias towards using these platforms, because these platforms are sometimes objects of study themselves. Differences in the degree of presence on social platforms in different fields of knowledge should be expected, as González-Díaz, Iglesias-García, and Codina (2015) have recently proved in their analysis of the discipline of Communication.

### 3.3. Comparing social platform metrics: from citations to followers

After analyzing the academic output and impact for the bibliometric community using *Google Scholar Citations*, and describing the preferences of the members of this scientific community for social interaction in the Web, in this section we are going to analyze the correlation between these metrics. Firstly, all metrics associated with *Google Scholar Citation* profiles, and secondly, all metrics associated and offered by each of the social platforms analyzed (*Mendeley*, *ResearcherID*, *ResearchGate*, and *Twitter*). Personal webpages have been excluded from this analysis.

By way of illustration, in Table 8 we show the median of the main metrics evaluated so that we can compare the performance or size of similar indicators in each web platform. In this sense, we highlight the following issues:

- Regarding the “Total citations received”, the higher median value corresponds to *Google Scholar* (156), followed by *ResearchGate* (85) and *ResearcherID* (63).
- As to the h-index, Google Scholar obtains a score of 6; whereas in *ResearcherID* this value is lower (4).
- Regarding academic output, *ResearchGate* achieves the first position (27), followed at a distance by *ResearcherID* (15) and *Mendeley* (9). The number of records stored in each *Google Scholar Citation* profile is not available in this work.

- Regarding the social interaction features ("following / followed by"), users in both *ResearchGate* and *Mendeley* show a slightly passive behavior: users tend to be followed by many people, but they do not follow many other users. Interestingly, the opposite behavior is found in *Twitter*, where scholars tend to follow many users, but it seems harder to be followed by others. Since *ResearchGate* and *Mendeley* deal exclusively with academic audiences, a logical explanation may be that respected scholars who create an account are widely followed, but they do not tend to follow other users. Nonetheless, in the open space defined by *Twitter*, the situation is just the opposite: gaining followers implies an active participation in the platform.

***Table 8. Median of principal metrics***

| SOURCE | METRIC | MEDIAN |
|---|---|---|
| **Google Scholar (n=811)** | Citations_total | **156** |
| | Citations_last5 | **117** |
| | H-index_total | **6** |
| | H-index_last5 | **5** |
| | i10_total | **4** |
| | i10_last5 | **3** |
| **ResearcherID (n=275)** | Total_articles | **15** |
| | Articles_cited | **11** |
| | Times_cited | **63** |
| | Average_citations | **5.75** |
| | H-index | **4** |
| **ResearchGate (n=515)** | RG Score | **13.82** |
| | Publications | **27** |
| | Impact_points | **12.97** |
| | Followers | **38** |
| | Following | **23** |
| | Downloads | **802** |
| | Views | **1845** |
| | Citations | **85** |
| | Profile_views | **696** |
| **Mendeley (n= 185)** | Publications | **9** |
| | Readers | **93** |
| | Followers | **3** |
| | Following | **2** |
| **Twitter (n=226)** | Tweets | **153.5** |
| | Followers | **99** |
| | Following | **130** |

In Table 9 we show all correlations achieved among each of the 31 metrics considered in this study (α= 0.05), whereas in Figure 5 we show the results of a Principal Component Analysis (PCA).

***Figure 5. Principal Component Analysis for 31 metrics associated with bibliometricians' social platform profiles***

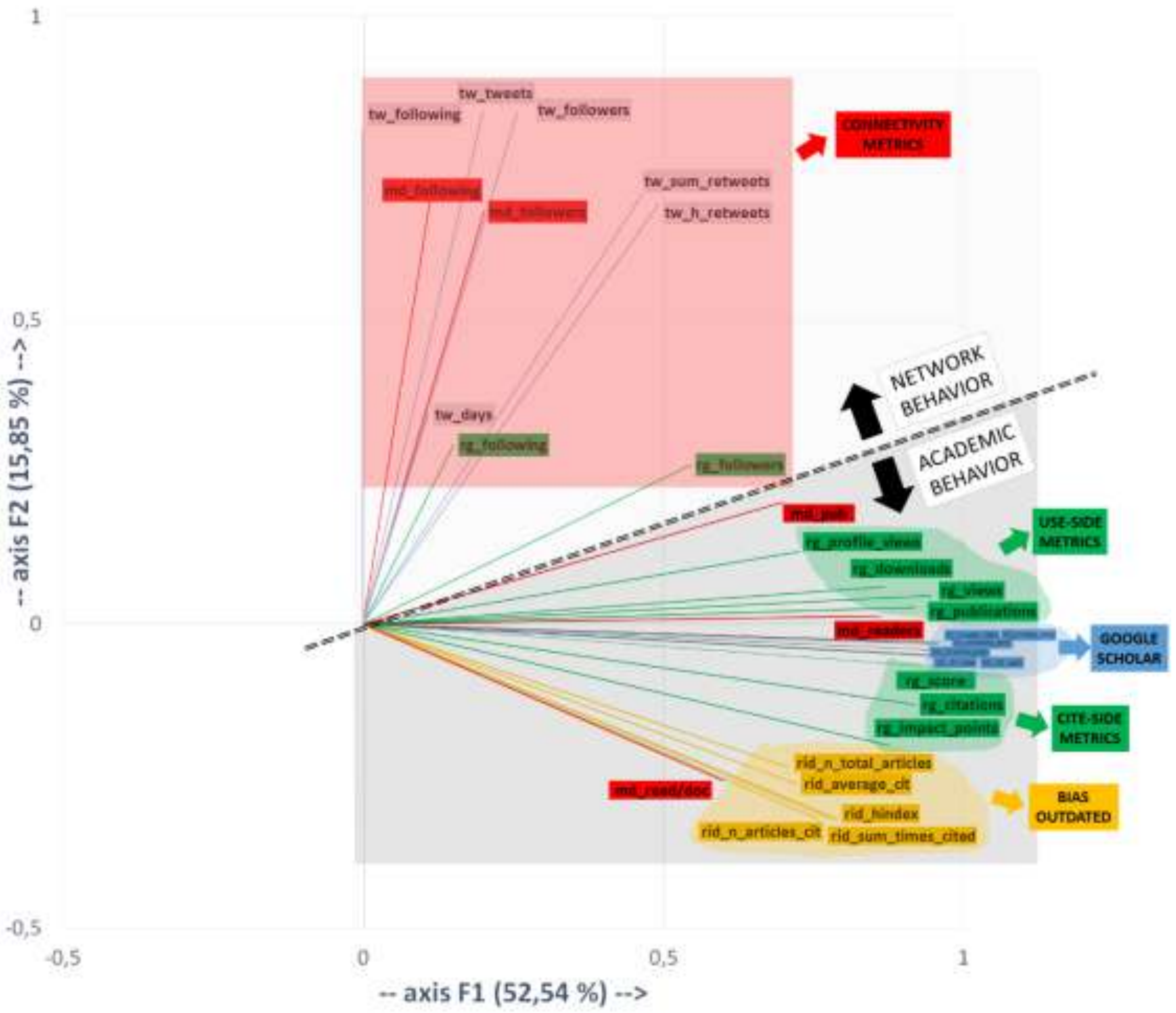

The main results are:

- We find two clear dimensions: at the top we can see all metrics related to connectivity and popularity (followers), and at the bottom, all metrics related to academic performance. This second group can further be divided into usage metrics (views and downloads) and citation metrics. *ResearchGate* provides examples for these two faces of academic performance, since *Google Scholar Citations* profiles do not offer data about downloads or reads.
- All metrics provided by *Google Scholar* (both citations and h-index) correlate strongly among themselves.
- We find a clear separation between the usage (views and downloads) and citation metrics (Citations, Impact Points) provided by *ResearchGate*. The RG Score for example displays a high correlation to metrics from *Google Scholar Citations*: i.e. total citations (r= 0.89) and the h-index (r= 0.92).
- The number of readers in *Mendeley* is connected to the usage metrics offered by *ResearchGate*, and strongly correlates to *Google's* total citations (r= 0.77), Google's h-index (r= 0.82), and the RG Score (r= 0.75). The number of documents in *Mendeley* is far from the *Mendeley* readers in this PCA, probably because *Mendeley* profiles aren't updated as regularly as GSC profiles. Of course, this also affects the combined metric "readers per document".

- Indicators from *ResearcherID* strongly correlate among themselves, but are slightly separated from other citation metrics (those from *Google Scholar* and *ResearchGate*). This issue can probably be explained by the low regularity with which *ResearcherID* profiles are updated. In view of the results, this isolation may be used as a mechanism to check the "currentness" (or lack thereof) of a profile in *ResearcherID*.
- All metrics associated with the number of followers (all *Twitter* metrics and their counterparts in *ResearchGate* and *Mendeley*) correlate among themselves, and are separated from the citation metrics. Curiously enough, the number of followers offered by *ResearchGate* is, within the group of connectivity metrics, the one which is closest to the usage metrics, serving in fact as a bridge between the two groups. This may mean that networking metrics from academic social networks correlate better with usage metrics than networking metrics from *Twitter*. *Mendeley*'s networking metrics, however, are placed closer to *Twitter*'s metrics.
- The impact of *Tweets* (measured by Retweets) is closer to the academic side. In any case, their correlation with impact measures is statistically significant ($\alpha$=0.05). The correlation of Sum Retweets and H-Retweets with *Google Scholar* total citations is 0.44 and 0.45 respectively.
- The number of days that a *Twitter* account has been active does not seem to correlate with any other *Twitter* metric. Unlike in online marketing, time is not a critic factor to achieve followers. Academic prestige and activity (number of *Tweets* tweeted) may be the most important parameters to achieve a great number of Twitter followers.

*Table 9. Correlation analysis (Spearman) for 31 metrics associated with bibliometricians' social platform profiles*

| PLATFORM | | GOOGLE SCHOLAR | | | | | | RESEARCHERID | | | | | RESEARCHGATE | | | | | | | | | MENDELEY | | | | | TWITTER | | | | | |
|---|---|---|---|---|---|---|---|---|---|---|---|---|---|---|---|---|---|---|---|---|---|---|---|---|---|---|---|---|---|---|---|---|
| | | 1 | 2 | 3 | 4 | 5 | 6 | 7 | 8 | 9 | 10 | 11 | 12 | 13 | 14 | 15 | 16 | 17 | 18 | 19 | 20 | 21 | 22 | 23 | 24 | 25 | 26 | 27 | 28 | 29 | 30 | 31 |
| GOOGLE SCHOLAR | 1 | 1.00 | 0.99 | 0.97 | 0.96 | 0.97 | 0.97 | 0.57 | 0.61 | 0.67 | 0.62 | 0.66 | 0.89 | 0.86 | 0.86 | 0.05 | 0.43 | 0.78 | 0.87 | 0.95 | 0.60 | 0.57 | 0.77 | 0.57 | 0.11 | 0.02 | 0.17 | 0.21 | -0.06 | 0.08 | 0.45 | 0.46 |
| | 2 | 0.99 | 1.00 | 0.97 | 0.97 | 0.96 | 0.97 | 0.56 | 0.62 | 0.67 | 0.63 | 0.67 | 0.91 | 0.88 | 0.88 | 0.06 | 0.46 | 0.81 | 0.90 | 0.94 | 0.63 | 0.58 | 0.79 | 0.61 | 0.13 | 0.05 | 0.18 | 0.21 | -0.04 | 0.07 | 0.44 | 0.46 |
| | 3 | 0.97 | 0.97 | 1.00 | 0.99 | 0.97 | 0.98 | 0.60 | 0.64 | 0.67 | 0.60 | 0.68 | 0.92 | 0.91 | 0.86 | 0.10 | 0.50 | 0.84 | 0.91 | 0.92 | 0.66 | 0.62 | 0.82 | 0.61 | 0.12 | 0.02 | 0.18 | 0.22 | -0.05 | 0.06 | 0.44 | 0.46 |
| | 4 | 0.96 | 0.97 | 0.99 | 1.00 | 0.97 | 0.98 | 0.57 | 0.63 | 0.66 | 0.59 | 0.67 | 0.93 | 0.90 | 0.87 | 0.07 | 0.50 | 0.84 | 0.91 | 0.92 | 0.66 | 0.59 | 0.81 | 0.63 | 0.12 | 0.03 | 0.17 | 0.22 | -0.05 | 0.07 | 0.43 | 0.45 |
| | 5 | 0.97 | 0.96 | 0.97 | 0.97 | 1.00 | 0.99 | 0.58 | 0.63 | 0.65 | 0.57 | 0.66 | 0.88 | 0.88 | 0.86 | 0.06 | 0.46 | 0.80 | 0.88 | 0.93 | 0.61 | 0.59 | 0.80 | 0.60 | 0.10 | 0.01 | 0.17 | 0.21 | -0.08 | 0.09 | 0.42 | 0.43 |
| | 6 | 0.97 | 0.97 | 0.98 | 0.98 | 0.99 | 1.00 | 0.56 | 0.62 | 0.65 | 0.58 | 0.66 | 0.90 | 0.87 | 0.87 | 0.05 | 0.47 | 0.81 | 0.88 | 0.94 | 0.62 | 0.59 | 0.80 | 0.62 | 0.10 | 0.01 | 0.17 | 0.21 | -0.08 | 0.08 | 0.42 | 0.44 |
| RESEARCHERID | 7 | 0.57 | 0.56 | 0.60 | 0.57 | 0.58 | 0.56 | 1.00 | 0.91 | 0.88 | 0.78 | 0.89 | 0.59 | 0.59 | 0.61 | 0.03 | 0.22 | 0.48 | 0.59 | 0.57 | 0.44 | 0.54 | 0.58 | 0.37 | 0.06 | 0.01 | -0.04 | 0.02 | -0.11 | 0.08 | 0.21 | 0.25 |
| | 8 | 0.61 | 0.62 | 0.64 | 0.63 | 0.63 | 0.62 | 0.91 | 1.00 | 0.96 | 0.85 | 0.97 | 0.67 | 0.65 | 0.71 | -0.02 | 0.23 | 0.55 | 0.65 | 0.62 | 0.45 | 0.53 | 0.59 | 0.40 | 0.01 | -0.04 | -0.08 | -0.02 | -0.18 | 0.11 | 0.14 | 0.20 |
| | 9 | 0.67 | 0.67 | 0.67 | 0.66 | 0.65 | 0.65 | 0.88 | 0.96 | 1.00 | 0.95 | 0.99 | 0.69 | 0.63 | 0.73 | -0.03 | 0.20 | 0.56 | 0.67 | 0.69 | 0.48 | 0.52 | 0.62 | 0.45 | 0.01 | -0.05 | -0.06 | 0.00 | -0.15 | 0.08 | 0.17 | 0.22 |
| | 10 | 0.62 | 0.63 | 0.60 | 0.59 | 0.57 | 0.58 | 0.78 | 0.85 | 0.95 | 1.00 | 0.93 | 0.60 | 0.54 | 0.65 | -0.04 | 0.14 | 0.50 | 0.61 | 0.65 | 0.46 | 0.50 | 0.58 | 0.41 | 0.03 | -0.01 | -0.03 | 0.02 | -0.10 | 0.09 | 0.21 | 0.25 |
| | 11 | 0.66 | 0.67 | 0.68 | 0.67 | 0.66 | 0.66 | 0.89 | 0.97 | 0.99 | 0.93 | 1.00 | 0.70 | 0.65 | 0.73 | -0.02 | 0.22 | 0.57 | 0.68 | 0.68 | 0.50 | 0.52 | 0.62 | 0.46 | 0.01 | -0.05 | -0.07 | 0.00 | -0.15 | 0.09 | 0.16 | 0.21 |
| RESEARCHGATE | 12 | 0.89 | 0.91 | 0.92 | 0.93 | 0.88 | 0.90 | 0.59 | 0.67 | 0.69 | 0.60 | 0.70 | 1.00 | 0.87 | 0.89 | 0.15 | 0.51 | 0.83 | 0.91 | 0.90 | 0.69 | 0.52 | 0.75 | 0.62 | 0.11 | 0.02 | 0.12 | 0.20 | -0.02 | 0.01 | 0.37 | 0.39 |
| | 13 | 0.86 | 0.88 | 0.91 | 0.90 | 0.88 | 0.87 | 0.59 | 0.65 | 0.63 | 0.54 | 0.65 | 0.87 | 1.00 | 0.78 | 0.26 | 0.63 | 0.89 | 0.94 | 0.83 | 0.70 | 0.67 | 0.77 | 0.43 | 0.19 | 0.12 | 0.18 | 0.20 | -0.04 | 0.10 | 0.38 | 0.40 |
| | 14 | 0.86 | 0.88 | 0.86 | 0.87 | 0.86 | 0.87 | 0.61 | 0.71 | 0.73 | 0.65 | 0.73 | 0.89 | 0.78 | 1.00 | -0.04 | 0.32 | 0.68 | 0.79 | 0.89 | 0.48 | 0.45 | 0.69 | 0.59 | 0.02 | -0.07 | 0.01 | 0.09 | -0.15 | 0.05 | 0.34 | 0.37 |
| | 15 | 0.05 | 0.06 | 0.10 | 0.07 | 0.06 | 0.05 | 0.03 | -0.02 | -0.03 | -0.04 | -0.02 | 0.15 | 0.26 | -0.04 | 1.00 | 0.70 | 0.34 | 0.26 | 0.06 | 0.42 | 0.30 | 0.09 | -0.24 | 0.17 | 0.14 | 0.16 | 0.13 | 0.25 | -0.12 | 0.09 | 0.11 |
| | 16 | 0.43 | 0.46 | 0.50 | 0.50 | 0.46 | 0.47 | 0.22 | 0.23 | 0.20 | 0.14 | 0.22 | 0.51 | 0.63 | 0.32 | 0.70 | 1.00 | 0.69 | 0.63 | 0.42 | 0.71 | 0.56 | 0.49 | 0.16 | 0.29 | 0.20 | 0.21 | 0.23 | 0.08 | -0.03 | 0.24 | 0.29 |
| | 17 | 0.78 | 0.81 | 0.84 | 0.84 | 0.80 | 0.81 | 0.48 | 0.55 | 0.56 | 0.50 | 0.57 | 0.83 | 0.89 | 0.68 | 0.34 | 0.69 | 1.00 | 0.95 | 0.75 | 0.82 | 0.64 | 0.74 | 0.44 | 0.25 | 0.15 | 0.16 | 0.20 | -0.01 | 0.02 | 0.32 | 0.34 |
| | 18 | 0.87 | 0.90 | 0.91 | 0.91 | 0.88 | 0.88 | 0.59 | 0.65 | 0.67 | 0.61 | 0.68 | 0.91 | 0.94 | 0.79 | 0.26 | 0.63 | 0.95 | 1.00 | 0.86 | 0.80 | 0.65 | 0.78 | 0.49 | 0.24 | 0.16 | 0.18 | 0.23 | 0.00 | 0.10 | 0.40 | 0.42 |
| | 19 | 0.95 | 0.94 | 0.92 | 0.92 | 0.93 | 0.94 | 0.57 | 0.62 | 0.69 | 0.65 | 0.68 | 0.90 | 0.83 | 0.89 | 0.06 | 0.42 | 0.75 | 0.86 | 1.00 | 0.58 | 0.53 | 0.78 | 0.61 | 0.07 | -0.02 | 0.07 | 0.13 | -0.12 | 0.06 | 0.35 | 0.36 |
| | 20 | 0.60 | 0.63 | 0.66 | 0.66 | 0.61 | 0.62 | 0.44 | 0.45 | 0.48 | 0.46 | 0.50 | 0.69 | 0.70 | 0.48 | 0.42 | 0.71 | 0.82 | 0.80 | 0.58 | 1.00 | 0.54 | 0.61 | 0.38 | 0.22 | 0.13 | 0.18 | 0.23 | 0.06 | 0.09 | 0.28 | 0.32 |
| MENDELEY | 21 | 0.57 | 0.58 | 0.62 | 0.59 | 0.59 | 0.59 | 0.54 | 0.53 | 0.52 | 0.50 | 0.52 | 0.52 | 0.67 | 0.45 | 0.30 | 0.56 | 0.64 | 0.65 | 0.53 | 0.54 | 1.00 | 0.83 | 0.27 | 0.43 | 0.36 | 0.24 | 0.21 | 0.12 | 0.06 | 0.35 | 0.39 |
| | 22 | 0.77 | 0.79 | 0.82 | 0.81 | 0.80 | 0.80 | 0.58 | 0.59 | 0.62 | 0.58 | 0.62 | 0.75 | 0.77 | 0.69 | 0.09 | 0.49 | 0.74 | 0.78 | 0.78 | 0.61 | 0.83 | 1.00 | 0.72 | 0.26 | 0.17 | 0.17 | 0.19 | 0.00 | 0.00 | 0.35 | 0.38 |
| | 23 | 0.57 | 0.61 | 0.61 | 0.63 | 0.60 | 0.62 | 0.37 | 0.40 | 0.45 | 0.41 | 0.46 | 0.62 | 0.43 | 0.59 | -0.24 | 0.16 | 0.44 | 0.49 | 0.61 | 0.38 | 0.27 | 0.72 | 1.00 | -0.10 | -0.17 | -0.05 | 0.04 | -0.15 | -0.06 | 0.14 | 0.14 |
| | 24 | 0.11 | 0.13 | 0.12 | 0.12 | 0.10 | 0.10 | 0.06 | 0.01 | 0.01 | 0.03 | 0.01 | 0.11 | 0.19 | 0.02 | 0.17 | 0.29 | 0.25 | 0.24 | 0.07 | 0.22 | 0.43 | 0.26 | -0.10 | 1.00 | 0.96 | 0.46 | 0.43 | 0.42 | 0.24 | 0.42 | 0.43 |
| | 25 | 0.02 | 0.05 | 0.02 | 0.03 | 0.01 | 0.01 | 0.01 | -0.04 | -0.05 | -0.01 | -0.05 | 0.02 | 0.12 | -0.07 | 0.14 | 0.20 | 0.15 | 0.16 | -0.02 | 0.13 | 0.36 | 0.17 | -0.17 | 0.96 | 1.00 | 0.46 | 0.41 | 0.45 | 0.27 | 0.41 | 0.41 |
| TWITTER | 26 | 0.17 | 0.18 | 0.18 | 0.17 | 0.17 | 0.17 | -0.04 | -0.08 | -0.06 | -0.03 | -0.07 | 0.12 | 0.18 | 0.01 | 0.16 | 0.21 | 0.16 | 0.18 | 0.07 | 0.18 | 0.24 | 0.17 | -0.05 | 0.46 | 0.46 | 1.00 | 0.87 | 0.77 | 0.29 | 0.71 | 0.69 |
| | 27 | 0.21 | 0.21 | 0.22 | 0.22 | 0.21 | 0.21 | 0.02 | -0.02 | 0.00 | 0.02 | 0.00 | 0.20 | 0.20 | 0.09 | 0.13 | 0.23 | 0.20 | 0.23 | 0.13 | 0.23 | 0.21 | 0.19 | 0.04 | 0.43 | 0.41 | 0.87 | 1.00 | 0.81 | 0.40 | 0.78 | 0.77 |
| | 28 | -0.06 | -0.04 | -0.05 | -0.05 | -0.08 | -0.08 | -0.11 | -0.18 | -0.15 | -0.10 | -0.15 | -0.02 | -0.04 | -0.15 | 0.25 | 0.08 | -0.01 | 0.00 | -0.12 | 0.06 | 0.12 | 0.00 | -0.15 | 0.42 | 0.45 | 0.77 | 0.81 | 1.00 | 0.18 | 0.55 | 0.53 |
| | 29 | 0.08 | 0.07 | 0.06 | 0.07 | 0.09 | 0.08 | 0.08 | 0.11 | 0.08 | 0.09 | 0.09 | 0.01 | 0.10 | 0.05 | -0.12 | -0.03 | 0.02 | 0.10 | 0.06 | 0.09 | 0.06 | 0.00 | -0.06 | 0.24 | 0.27 | 0.29 | 0.40 | 0.18 | 1.00 | 0.30 | 0.32 |
| | 30 | 0.45 | 0.44 | 0.44 | 0.43 | 0.42 | 0.42 | 0.21 | 0.14 | 0.17 | 0.21 | 0.16 | 0.37 | 0.38 | 0.34 | 0.09 | 0.24 | 0.32 | 0.40 | 0.35 | 0.28 | 0.35 | 0.35 | 0.14 | 0.42 | 0.41 | 0.71 | 0.78 | 0.55 | 0.30 | 1.00 | 0.98 |
| | 31 | 0.46 | 0.46 | 0.46 | 0.45 | 0.43 | 0.44 | 0.25 | 0.20 | 0.22 | 0.25 | 0.21 | 0.39 | 0.40 | 0.37 | 0.11 | 0.29 | 0.34 | 0.42 | 0.36 | 0.32 | 0.39 | 0.38 | 0.14 | 0.43 | 0.41 | 0.69 | 0.77 | 0.53 | 0.32 | 0.98 | 1.00 |

| COD | METRIC | COD | METRIC | COD | METRIC |
|---|---|---|---|---|---|
| 1 | GS_citations_total | 12 | RG_score | 23 | MEND_readers / document |
| 2 | GS_citations_last5 | 13 | RG_publications | 24 | MEND_followers |
| 3 | GS_hindex_total | 14 | RG_impact_points | 25 | MEND_following |
| 4 | GS_hindex_last5 | 15 | RG_following | 26 | TW_tweets |
| 5 | GS_i10_total | 16 | RG_followers | 27 | TW_followers |
| 6 | GS_i10_last5 | 17 | RG_downloads | 28 | TW_following |
| 7 | RID_n_total_articles | 18 | RG_views | 29 | TW_dias |
| 8 | RID _n_articles_cit | 19 | RG_citations | 30 | TW_sum_retweets |
| 9 | RID _sum_times_cited | 20 | RG_profile_views | 31 | TW_h_retweets |
| 10 | RID _average_cit | 21 | MEND_pub | | |
| 11 | RID _hindex | 22 | MEND_readers | | |

## 3.4. Data reliability

After describing the multifaceted presence (authors, documents, and sources) of the bibliometric community in *Google Scholar Citations*, describing the presence of the authors of this community in other social platforms, and analyzing the possible correlation between all metrics offered by these platforms, it is absolutely essential to face the discussion about the reliability of these metrics and platforms. In Science, if the data source and the instrument (that stores that data and computes the measures) are not reliable, the results achieved are meaningless and scientifically irrelevant; such groundless results should not be considered as proper scientific results until their validity is proven.

In Bibliometrics, there is a large tradition of studies addressing the errors related to the correct assignment of citations to documents in bibliometric databases, as well as the deficiencies in the design or application of bibliometric indicators (Sher, Garfield & Elias, 1966; Poyer, 1979; Garfield, 1983; Moed & Vriens, 1989; Garfield, 1990; Garcia-Perez, 2010; Franceschini, Maisano & Mastrogiacomo, 2015).

Since these platforms are quite new, there are still few in-depth empirical studies using representative samples which may allow us to make informed assertions about the reliability of these platforms. So far, there are only a few isolated analyses pointing out errors, inaccuracies and inconsistencies. Regrettably, there are not many of these interesting works, and they don't often go beyond reporting a few anecdotal issues. In this respect, we must highlight the great impact of Peter Jacsó's works, who analyzed the strengths and specially the weaknesses of Google Scholar (Jacsó 2005; 2006a; 2006b; 2008; 2010).

In order to contextualize all the data offered previously in this work, we present a final section providing insights about the different kinds of errors found in each of the platforms, with a special emphasis in *Google Scholar*, since it has been our main source of data.

### *3.4.1. The uncontrolled giant: Google Scholar & Google Scholar Citations*

The errors that can compromise the metric portrait of an author offered by *Google Scholar* can be grouped into two main sections. First, the errors *Google Scholar* sometimes makes when it indexes a document or when it assigns citations to it. Second, the specific errors that are sometimes made during the creation of a *Google Scholar Citations* profile.

The former are a logical consequence of the tricky and complex task that is automatically searching the current academic papers available in the net. This task also involves merging in only one record all possible versions of the same work, and linking to it all documents in which it is cited (keeping in mind that these documents and references can be presented in the most varied formats). The latter are the ultimate responsibility of the author, who must periodically revise his/her profile in order to eliminate misattributed documents which might been included in the automatic weekly updates, clean the records by merging different versions of the same document when Google Scholar's algorithms are not able to detect their similarity, as well as improve and complete the bibliographic references of these documents (filling in blank fields in a document when Google Scholar hasn't been able to find that information).

Next, we classify, describe, and illustrate some of the most common mistakes in Google Scholar:

### *a) Incorrect identification of the title of the document*

Google Scholar always tries to extract bibliographic information from the HTML Meta tags in a webpage. When there are no Meta tags available, it parses the webpage itself (the HTML code of the page, or even PDFs themselves). Even though its spiders are able to successfully parse pages with a quite broad range of different structures, and despite the fact that they have published a very clear set of inclusion guidelines, some parsing errors occasionally arise for documents extracted from websites with unusual layouts. It is not rare in these cases that an incorrect text string is selected as the title of the document. In Figure 6 we illustrate an example in which an incorrect string ("www.redalyc.org") has been selected as the title of the document in several records, probably because it is the string that is featured with a higher font size in the first page of the PDF document from which Google Scholar has parsed the bibliographic information. Note that the authors and the source publications are correctly assigned.

***Figure 6. Document titles improperly identified in Google Scholar: URLs***

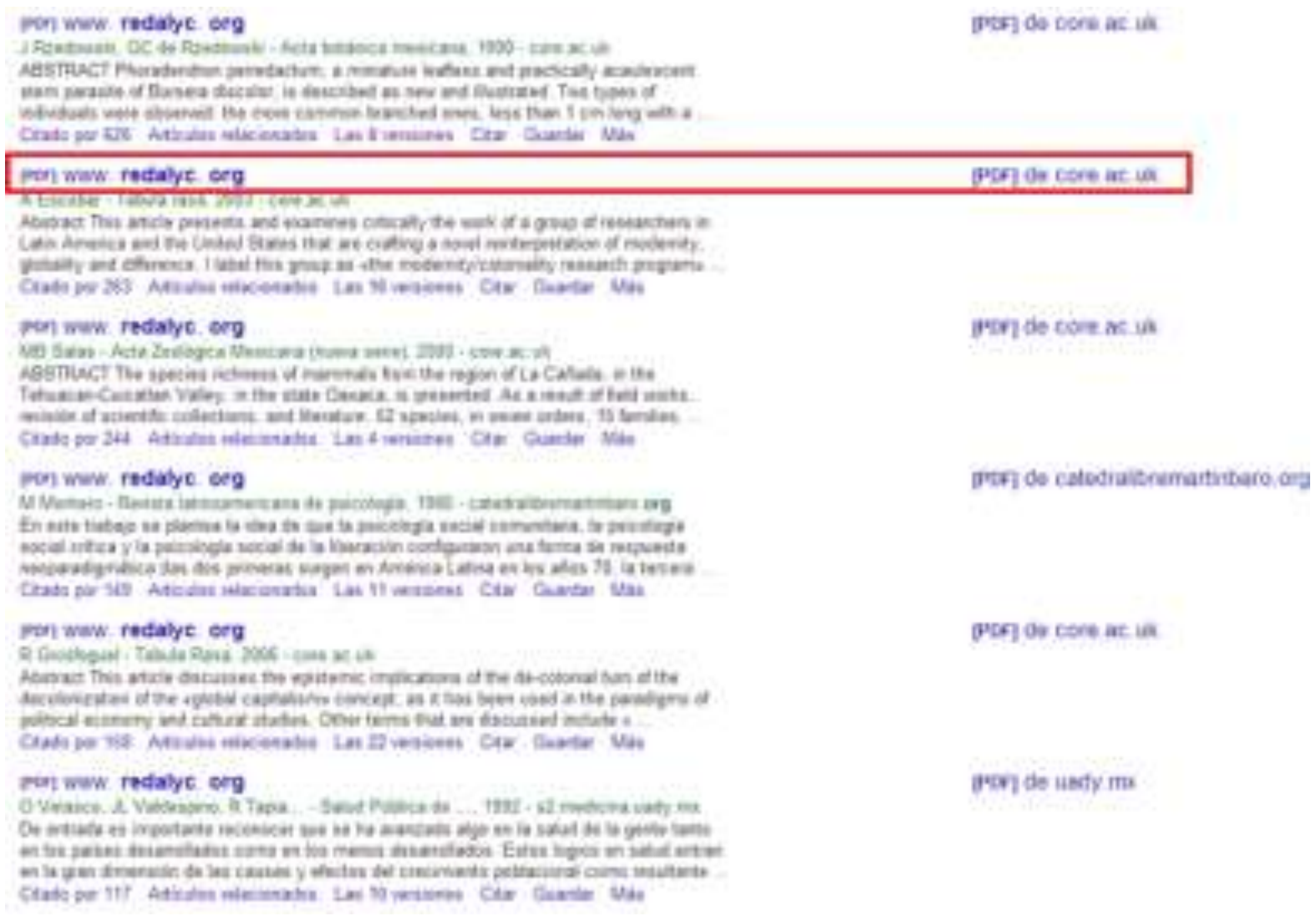

In many other occasions, other text strings, such as the author's name and/or the year of publication, are incorrectly selected as the title of the document. In Figure 7 we can observe how "de Solla" has been selected as the title in many records.

***Figure 7. Author names incorrectly selected as document titles in Google Scholar***

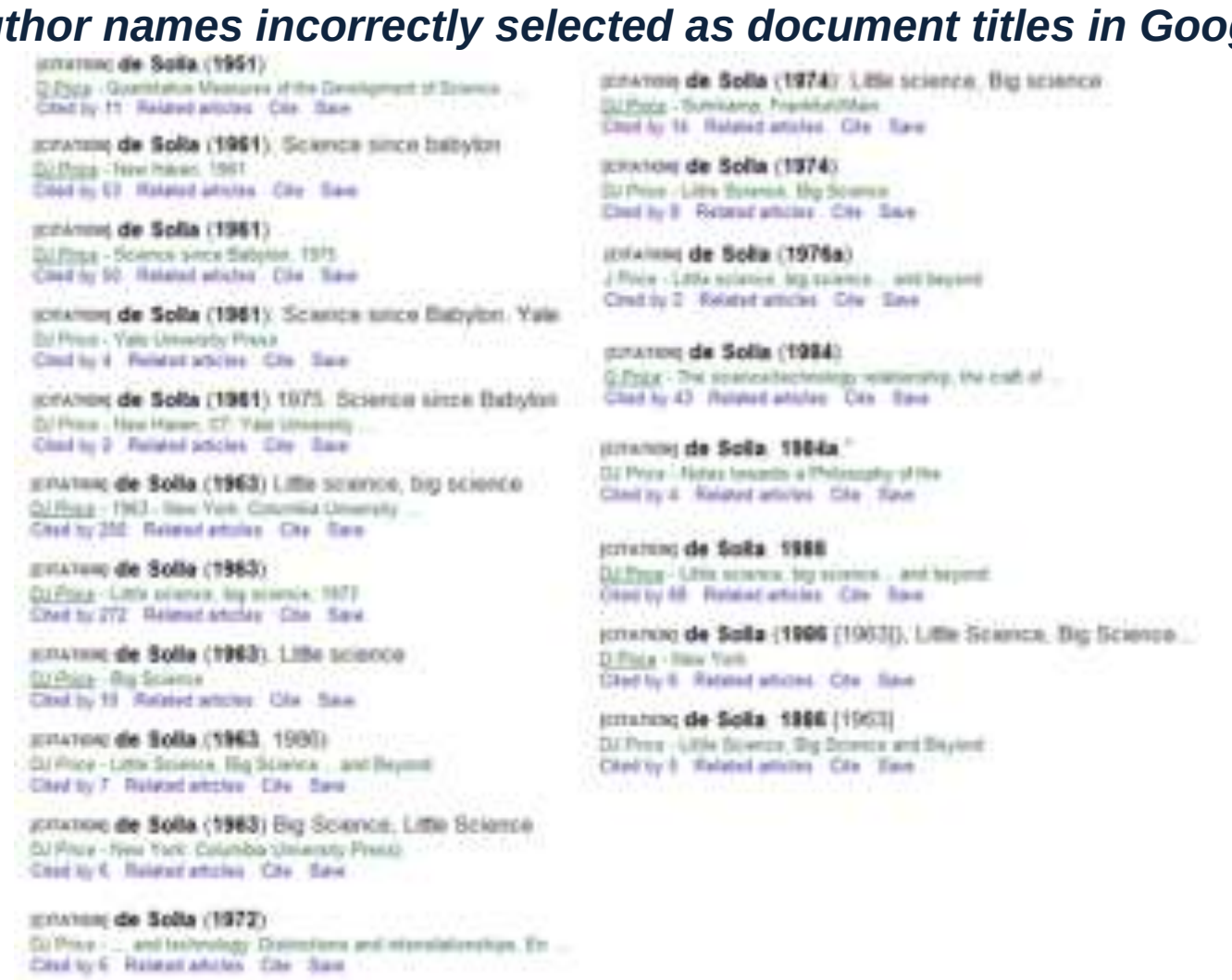

Source: [https://scholar.google.com/scholar?start=0&q=allintitle:+%22de+solla%22+-Moravcsik+-gulls+-comments+-1922+-foreword+-Toward+-tribute+-space+-pensamento+-address+-appreciation&hl=en&as_sdt=0,5](https://scholar.google.com/scholar?start=0&q=allintitle:+%22de+solla%22+-Moravcsik+-gulls+-comments+-1922+-foreword+-Toward+-tribute+-space+-pensamento+-address+-appreciation&hl=en&as_sdt=0,5)

### *b) Ghost authors*

The topic of ghost authors, citations, and documents was approached by Jacsó in numerous works, mostly before *Google Scholar Citations* was launched. Although profiles have served to filter and correct many mistakes, some of them still persist, especially if authors do not clean their personal profiles. In Figure 8 we can see one such example. In this case, the record only displays one person as the author of the article (Carmen Martín Moreno), when in fact the article was written by two authors (Elías Sanz-Casado and Carmen Martín Moreno).

In this case, Google Scholar extracted the bibliographic information from the HTML Meta tags in the website of the journal where the article was published, but, as we can see in Figure 8 (bottom image), these metadata were already incorrect (the title should read “Técnicas bibliométricas aplicadas a **los** estudios de usuarios”), and incomplete (Elías Sanz-Casado is missing from the record). Nonetheless, thanks to Google Scholar Citations, Elías was able to add the document to his profile, even if his name is still missing from the authors field (Figure 8, top left).

***Figure 8. Missing authors in primary versions of documents in Google Scholar***

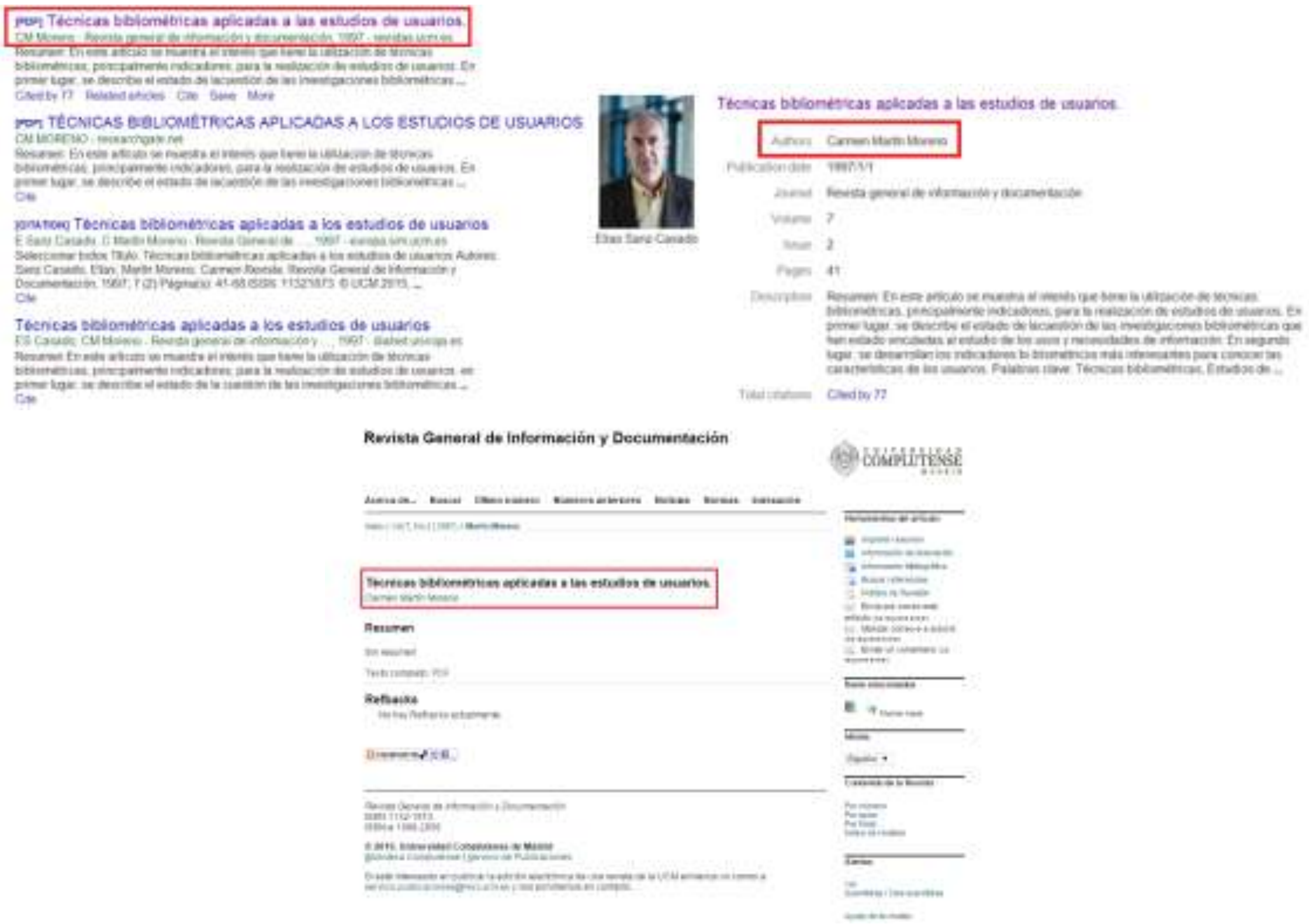

## *c) Book reviews indexed as books*

Among the most common mistakes in document identification is mistaking the review of a book for the book itself. In Figure 9 we show two different records which correspond with book reviews of the work “Introduction to informetrics. Quantitative methods in Library, Documentation and Information Science” by Egghe and Rousseau. At a first glance the first record (Figure 9; top) looks like a normal record, since the title and authors of the book have been correctly identified. However, the record actually points to a review of the book published in *Revista Española de Documentación Científica*. The second record (Figure 9; bottom), is also a review of the book which was published in *Aslib Proceedings*. In this case, the author of the review is the one who appears in the GS record (Brookes).

***Figure 9. Authorship and attribution of book reviews***

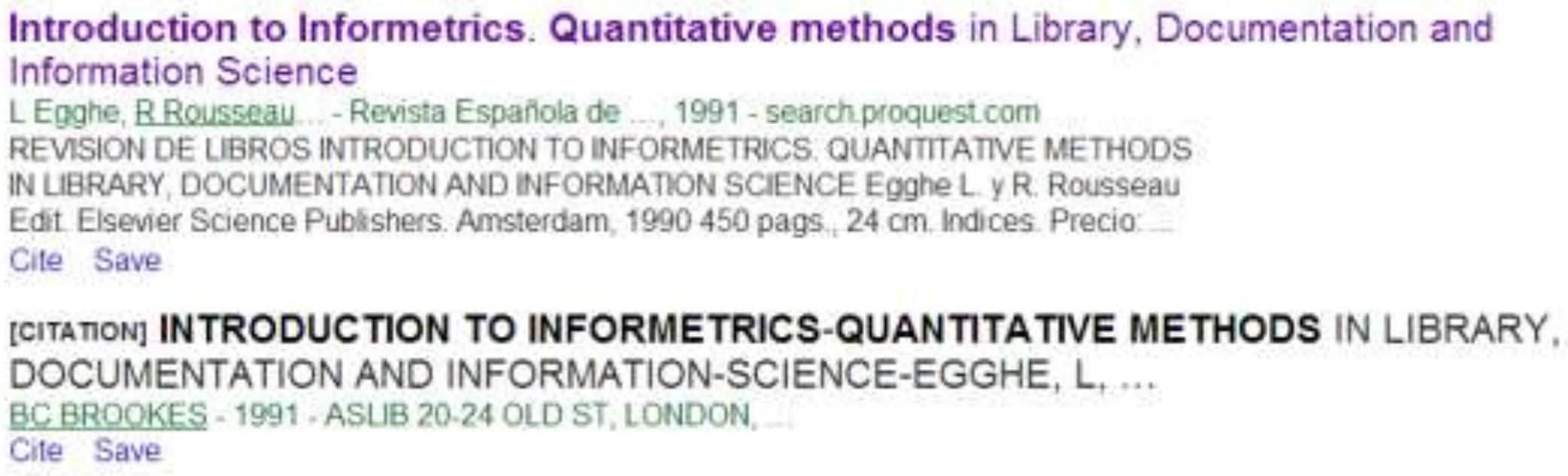

### *c) Incorrect attribution of documents to authors*

Somewhat related to the previous error is the attribution of a document to the wrong authors. In Figure 10 we observe a special case: the book “Introduction to informetrics. Quantitative methods in Library, Documentation and Information Science” by Egghe and Rousseau, is wrongly attributed to Tague-Sutcliffe, probably because this author has a short publication in the journal Information Processing & Management (Figure 10; bottom) with a similar title (“An introduction to informetrics”).

***Figure 10. Authorship improperly assigned in Google Scholar***

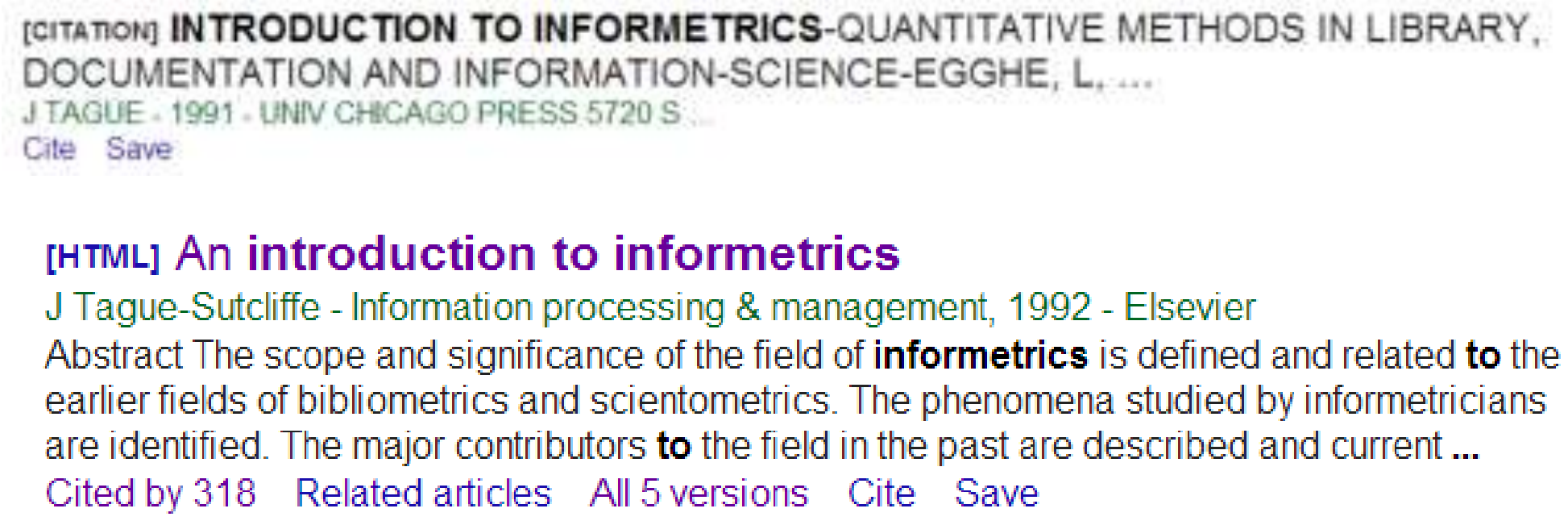

### *d) Failing to merge all versions of a same document into one record*

Although the algorithms for grouping versions work well in most cases, Google Scholar sometimes fails to realize that two or more records it has indexed actually represent the same document. This happens when there are enough formal differences between the metadata of the two versions (differences in the way the name of the authors have been stored, in the title, the year of publication…), that Google Scholar judges they’re not similar enough to be the same document. This issue mostly affects document types other than journal articles (books, book chapters, reports), but duplicate articles also exist. Articles translated into one or more languages are an extreme example: in those cases, the title of the original version is completely different to that of the translated version, so it is understandable that Google Scholar doesn’t realize they are the same document. From a bibliometric perspective, however, their citation counts shouldn’t be split.

This issue obviously affects the citation count of some documents. In Figure 11 we can observe how this phenomenon affects a book chapter: “Measuring science”, by Van Raan.

*Figure 11. Versions of book chapters improperly tied in Google Scholar*

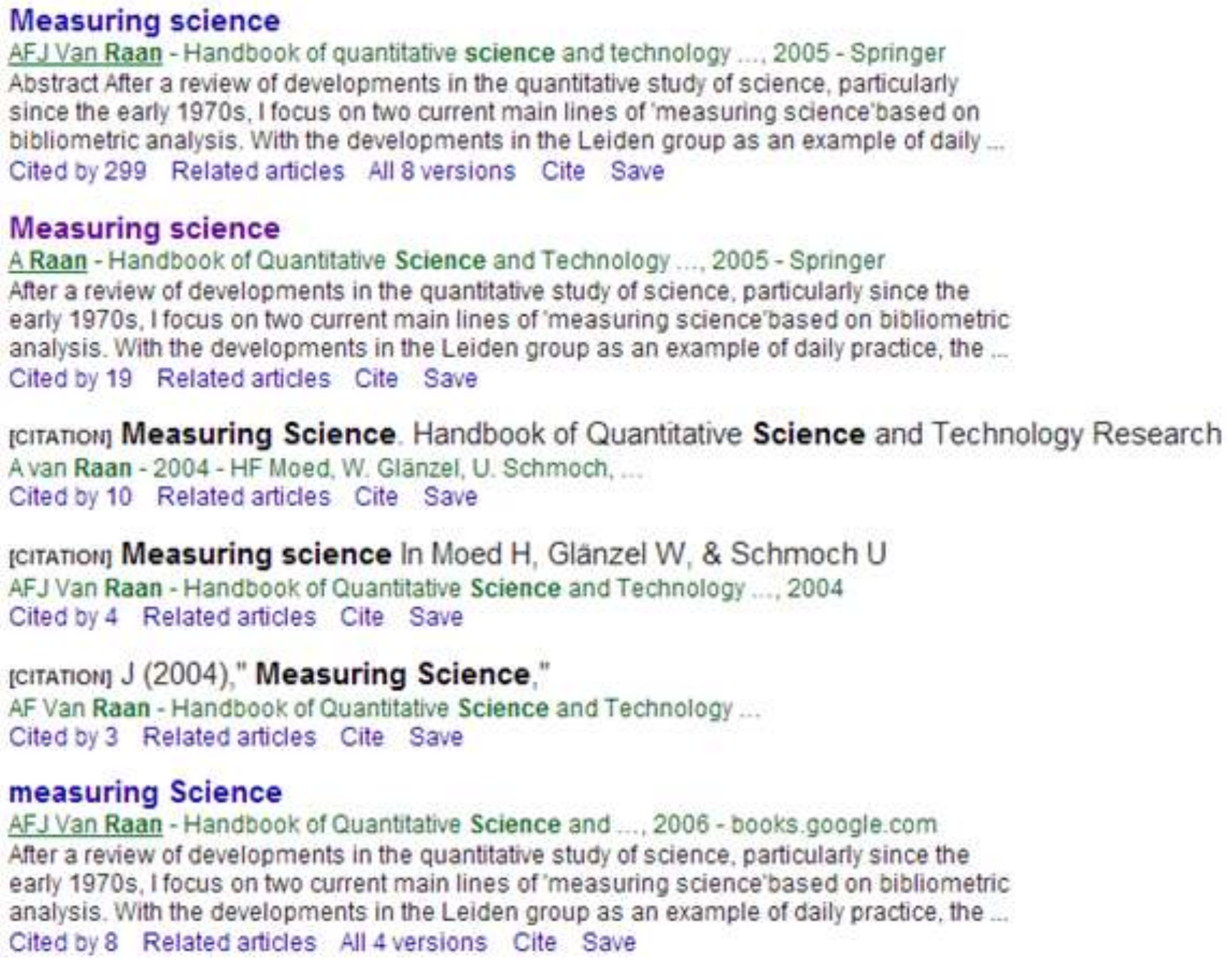

**Measuring science**
AFJ Van Raan - Handbook of quantitative science and technology ..., 2005 - Springer
Abstract After a review of developments in the quantitative study of science, particularly since the early 1970s, I focus on two current main lines of 'measuring science'based on bibliometric analysis. With the developments in the Leiden group as an example of daily ...
Cited by 299 Related articles All 8 versions Cite Save

**Measuring science**
A Raan - Handbook of Quantitative Science and Technology ..., 2005 - Springer
After a review of developments in the quantitative study of science, particularly since the early 1970s, I focus on two current main lines of 'measuring science'based on bibliometric analysis. With the developments in the Leiden group as an example of daily practice, the ...
Cited by 19 Related articles Cite Save

[CITATION] **Measuring Science**. Handbook of Quantitative **Science** and Technology Research
A van Raan - 2004 - HF Moed, W. Glänzel, U. Schmoch, ...
Cited by 10 Related articles Cite Save

[CITATION] **Measuring science** In Moed H, Glänzel W, & Schmoch U
AFJ Van Raan - Handbook of Quantitative Science and Technology ..., 2004
Cited by 4 Related articles Cite Save

[CITATION] J (2004)," **Measuring Science**,"
AF Van Raan - Handbook of Quantitative Science and Technology ...
Cited by 3 Related articles Cite Save

**measuring Science**
AFJ Van Raan - Handbook of Quantitative Science and ..., 2006 - books.google.com
After a review of developments in the quantitative study of science, particularly since the early 1970s, I focus on two current main lines of 'measuring science'based on bibliometric analysis. With the developments in the Leiden group as an example of daily practice, the ...
Cited by 8 Related articles All 4 versions Cite Save

### *d) Grouping different editions of the same book in a single record*

Conversely to the previous error, Google Scholar sometimes groups together records that should stay separate, for example in the cases when there are different editions of the same book (a new book edition provides new content, contrary to a reprinting of a book, which is identical to the previous printing). Figure 12 illustrates the case of “Little Science, big Science”, written by Price. This book was first published in 1963 by Columbia University Press, and reedited in 1986 under the title “Little science, big science… and beyond”, an edition that contained the original text of the book, as well as seven of his most famous articles.

*Figure 12. Different book editions tied in Google Scholar*

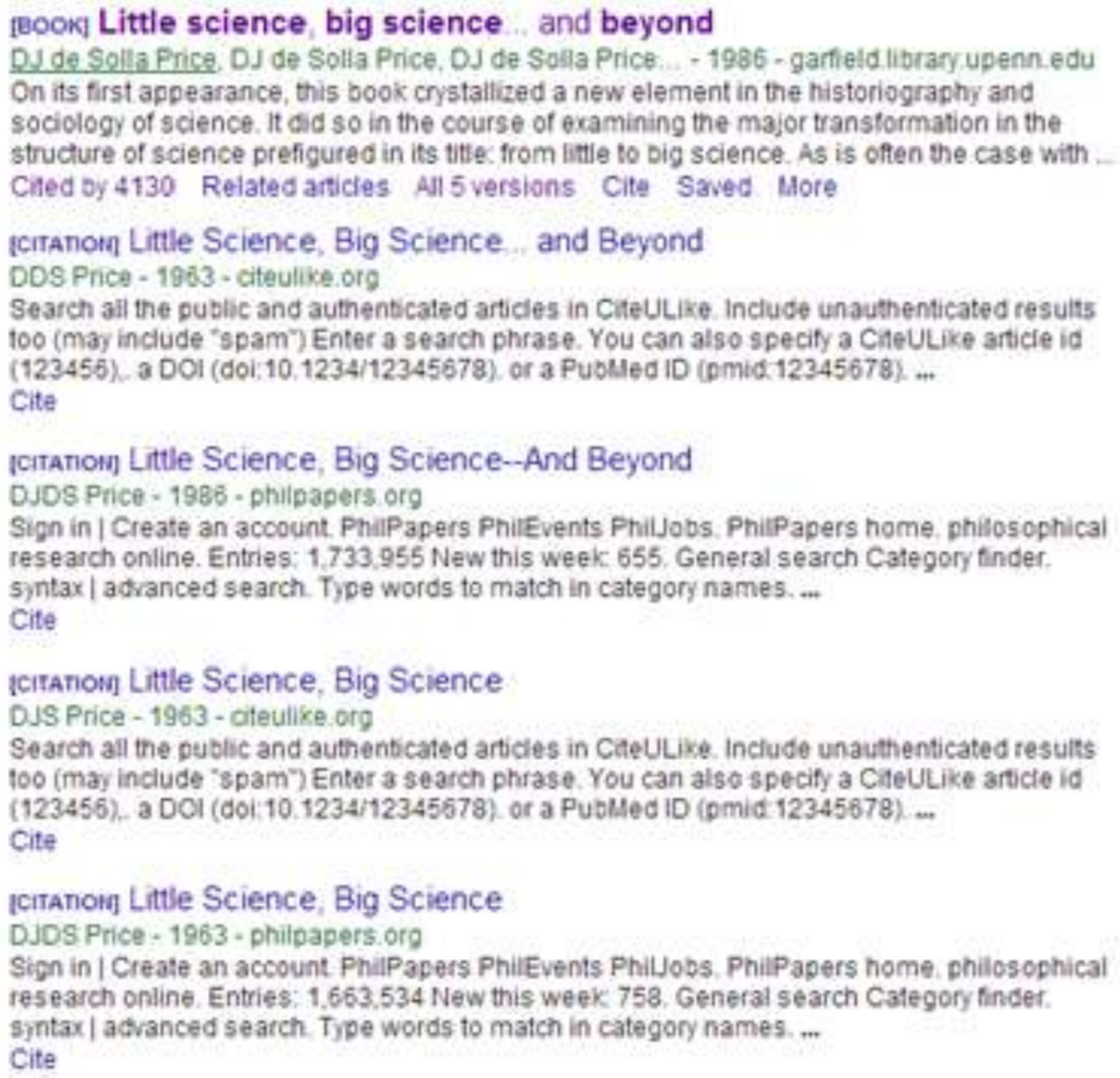

[BOOK] **Little science, big science... and beyond**
DJ de Solla Price, DJ de Solla Price, DJ de Solla Price... - 1986 - garfield.library.upenn.edu
On its first appearance, this book crystallized a new element in the historiography and sociology of science. It did so in the course of examining the major transformation in the structure of science prefigured in its title: from little to big science. As is often the case with ...
Cited by 4130 Related articles All 5 versions Cite Saved More

[CITATION] Little Science, Big Science... and Beyond
DDS Price - 1963 - citeulike.org
Search all the public and authenticated articles in CiteULike. Include unauthenticated results too (may include "spam") Enter a search phrase. You can also specify a CiteULike article id (123456),. a DOI (doi:10.1234/12345678). or a PubMed ID (pmid:12345678). ...
Cite

[CITATION] Little Science, Big Science--And Beyond
DJDS Price - 1986 - philpapers.org
Sign in | Create an account. PhilPapers PhilEvents PhilJobs. PhilPapers home. philosophical research online. Entries: 1,733,955 New this week: 655. General search Category finder. syntax | advanced search. Type words to match in category names. ...
Cite

[CITATION] Little Science, Big Science
DJS Price - 1963 - citeulike.org
Search all the public and authenticated articles in CiteULike. Include unauthenticated results too (may include "spam") Enter a search phrase. You can also specify a CiteULike article id (123456),. a DOI (doi:10.1234/12345678). or a PubMed ID (pmid:12345678). ...
Cite

[CITATION] Little Science, Big Science
DJDS Price - 1963 - philpapers.org
Sign in | Create an account. PhilPapers PhilEvents PhilJobs. PhilPapers home. philosophical research online. Entries: 1,663,534 New this week: 758. General search Category finder. syntax | advanced search. Type words to match in category names. ...
Cite

The primary version (which has received 4,130 citations) is the edition from 1986, but among its versions are several records pointing to the version from 1963. Different editions of the same book should be treated as separate documents when computing citations because their content may be very different.

Of course, aumatically detecting and managing these details is a very complex task, and only a very tiny fraction of the documents indexed in Google Scholar (the most influential manuals and seminal works) would benefit from this thorough treatment. We must not forget that *Google Scholar* is, first of all, a search tool devoted to helping researchers find academic information. A great percentage of users probably don't care about the different editions of a book, and those who do probably just want the most recent one. That may be the reason why Google Scholar usually displays the most recent edition of a book as the primary version. The use of separate entries for different editions is something just a few people, like librarians, would be interested in.

In any case, this may have an important effect in citation counts because citations to different editions (providing different content) are added together. In Figure 13 we can see how the 1986 edition of the book is receiving citations that were actually made to the original work published in 1963.

***Figure 13. Citations to different book editions tied in Google Scholar***

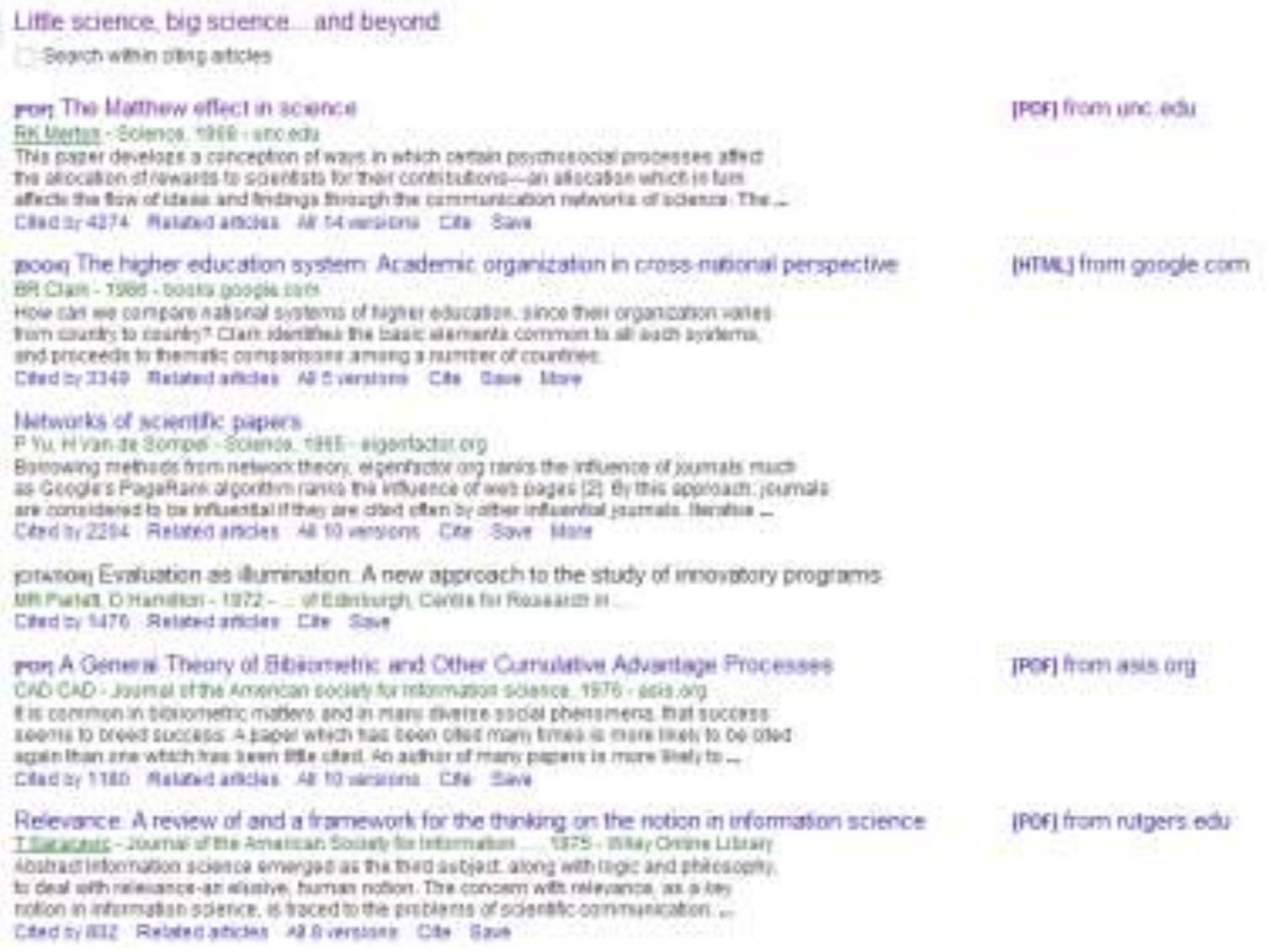

### *e) Improper attribution of citations to a document*

Document citation counts in Google Scholar are also affected by the attribution of "ghost" citations to documents, that is, citations that aren't actually there when we examine the citing document. Figure 14 shows an example of this issue: the work "Le transfert de l'information scientifique et technique: le rôle des nouvelles technologies de l'information face à la crise du modèle actuel de communication écrite" has allegedly received eight citations, but if we manually examine the second document in the list (marked in red), we can't find any mention of the cited work. This phenomenon has been frequently observed in documents stored in the E-LIS repository.[3]

*Figure 14. Appearance of false citations*

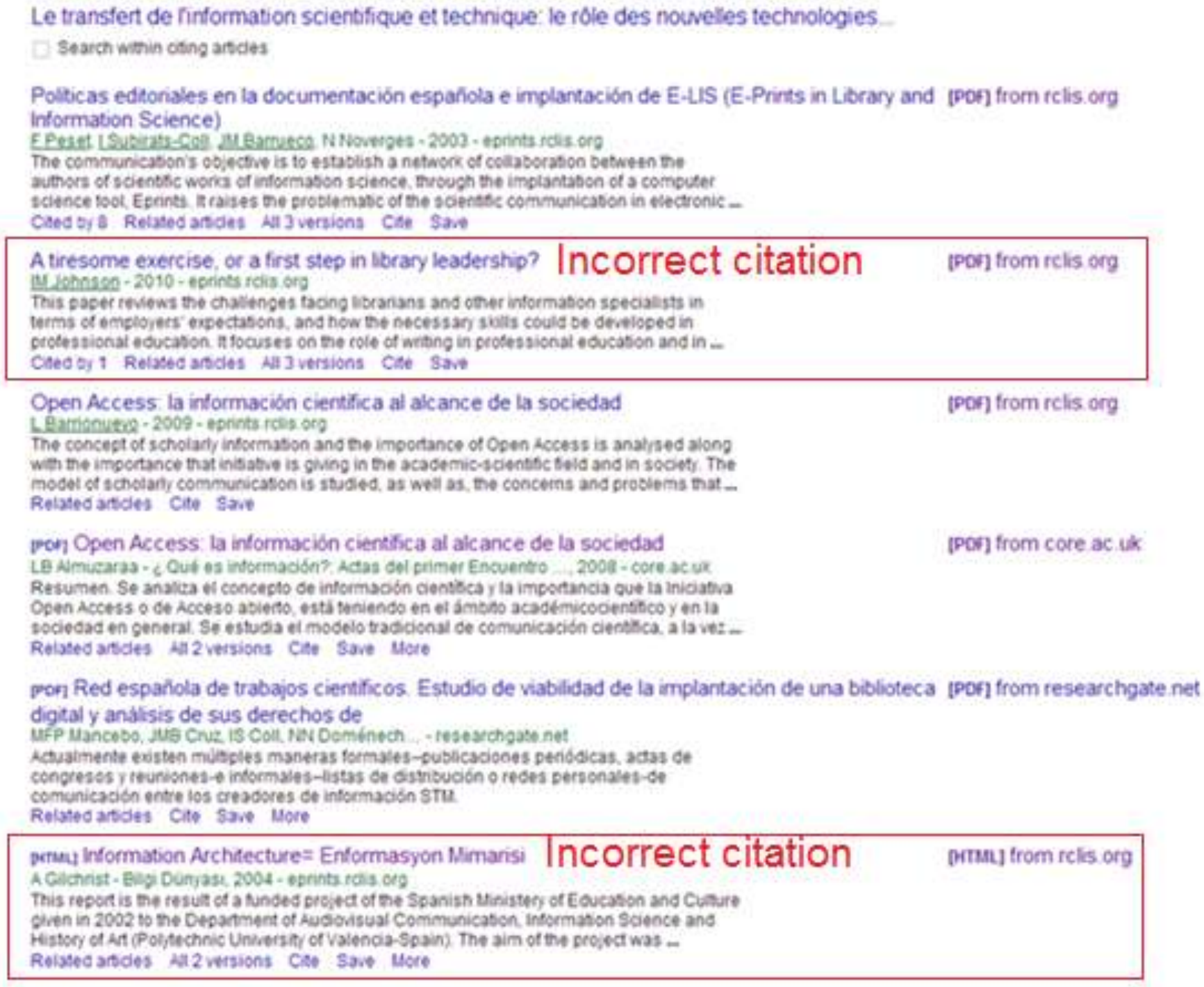

## *f) Duplicate citations*

This phenomenon is a consequence of an issue previously discussed. When Google Scholar fails to realise that two records are actually versions of the same document, these versions are stored as if they were different documents. Therefore, each of them provides its own set of citations to the citation pool. Since the two sets of citations are probably identical, each cited document will receive two citations from what is actually only one document, thus falsely inflating their citation counts.

In Figure 15 we observe a double example of this phenomenon. In the first case (first red rectangle), there are three versions of the same document. Note the differences in the way the authors' names are stored, since this is probably the reason why the records weren't merged into one. In the second case (second red rectangle), the two records refer to the same document (the first one is the English version of the article, and the second one is the Spanish version).

[3] http://eprints.rclis.org

***Figure 15. Duplicate citations in Google Scholar***

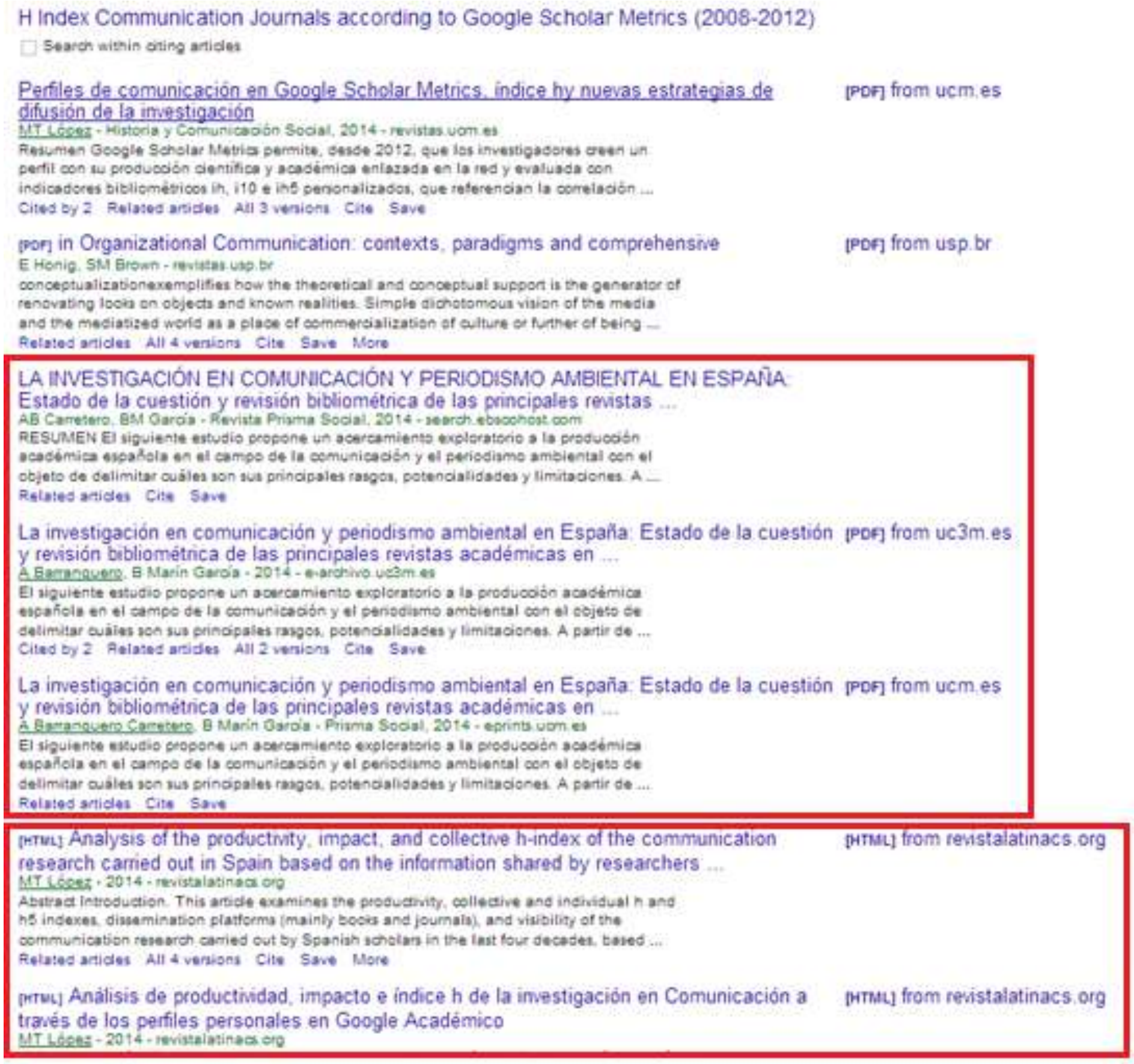

### *g) Missing citations*

There are cases when Google Scholar's parser fails to match a cited reference inside document, with the record of the document it is citing. When Google Scholar parses the reference section within an article, it tries to find a match for these references in its records, but if for some reason the reference hasn't been correctly recorded (authors of the citing article may have made a mistake when citing it or used an uncommon reference format Google Scholar doesn't understand) the system will be unable to make the connection between the two documents.

However, we also find examples in which no apparent mistake has been made in the citing document, but still the citation isn't attributed to the cited document.

In order to illustrate this issue, in Figure 16 we show how a document ("How to cook the university rankings") is citing in its reference section other document (a doctoral thesis). However, this citation doesn't appear as one of the 13 citations that the thesis has received according to *Google Scholar*. The reason is unknown. At the time the citing document was first indexed, the connection wasn't made for some reason, and this error hasn't been solved since. Typos in the PDF can also generate this kind of error.

*Figure 16. Citations unrevealed in Google Scholar*

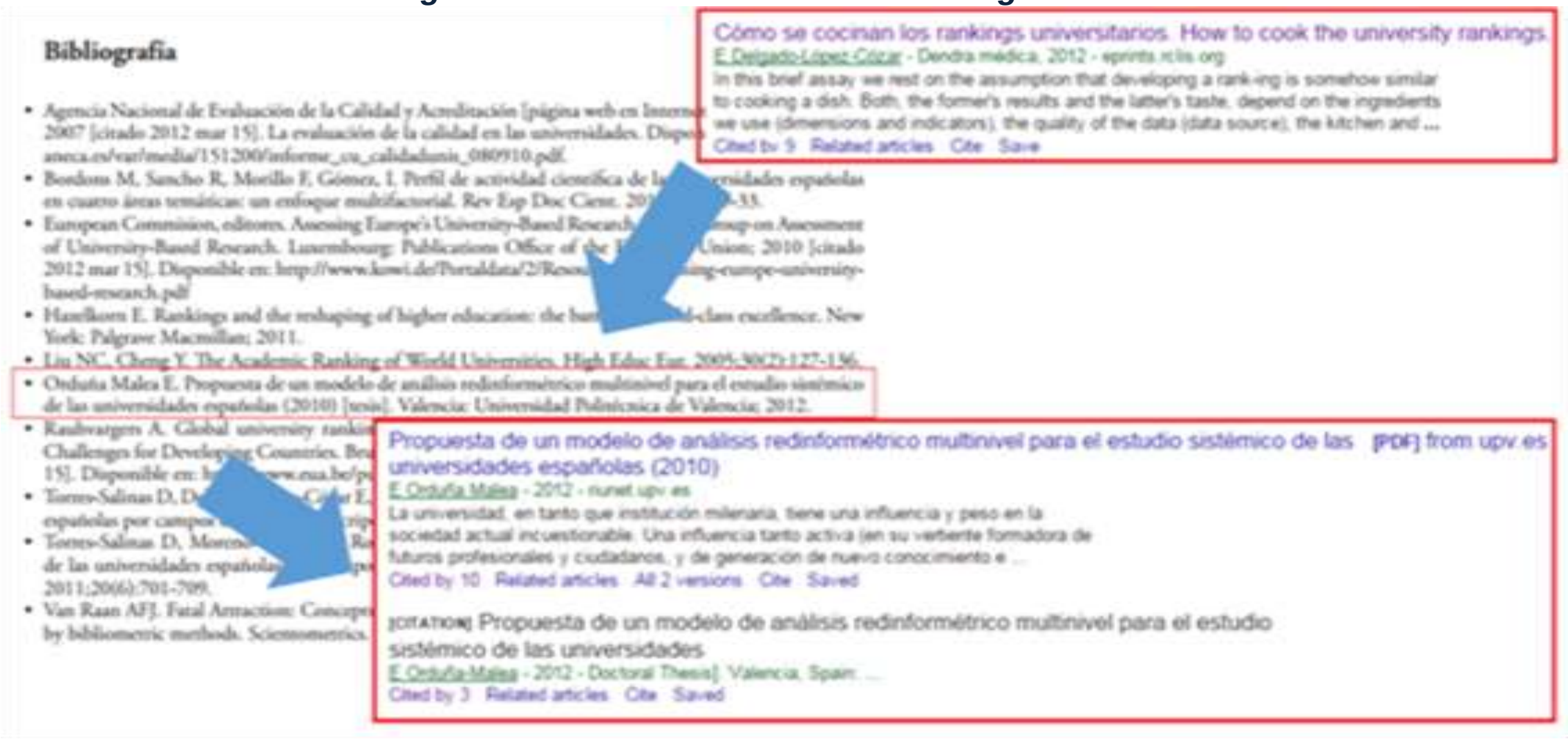

All the errors previously described are related directly with the *Google Scholar* database (and are concerned with how the automatic parser works). Next we show some of the mistakes identified in the elaboration of bibliographic profiles through *Google Scholar Citations*:

### a) Duplicate profiles

Since the only restriction to create a public academic profile in *Google Scholar Citations* is to provide a valid email, an author (or anyone really) may create as many profiles as he/she wants. This opens the door to the existence of duplicate profiles, that is, different profiles about the same person. In Figure 17 we present some examples of duplicate profiles of authors related to the field of Bibliometrics. The differences in citation counts between profiles are sometimes quite high (for example, one of the profiles belonging to Ruiz-Castillo achieves 1,843 citations whereas in the second profile the figure goes up to 2,430).

*Figure 17. Duplicate profiles in Google Scholar Citations*

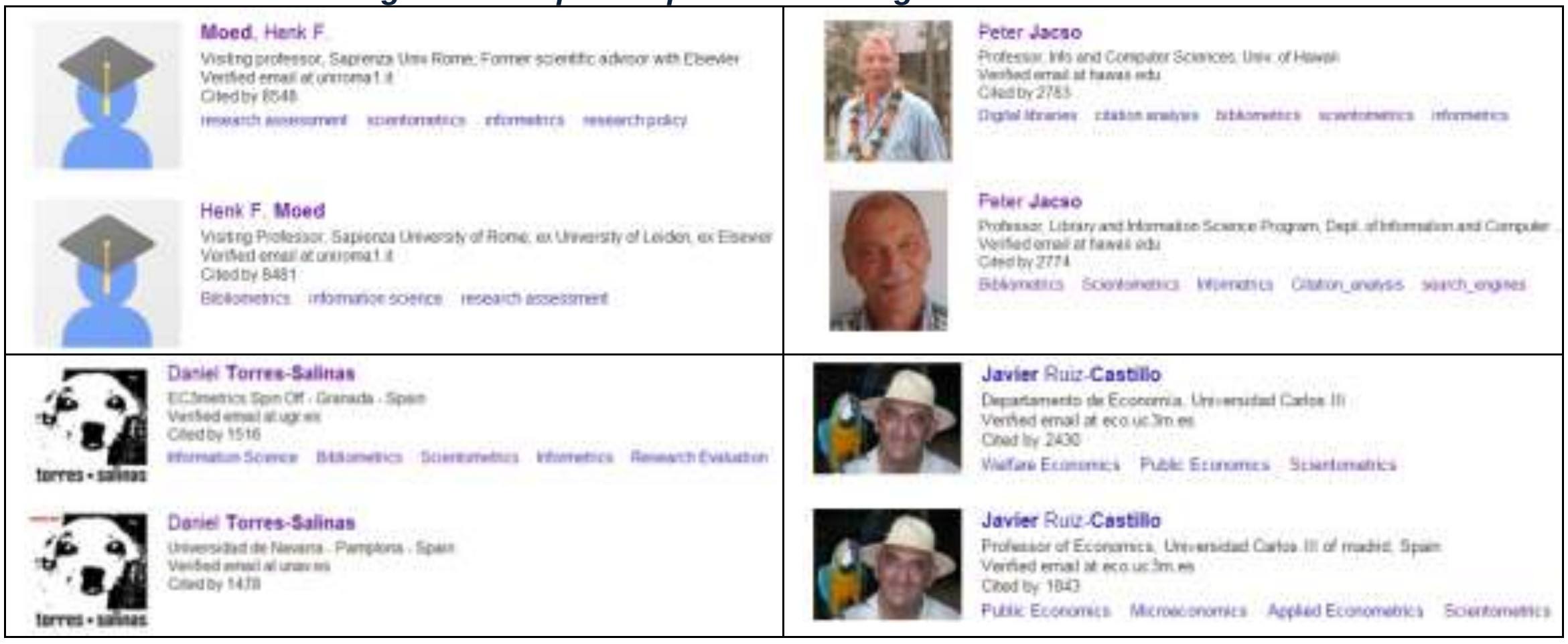

A real problem can arise when one of the profiles has been created by someone other than the author the profile is about. The author may send a request to Google Scholar to delete the profile, but this kind of requests might take a while to be processed, generating a feeling of helplessness in the author.

### b) Variety of document types (including non-academic documents)

One of the main criticisms to the profiles in *Google Scholar Citations* (when considering whether they're suited for evaluation purposes) is the inclusion of a wide variety of document types: from peer-reviewed articles to posters. An author can add any kind of work to his profile, and sometimes they aren't even academic works: teaching materials, software, online resources, etc. (Figure 18).

While this is a true shortcoming from the research evaluation perspective, these profiles are designed to showcase any material that the author considers appropriate, especially if these materials could potentially generate some kind of impact through citations. The possibility to select the document typology (as *ResearchGate* does) may help solve this problem. However, the selection of document type is only an internal mechanism not reflected in the public profile.

***Figure 18. Teaching materials in Google Scholar Citations***

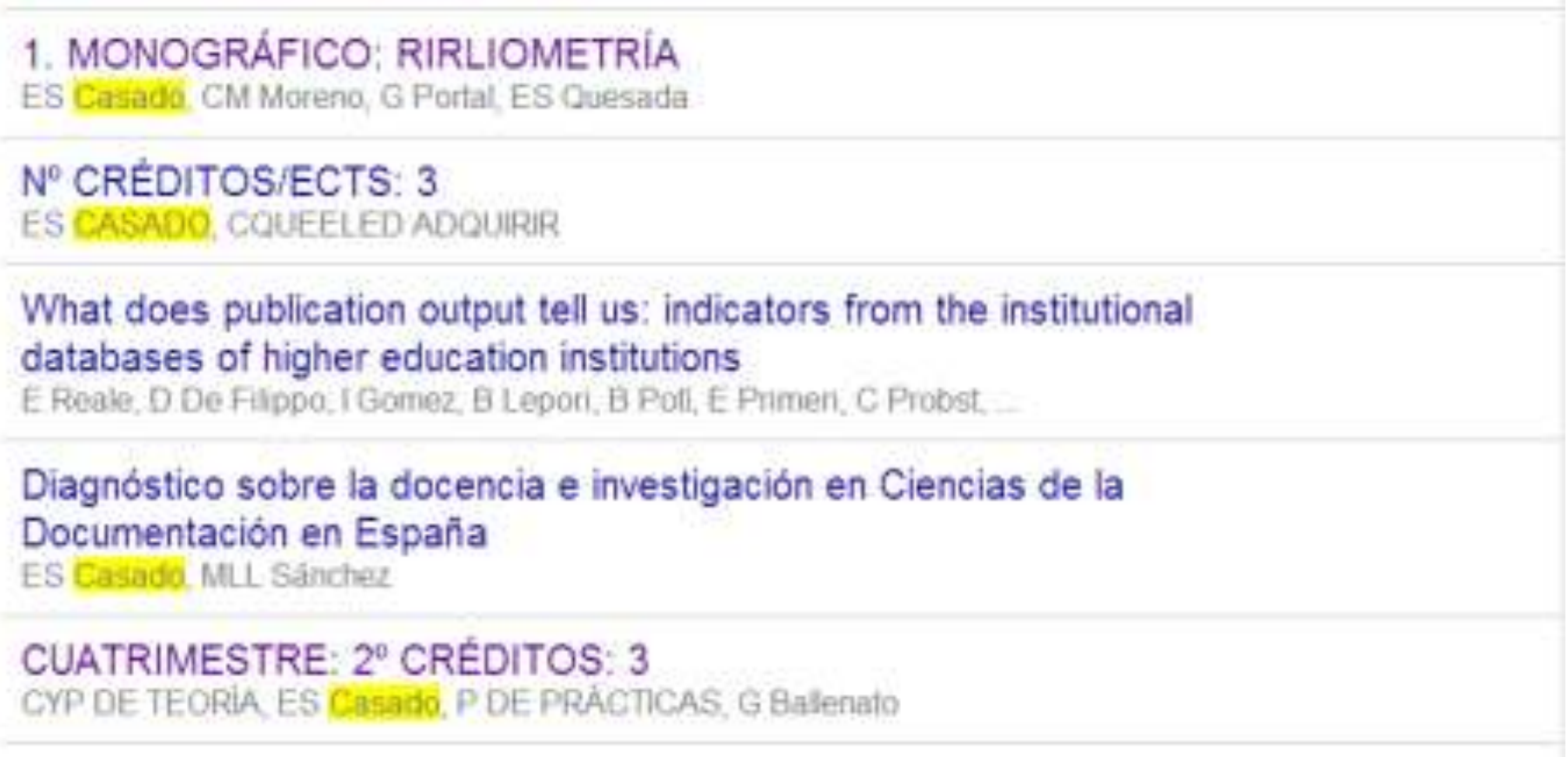

### c) Inclusion of missattributed documents in the profile

The Google Scholar team doesn't oversee the validity of all the information available in Google Scholar Citations. Therefore, it is the sole responsibility of the author that the information visible in his/her profile is accurate. Profiles can be set to be updated automatically (when the system finds an article that it's reasonably sure it's yours, it is automatically added to your profile), or by asking the author for confirmation first when the system thinks an addition or a change should be made. If the user selects the automatic updates, there is a risk that the system will add documents to the profile that the author hasn't actually written, thus falsely increasing the author's bibliometric indicators. The author will probably be completely oblivious to this issue if he or she doesn't check the profile regularly. If that is the case, it shouldn't be considered an active attempt to fake one's bibliometric indicators, but it is still a matter that should be fixed as soon as it comes to the author's knowledge. In Figure 19 we can see an example: the third document (marked in red), which has received 40 citations, hasn't been written by the owner of the profile (Imma Subirats-Coll).

*Figure 19. Misattributed documents in Google Scholar Citations*

| Title 1–20 | Cited by | Year |
|---|---|---|
| Open archives initiative. Protocol for metadata harvesting (OAI-PMH): descripción, funciones y aplicaciones de un protocolo<br>IS Coll, JMB Cruz<br>El profesional de la información 12 (2), 99-106 | 79 | 2003 |
| Open data y Linked open data: su impacto en el área de bibliotecas y documentación<br>MF Peset Mancebo, A Ferrer Sapena, I Subirats-Coll<br>El profesional de la información 20 (2), 165-173 | 48 | 2011 |
| Sistemas integrales para la automatización de bibliotecas basados en software libre<br>O Arriola Navarrete, K Butrón Yáñez<br>Acimed 18 (6), 0-0 | 40 | 2008 |
| E-LIS: an international open archive towards building open digital libraries<br>A De Robbio, I Subirats Coll<br>High energy physics libraries webzine 11, 2005 | 29 | 2005 |

We can find examples where the owner of the profile has participated as a translator or editor of a work (Figure 20). The assignation of the citation counts of a work to the people who have fulfilled this kind of roles is controversial. At the very least, they should make sure that their role is clearly stated and visible in the profile.

*Figure 20. Edition and translation roles in Google Scholar Citations*

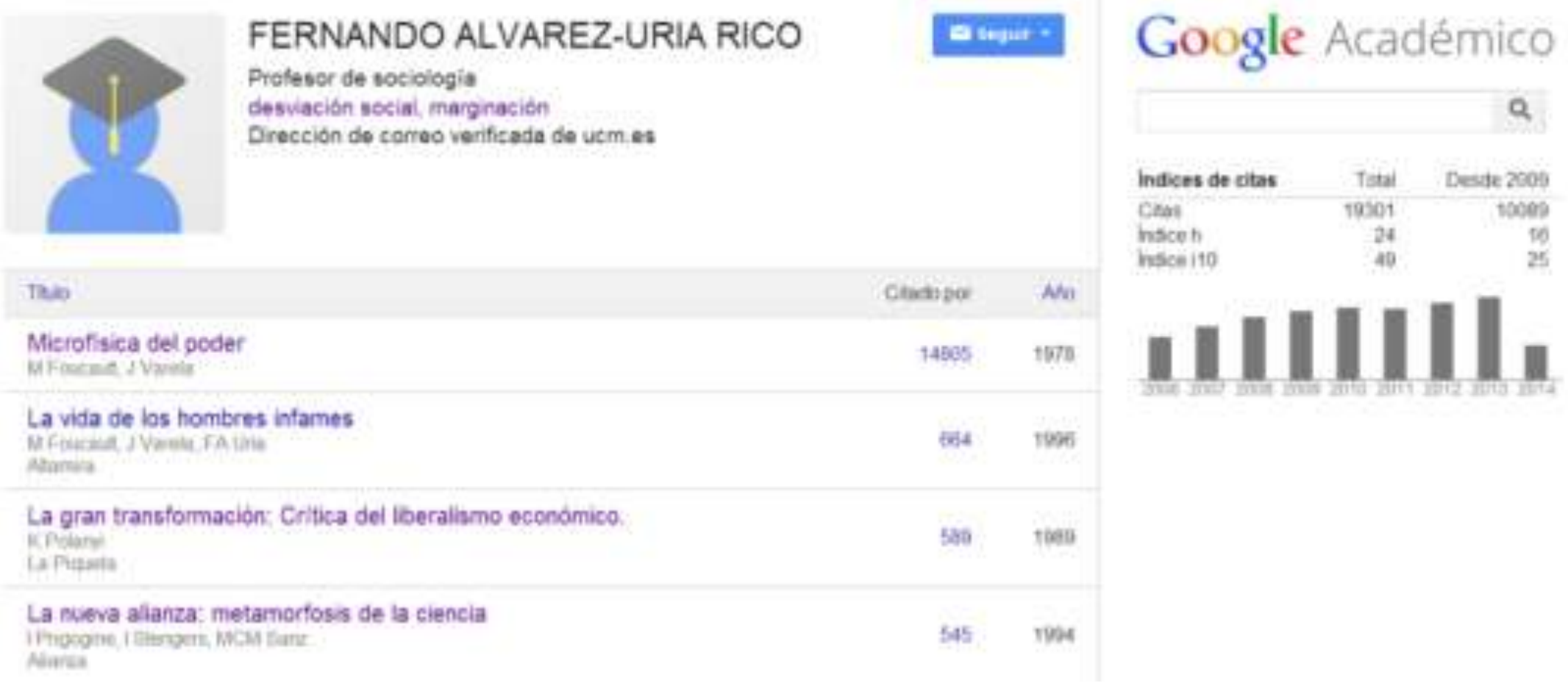

## *d) Deliberate manipulation of documents and citations in Google Scholar*

Another issue is that of the conscious manipulation of profiles by their owners. The fact that anyone, without advanced technical skills, can manipulate his/her own bibliometric indicators, or other people's (Delgado López-Cózar, Robinson-García & Torres-Salinas, 2014) may affect the credibility of GSC academic profiles if no action to control this issue is taken by the *Google Scholar* team. In Figure 21 we observe how uploading a set of fake documents to a repository (with nonsensical text, and a list of references which include the set of documents whose impact you want to boost) will, in just a few days, cause the desired adulteration of citation scores in the profiles of the authors of the referenced documents.

***Figure 21. Effect of data manipulation in Google Scholar Citations***

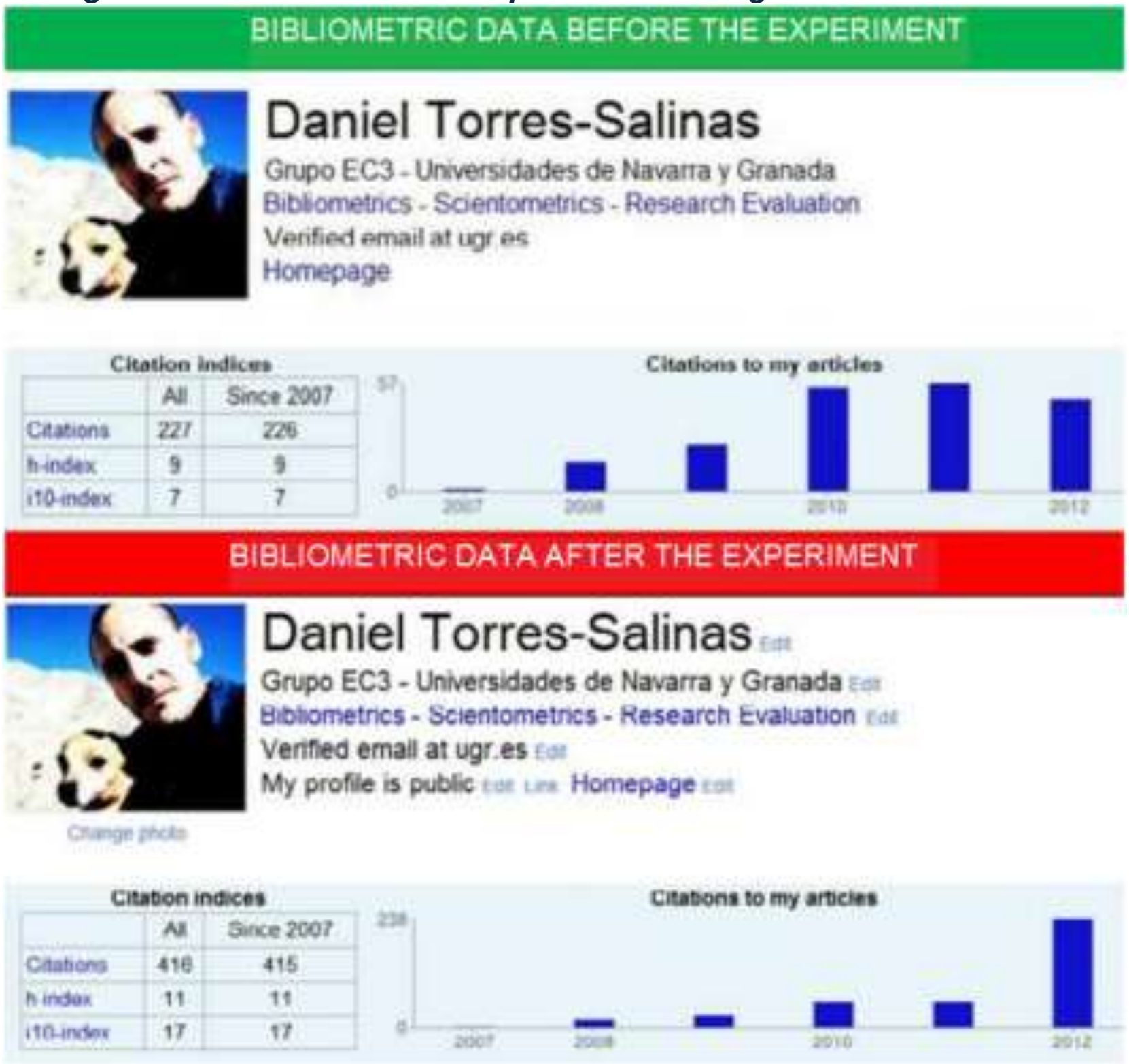

*Source: Delgado López-Cózar, Robinson-García & Torres-Salinas, 2014*

### *e) Duplicate documents in profiles*

This is also a side effect of the cases when Google Scholar fails to group together different versions of the same document. The consequence for the profiles is that the different versions will also be added as different records in the profile, which might affect (positively or negatively) indicators like the h-index and the i-index, which are computed automatically. Fortunately, profile users can manually merge records in their profile, which will solve this issue (Figure 22). This merge only affects the author's profile. It doesn't alter Google Scholar search query results in any way, that is, there will still be two (or more) records for that document in Google Scholar's index, at least until the error gets fixed in a future update.

***Figure 22. Versions not tied in Google Scholar Citations***

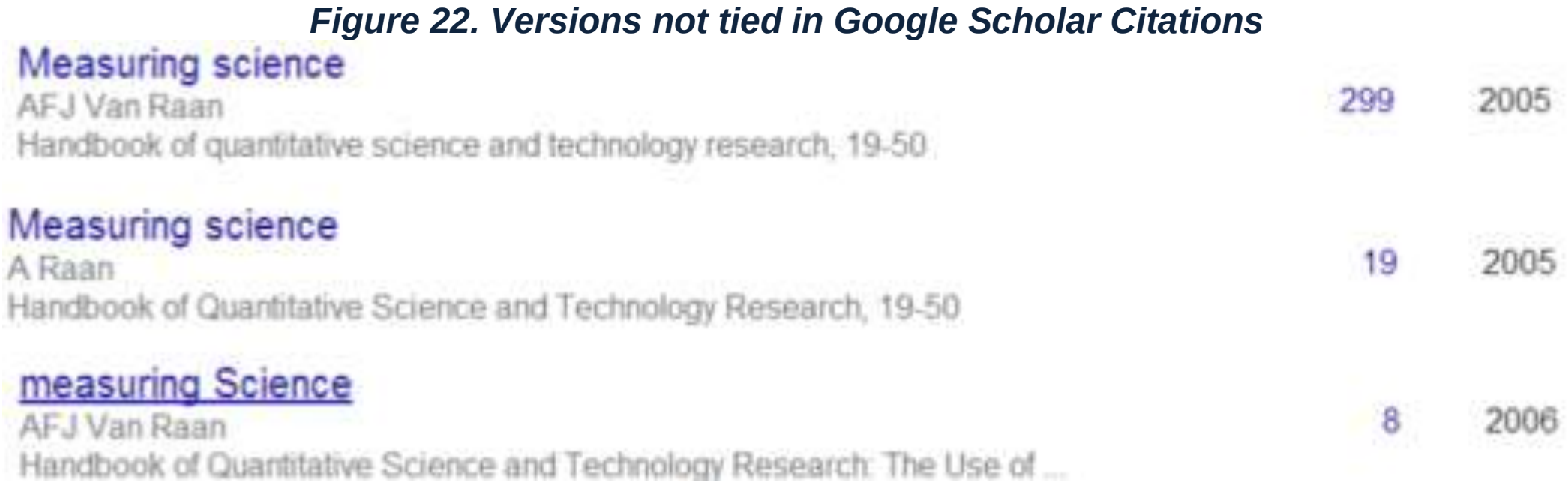

### f) Incorrectly merged documents

The downside to the fact that an author can freely merge documents in his/her profile is, obviously, that incorrect merges (of different documents) can also be made. As we discussed before, Google Scholar doesn't run any validity or accuracy checks on the information displayed in these profiles. Of course, this can also have a distorting effect on the automatically generated author-level indicators.

***Figure 23. Incorrectly merged records in Google Scholar Citations***

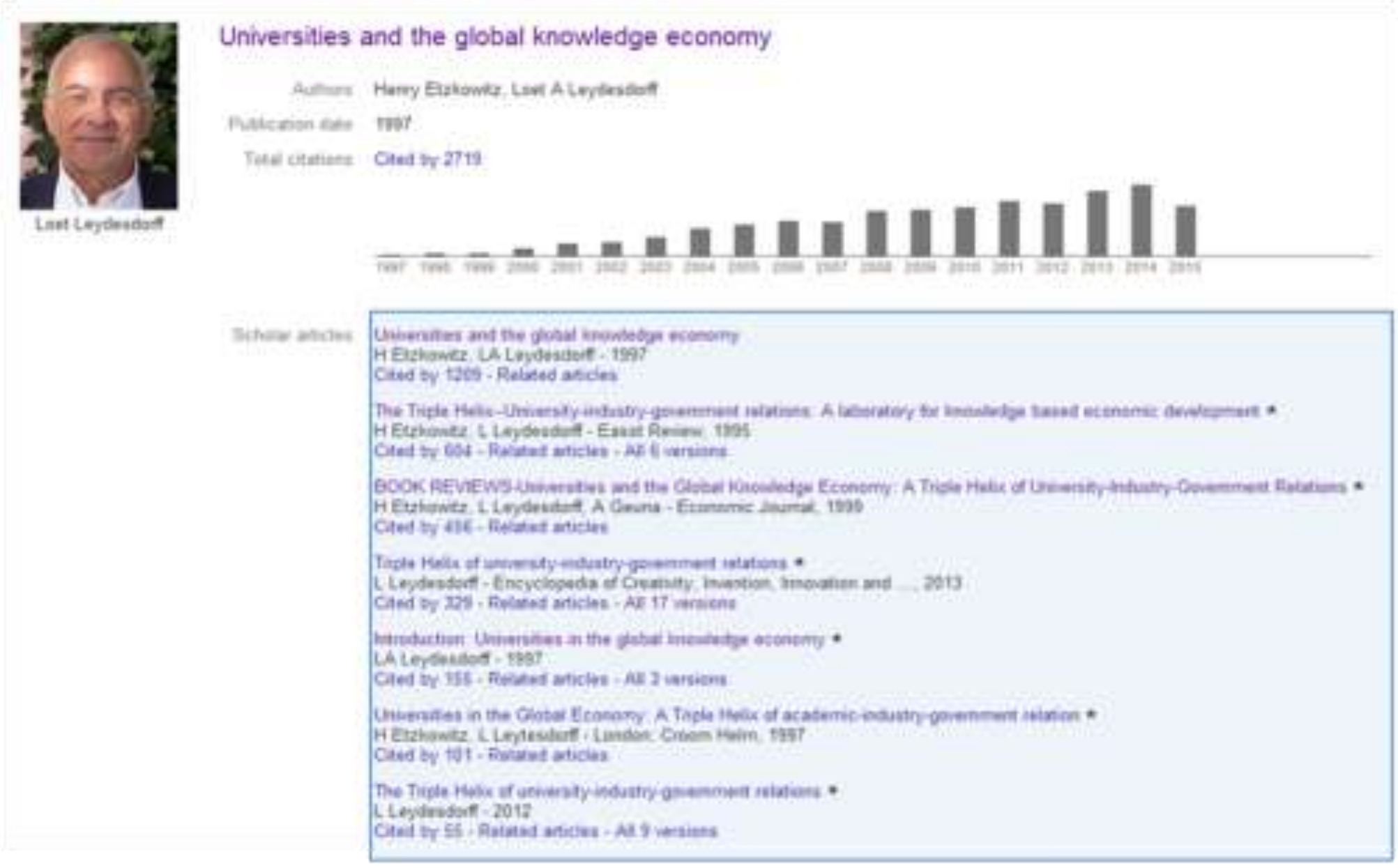

### g) Unclean document titles

This error is also inherited from Google Scholar's metadata parsing errors. Google Scholar Citations allows authors to modify almost all aspects of a record in their profile, including the title of the documents. Unfortunately, not all authors pay attention to such details, and so these errors persist (Figure 24).

***Figure 24. Parse errors in identifying document titles in Google Scholar Citations***

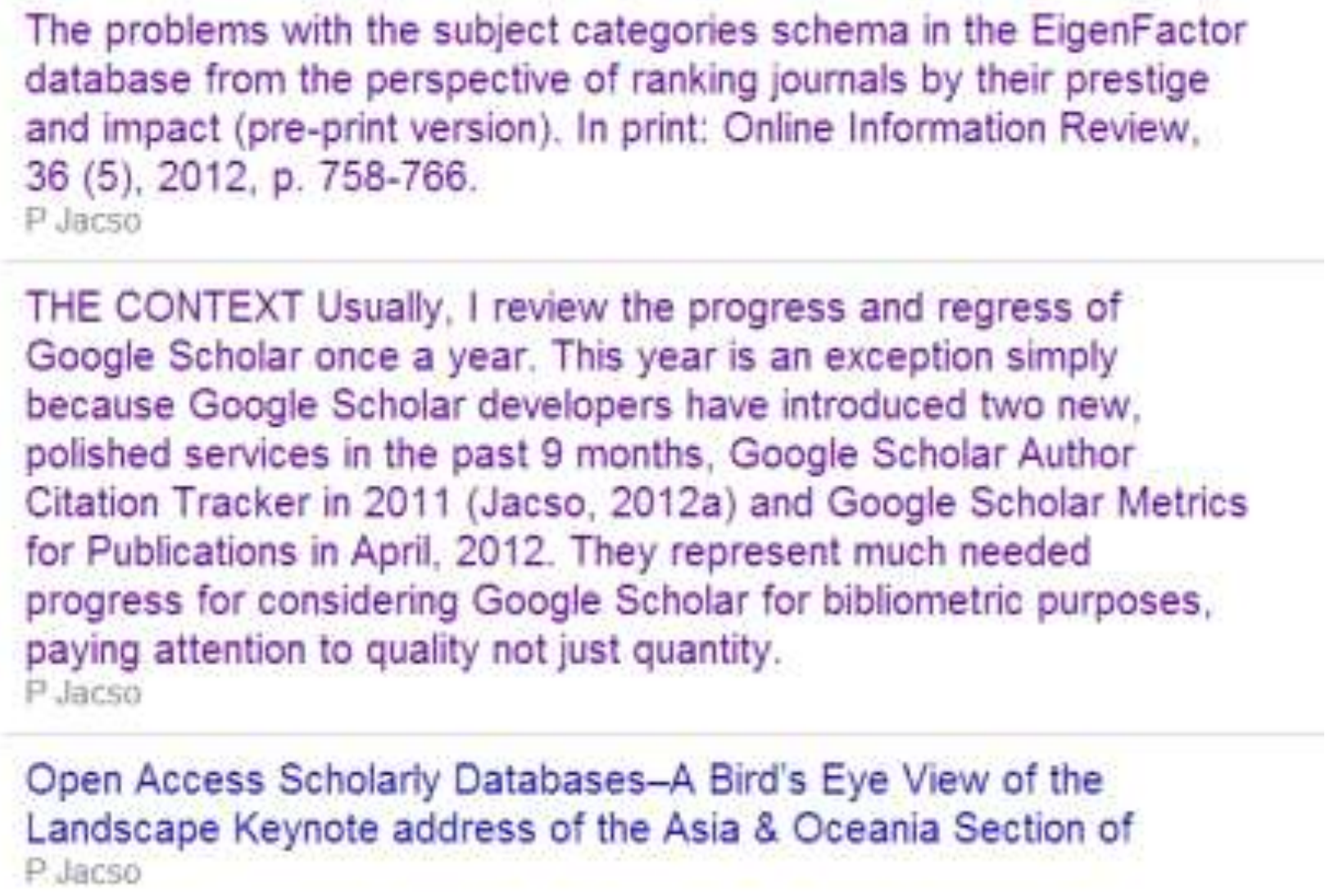

The problems with the subject categories schema in the EigenFactor database from the perspective of ranking journals by their prestige and impact (pre-print version). In print: Online Information Review, 36 (5), 2012, p. 758-766.
P Jacso

THE CONTEXT Usually, I review the progress and regress of Google Scholar once a year. This year is an exception simply because Google Scholar developers have introduced two new, polished services in the past 9 months, Google Scholar Author Citation Tracker in 2011 (Jacso, 2012a) and Google Scholar Metrics for Publications in April, 2012. They represent much needed progress for considering Google Scholar for bibliometric purposes, paying attention to quality not just quantity.
P Jacso

Open Access Scholarly Databases–A Bird's Eye View of the Landscape Keynote address of the Asia & Oceania Section of
P Jacso

### *h) Missing or uncommon areas of interest*

One last limitation that may affect the results of this Working Paper is related to the areas of interest declared by the authors in their profiles (a maximum of five areas can be provided). Researchers in bibliometrics with a public profile in Google Scholar Citations, but haven't declared any area of interest (Figure 24, top), those who use uncommon keywords, or keywords in a language other than English (Figure 25, bottom) may have been overlooked.

***Figure 24. Missing (top) and uncommon (bottom) areas of interest in Google Scholar Citations***

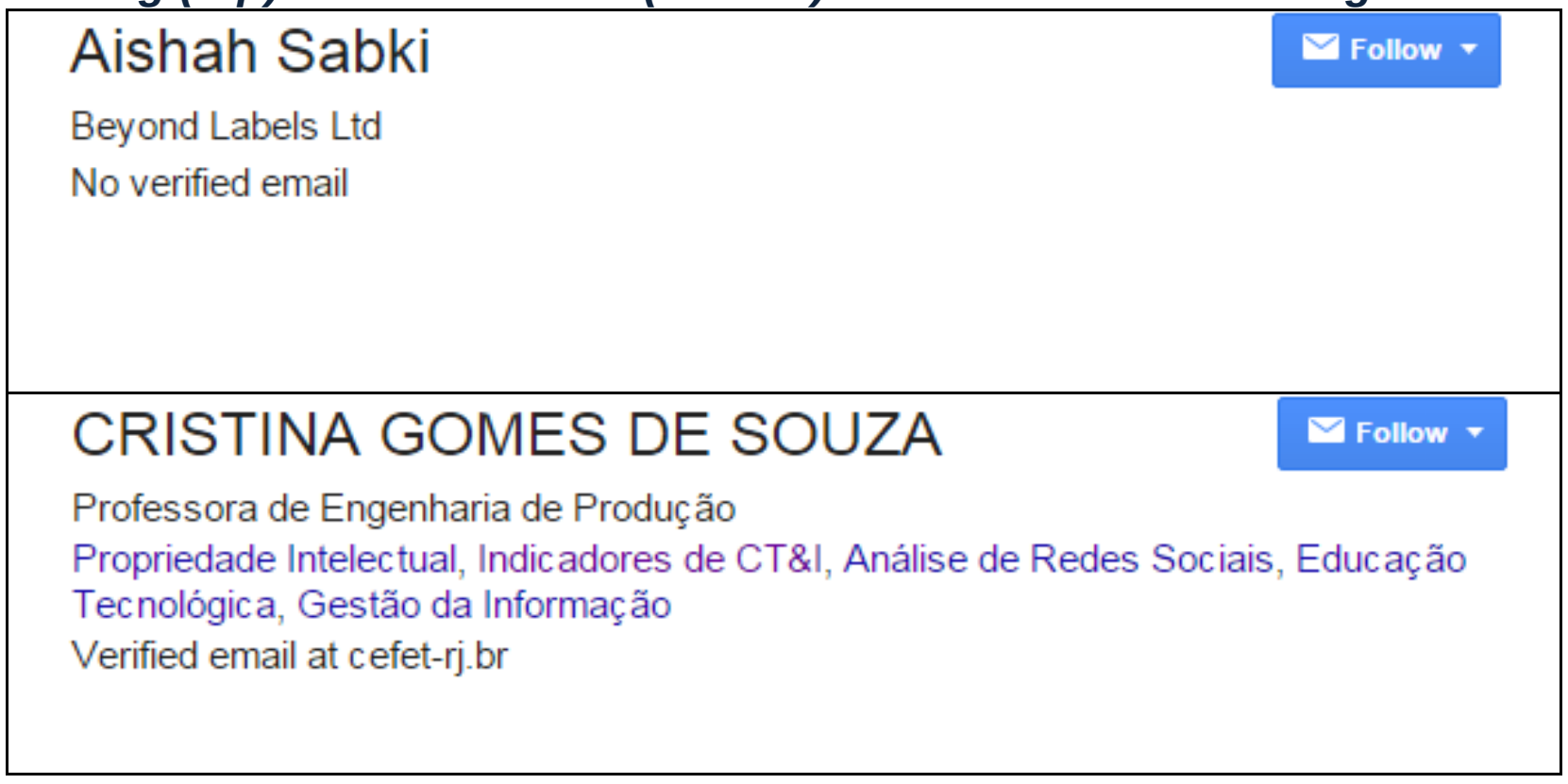

## *3.4.2. ResearcherID*

One of the main shortcomings that characterize *ResearcherID* is the need to manually update the profiles. An author needs to synchronize his/her account with a search in *Web of Science Core Collection* in order to update the list of publications, unlike in Google Scholar and ResearchGate, where the process is largely carried out by the system, and authors only need to confirm new additions or modifications when the system prompts them to do so.

The fact that active manual intervention is needed on the author's part to keep the profile up to date results in a very inconsistent set of data. Authors concerned with online visibility will regularly update their profile, but in the majority of cases, authors will rarely visit their profile again after setting it up the first time. This may explain the results previously shown in Figure 5.

Moreover, we have found additional shortcomings in the system, caused by incorrectly attributed citations in Web of Science, which affect *ResearcherID* profiles.

Let's illustrate this issue with an example in which Dr. Eugene Garfield will be our test subject. In figure 25 we can see the citation metrics for Eugene Garfield's academic profile according to *ResearcherID*, which displays the number of articles published, the sum of times cited, the h-index, and other bibliometric indicators based on data from Web of Science Core Collection. Since Dr. Garfield hasn't created a Google Scholar Citations profile for himself, we generated a private profile in GSC (only accessible by us) in order to compare the indicators provided by the two profile platforms. A screenshot of this profile can be seen in Figure 26.

*Figure 25. Eugene Garfield's academic profile in ResearcherID*

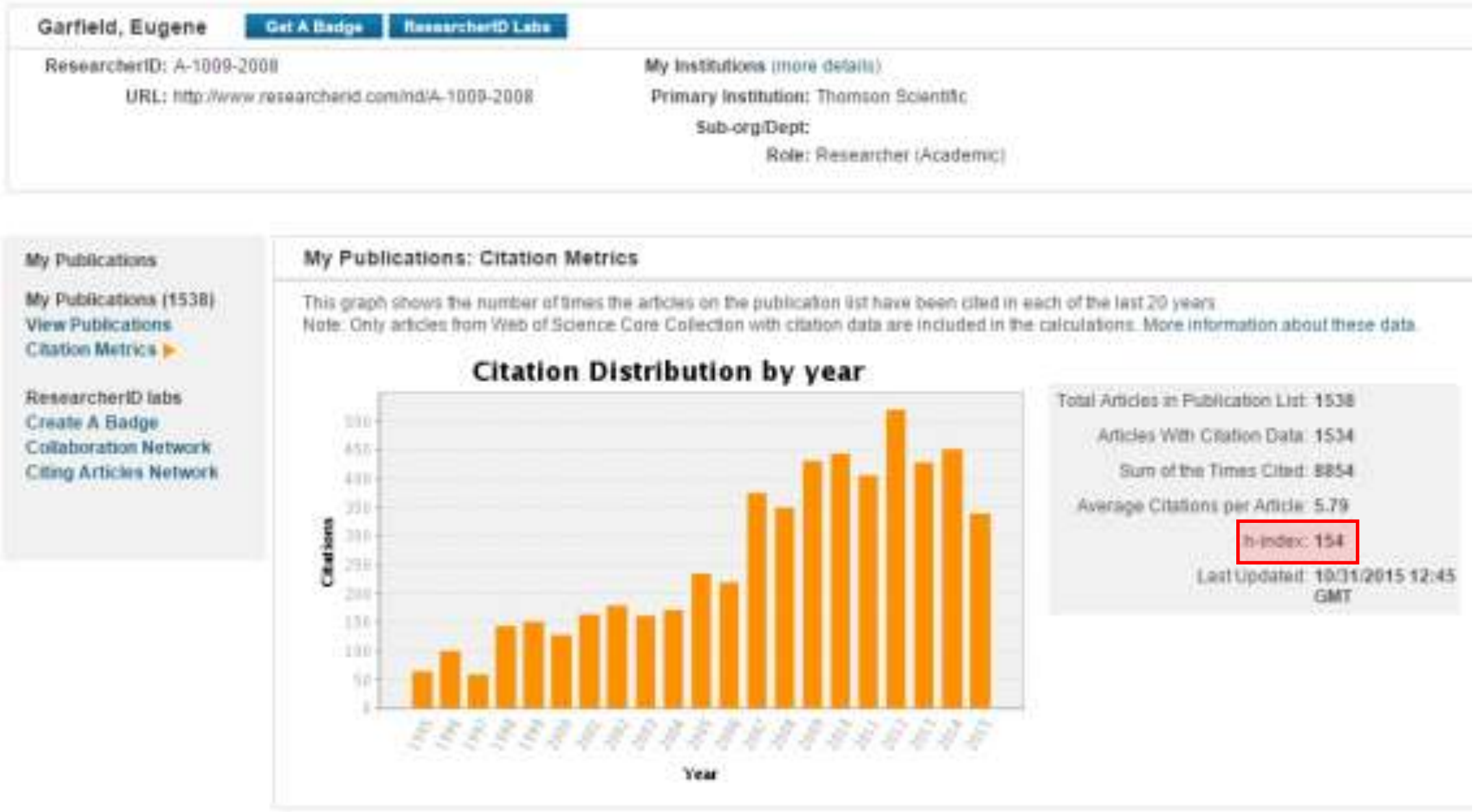

*Figure 26. Eugene Garfield's academic profile in Google Scholar Citations*

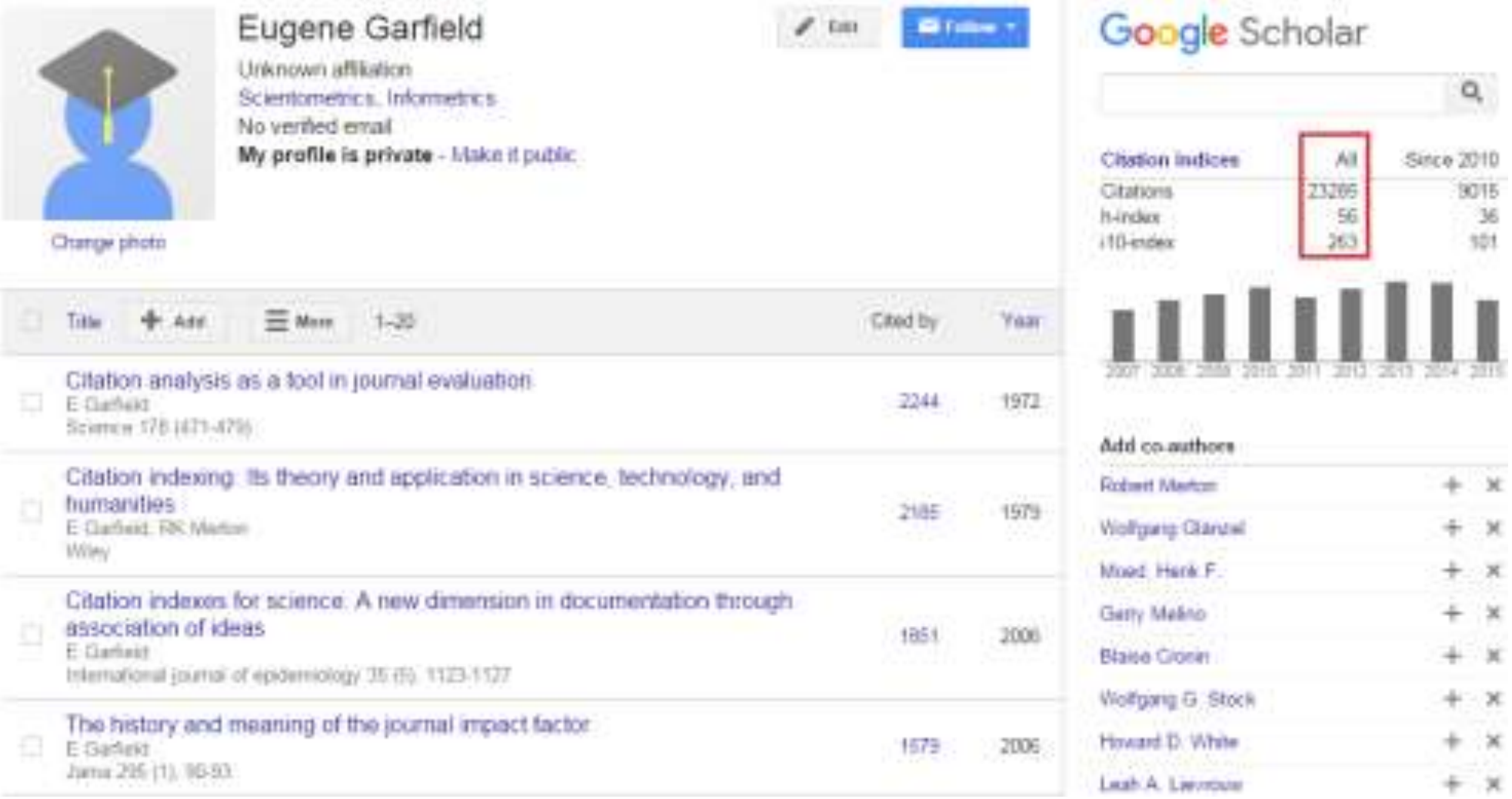

As we can see, there is a huge difference between Dr. Eugene Garfield's h-index according to *ResearcherID* (154) and his h-index according to *Google Scholar* (55). This is caused by a technical error in the data provided by Web of Science. Dr. Garfield's *ResearcherID* profile contains a great number of works published in *Current Contents*, many of them with exactly 200 citations (Figure 27), an odd phenomenon. There is another large group of documents with exactly 155 citations, and other groups of documents which also share the same number of citations.

***Figure 27. Eugene Garfield's publication view in ResearcherID***

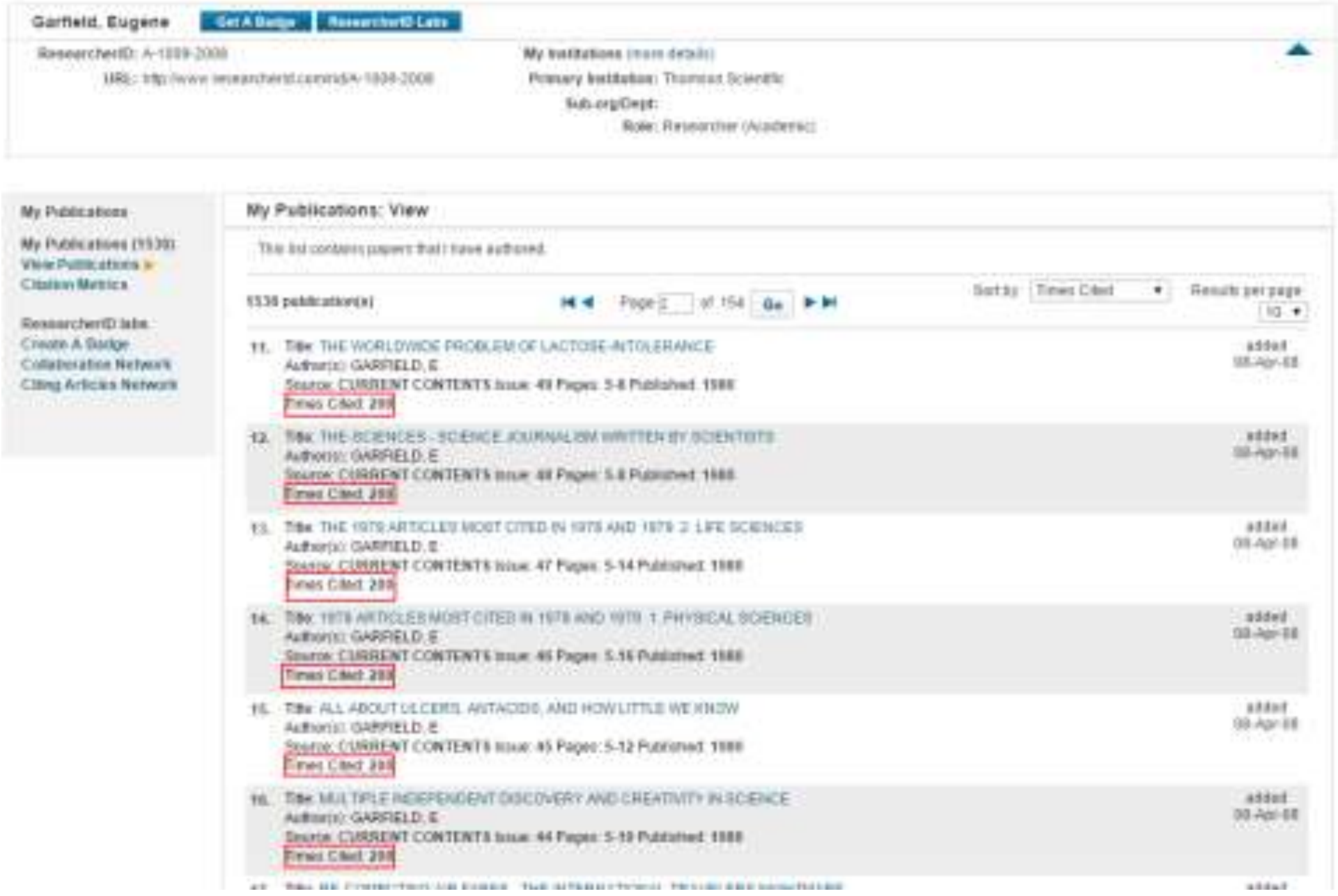

The examination of any of these documents on the *Web of Science* database reveals that all these citations have been incorrectly attributed. In fact, there are some cases where, according to *Web of Science*, a document cites itself (Figure 28). The cause for this error is yet unknown to us and further research is needed to ascertain how often this kind of error occurs throughout the *Web of Science* database.

***Figure 28. Eugene Garfield's citing articles in Web of Science***

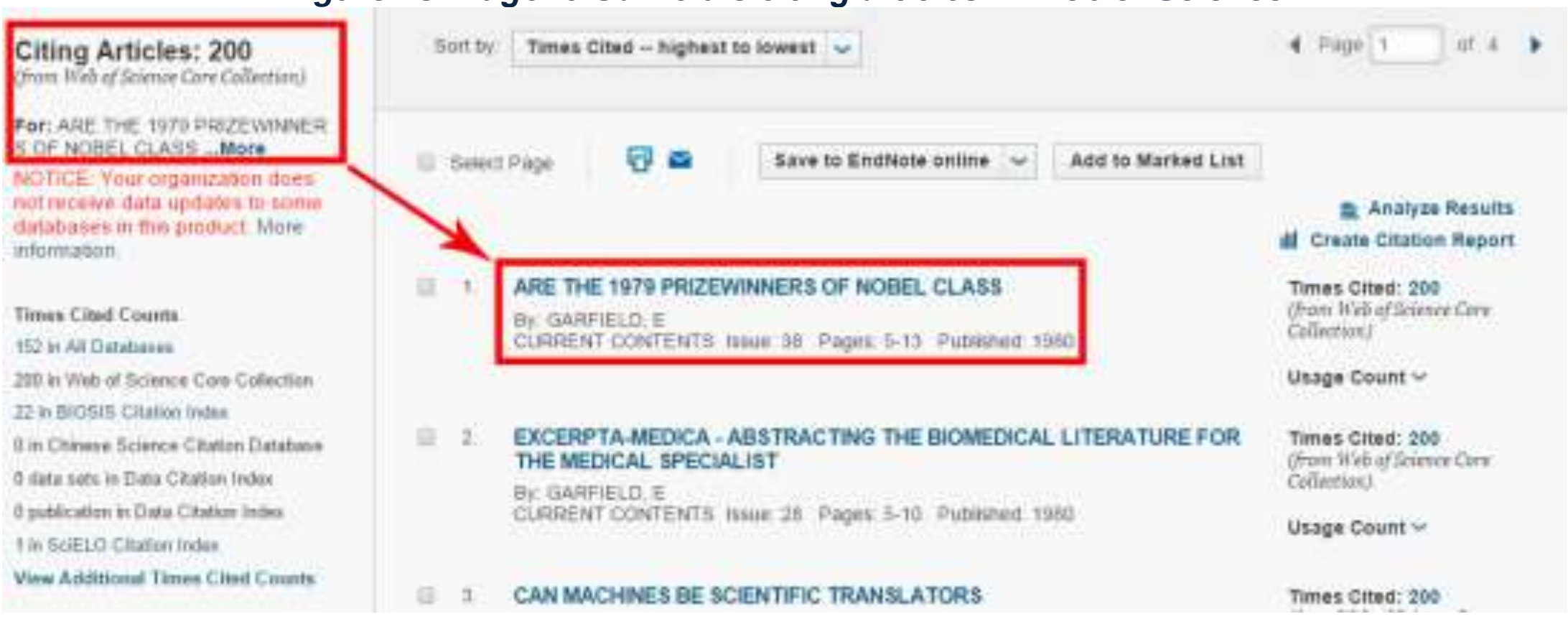

## 3.4.3. Mendeley

An unusual phenomenon was detected while perusing some bibliometricians' profiles in *Mendeley*: many papers published in the *Journal of the American Society for Information Science and Technology* had abnormally high reader counts (number of *Mendeley* users who have saved a certain paper to their collection of references). On November 6th, 2015, a group of *JASIST* articles all exhibited exactly 5,074 readers. Figure 29, a snapshot taken from Mike Thelwall's Mendeley profile, illustrates this phenomenon.

***Figure 29.***

***Mike Thelwall's publications with incorrect reader counts in Mendeley***

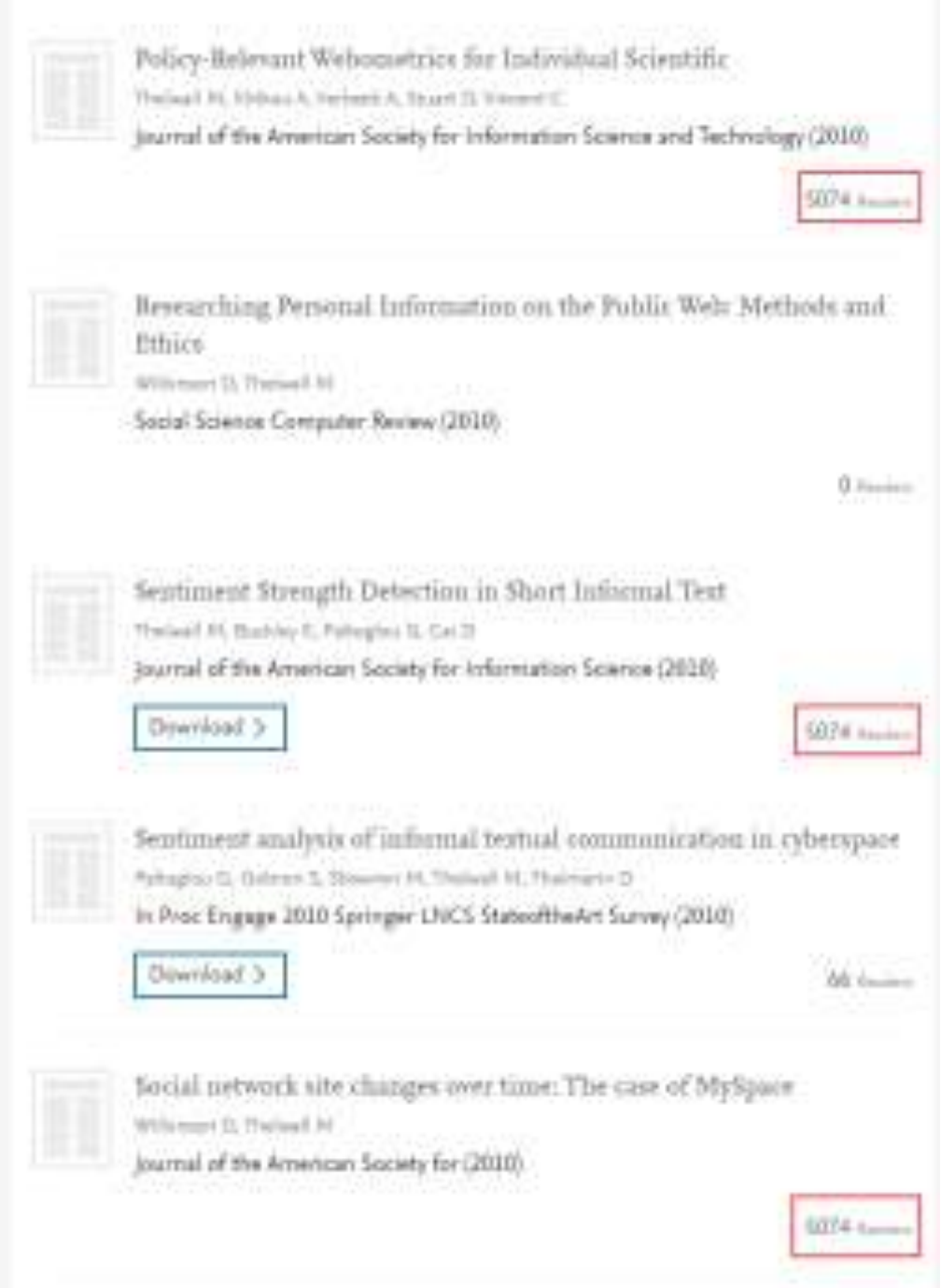

The immediate cause of this issue seems to be that all of these articles had been incorrectly linked to the same paper (Figure 30), which had precisely 5,074 readers. This paper - which doesn't have anything to do with the *JASIST* articles shown previously - could be accessed by clicking on any of the titles of the *JASIST* papers from their authors' profiles. The technical reason why this could've happened is yet unknown.

***Figure 30. Publication causing readership metrics misleading in Mendeley***

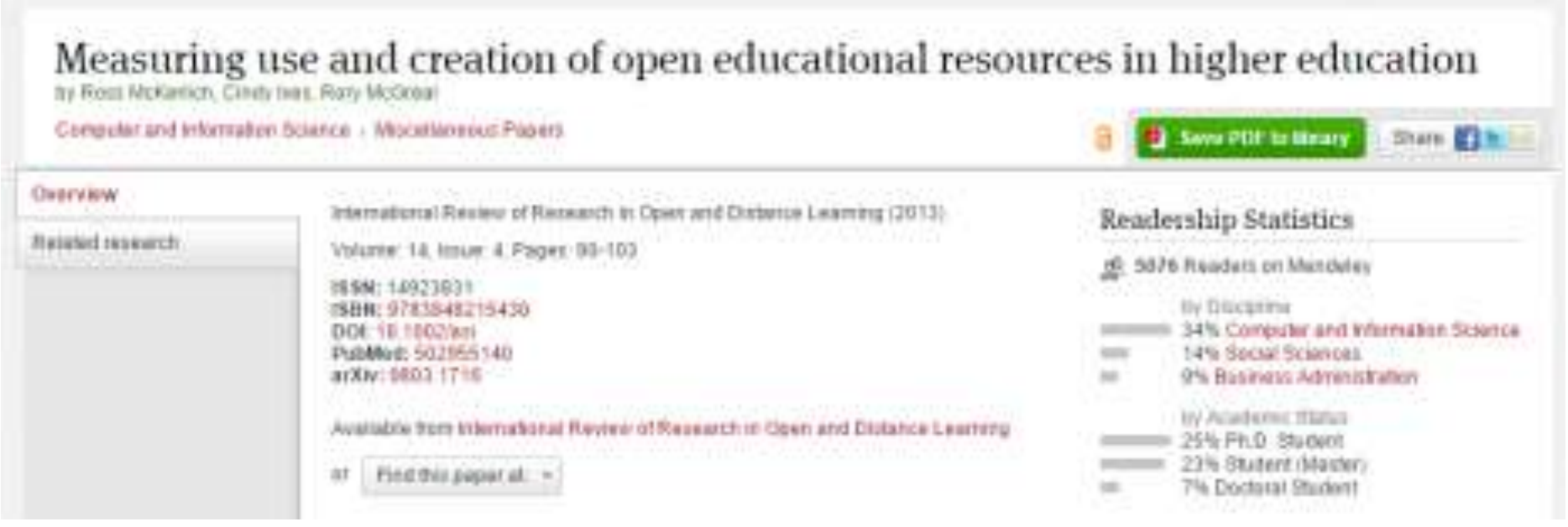

The fact is that this phenomenon has affected several researchers in our study, greatly distorting their aggregate reader counts. The most noticeable case is that of Dr. Mike Thelwall, who has 23 articles affected by this issue in his personal profile, rising his aggregate reader count to 118,046 readers on November 6th (Figure 31), much higher than the count we collected on September (7,423). The error hasn't been fixed yet, and this count keeps growing every day (144,319 by January the 14th, 2016).

***Figure 31. Mike Thelwall's personal profile metrics in Mendeley (6th November 2015)***

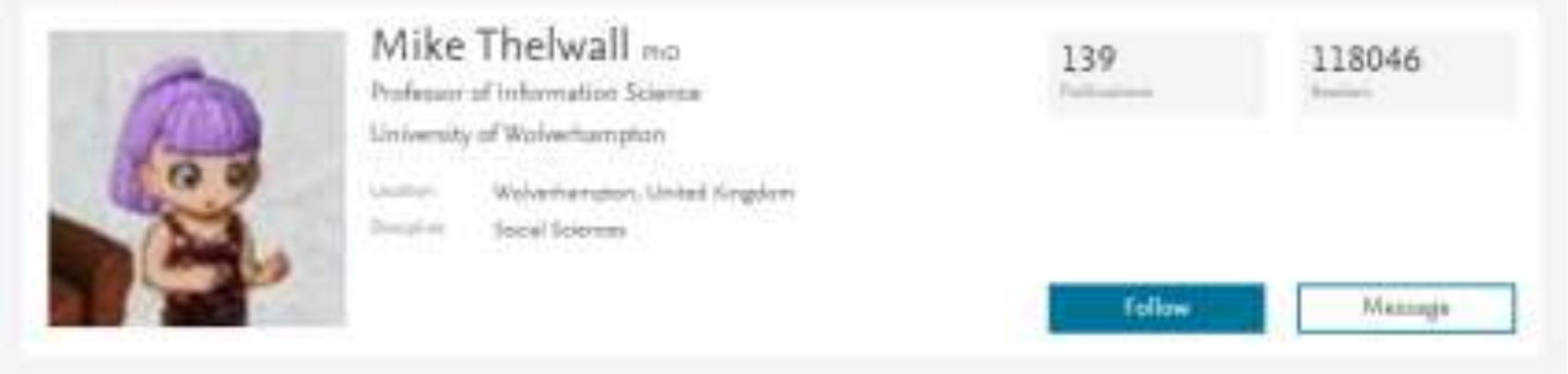

Lastly, it is important to note that if you search any of these documents directly on *Mendeley'*s search feature, the results show the correct (or at least more plausible) reader count for the articles (Figure 32).

***Figure 32. Direct search of documents in Mendeley***

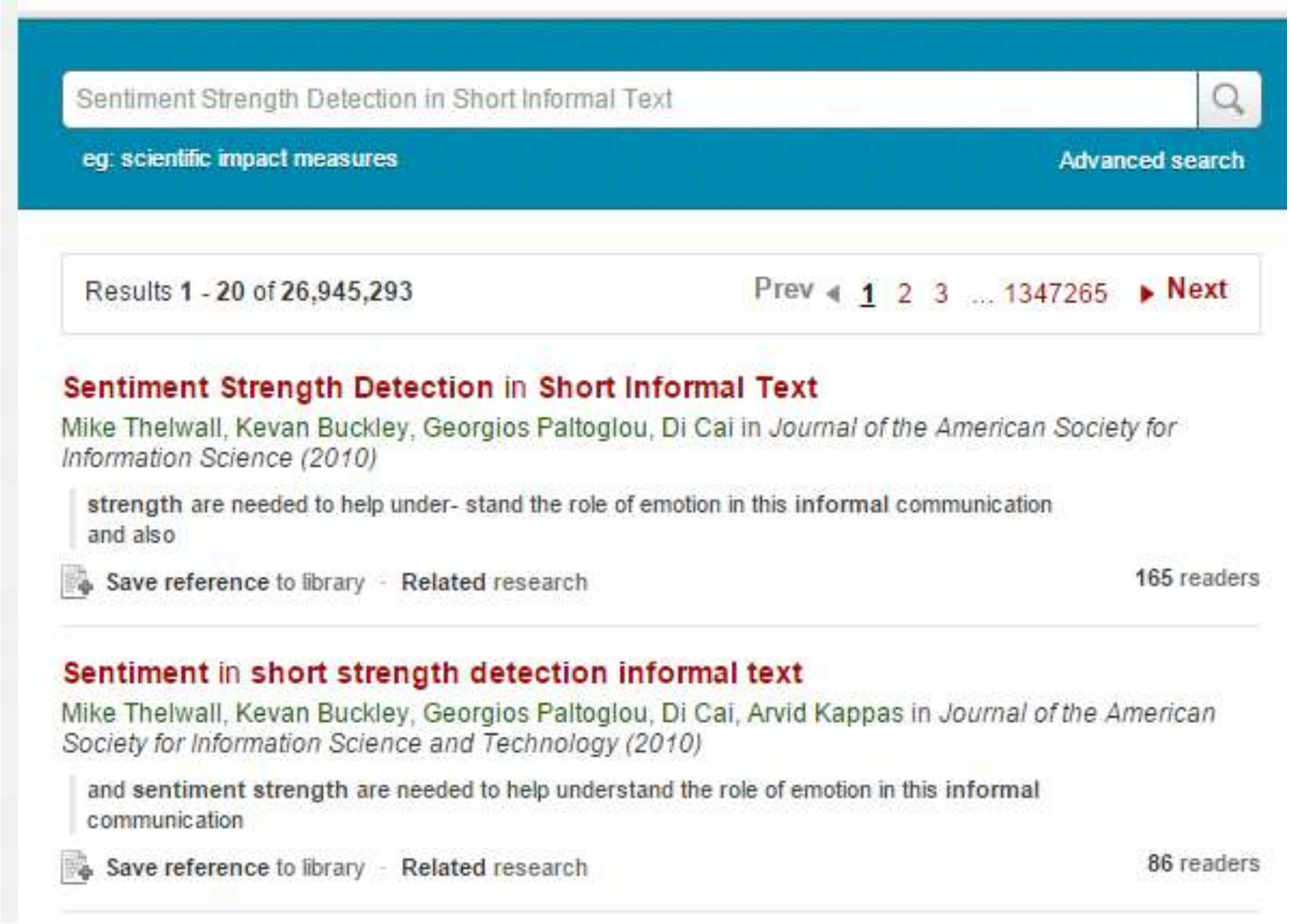

Apart from these anomalous readership metrics in *Mendeley* (that should be understood as an anecdotal mistake that Mendeley will fix soon), we have found other malfunctions caused by errors in the metadata of the references added to the platform, which also affect readership metrics.

In Figure 32 we can see how one author (Arvid Kappas) is missing from one of the two versions of the article "Sentiment in short strength detection informal text". Probably for this reason, *Mendeley* didn't consider them to be the same document, and thus, at some point it created a second record for the document instead of merging it with the version it already had. This, in turn, meant that the reading counts would be split between the two versions of the document (a similar scattering effect to the one found *Google Scholar Citations* with versions and citations, as we previously described).

Not only incorrect metadata can lead to erroneous reader counts, missing metadata can also be dangerous. In Figure 33, taken from Zhigang Hu's *Mendeley* profile on November the 6th, 2015, there are examples of both incorrect metadata (the first article) and missing metadata (the second article) leading to inaccurate reader counts.

***Figure 33. Documents with incorrect or missing metadata affecting Mendeley reader counts***

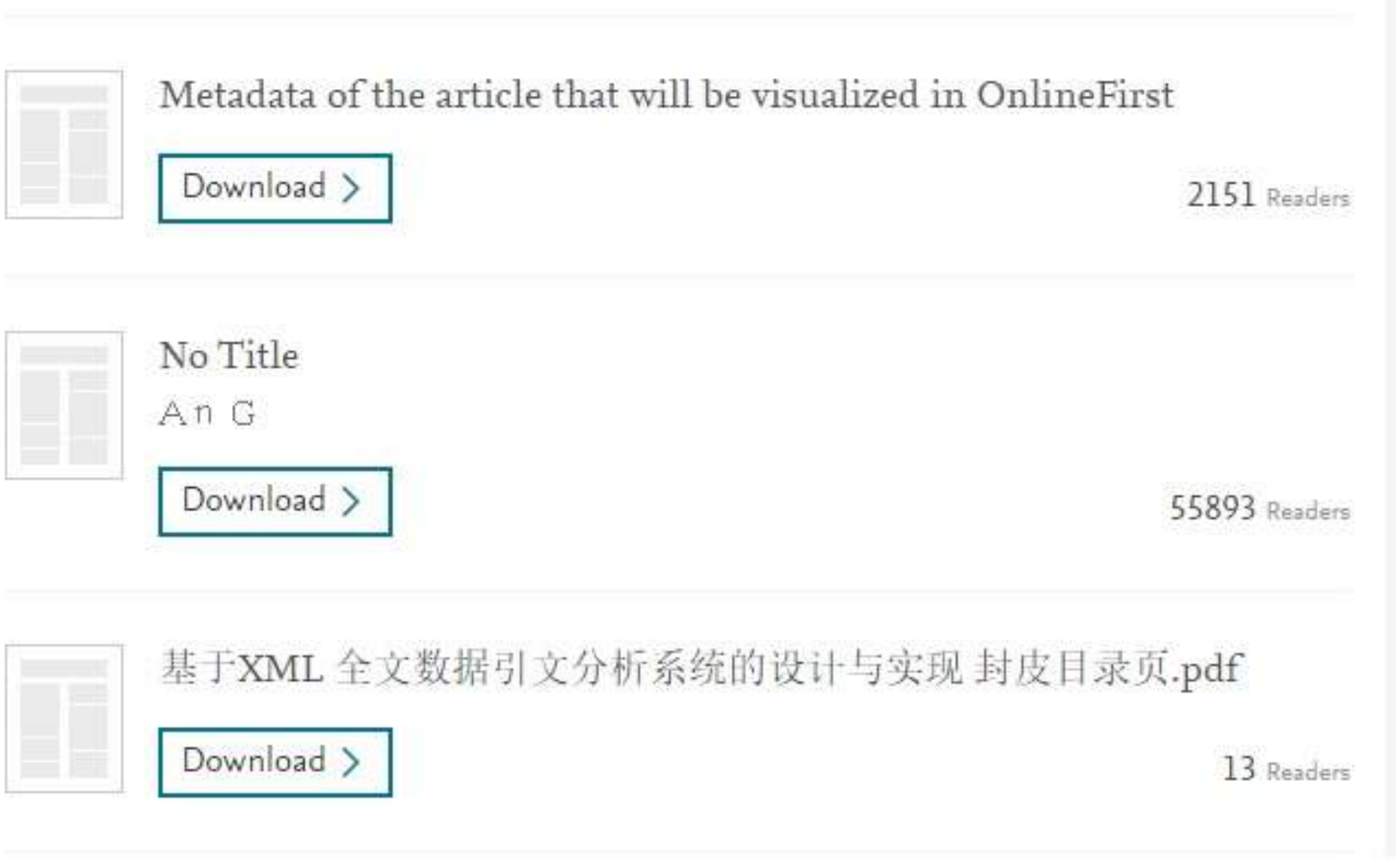

In the first case, the title of one of this researcher's articles wasn't correctly parsed from the PDF of the article, and an incorrect string was selected as the title instead. This is a relatively common issue, so all the articles which have been incorrectly parsed in a similar way and share the same incorrect title "Metadata of the article that will be visualized in *OnlineFirst*" have been lumped together by *Mendeley*, which explains the high reader count for that article. The same explanation could probably be applied to the second document. All documents with a missing title or with the incorrect title "No Title" must have been merged by *Mendeley* to obtain such a high reader count (55,893).

### 3.4.4. ResearchGate

*ResearchGate* (RG), the academic profiling and sharing platform created by Dr. Ijad Madisch[4] and Dr. Sören Hofmayer[5] in 2008, is currently gaining momentum as one most used services of this kind among researchers. In May 2015 they announced they had reached 7 million users,[6] and just five months later, in October, they claimed to have reached 8 million.[7]

The reasons behind the success of this platform are undoubtedly related to the constant stream of new (and usually very convenient) features the platform has been introducing during the past months, but probably also to the constant flow of ego-boosting e-mails that users receive informing them about the great impact their work is having on the scientific community.

4 https://www.researchgate.net/profile/Ijad_Madisch
5 https://www.researchgate.net/profile/Soeren_Hofmayer
6 https://www.researchgate.net/blog/post/celebrating-seven-million-members-and-seven-years-of-researchgate
7 https://www.researchgate.net/blog/post/8-out-of-8-million

Like the rest of platforms fulfilling similar needs, RG computes a set of indicators which are designed to measure the popularity, impact, and degree of use of the documents a researcher uploads to the system (Thelwall and Kousha, 2015). In section 3.3 we observed how these metrics (especially the RG Score) achieved a high correlation with impact metrics provided by *Google Scholar Citations* (especially total citations and h-index). Moreover, this platform was, at the moment we collected the data, the only one who provided both citation and usage metrics for articles (until the Web of Science began to offer usage metrics in November 2015). All these impressive results are partly a consequence of this momentum in terms of user growth.

However, we must point out some important shortcomings related to the lack of transparency in the way all these metrics are computed, a lack of transparency that makes them currently unsuitable for scientific evaluation. It looks like *ResearchGate* is acting like a modern "alchemist", in the sense that it produces its own "concoctions", but without revealing their ingredients and method of preparation to anyone, an issue that, of course, has not gone unnoticed by the scientific community.[8]

First, we may consider the RG Score, which is the indicator they display more prominently in the researchers' profiles, situated right next to the name of the researcher. According to *ResearchGate*[9], this author-level indicator measures "scientific reputation based on how all of your research is received by your peers". The main concern with this indicator - in terms of usefulness for scientific evaluation - is that the way it's calculated hasn't been made public. Therefore, even though this indicator may be a good way to attract researchers who enjoy going on ego trips once in a while, the fact that only *ResearchGate* knows how to calculate it renders it ill-suited for research assessment before the discussion about its intrinsic merits and defects can even begin.

Another matter is that, at the end of September 2015, that is, a few weeks after we collected our data about bibliometric researchers (results offered in sections 3.1 and 3.2), *ResearchGate* combined two of the indicators they used to display on its users' profiles (document views and downloads) into one (Reads).[10]

According to them, "a read is counted each time someone reads the summary or full-text, or downloads one of your publications from *ResearchGate*". However, the "document views" and "download counts" collected in September don't match the "read counts" available after that change (Table 10). We can easily see how "Reads" are clearly lower than the combination of downloads and views. The separation of summary views and document views may have something to do with this issue, and it's a matter that should be further analyzed.

---

[8] http://blogs.lse.ac.uk/impactofsocialsciences/2015/12/09/the-researchgate-score-a-good-example-of-a-bad-metric

[9] https://www.researchgate.net/publicprofile.RGScoreFAQ.html

[10] https://www.researchgate.net/blog/post/introducing-reads

*Table 10. Top 10 authors with the highest Reads counts on ResearchGate (9th of November, 2015), compared to their Downloads and Views counts on the 10th of September, 2015.*

| AUTHOR NAME | SEPTEMBER 10th (2015) | | NOVEMBER 9th (2015) | MISMATCH (%) |
|---|---|---|---|---|
| | DOWNLOADS | VIEWS | READS | |
| Loet Leydesdorff | 32,165 | 42,926 | 21,013 | 27.98 |
| Mike Thelwall | 24,989 | 34,376 | 17,748 | 29.90 |
| Chaomei Chen | 31,579 | 26,734 | 13,452 | 23.07 |
| Nader Ale Ebrahim | 31,853 | 23,144 | 10,282 | 18.70 |
| Lutz Bornmann | 13,556 | 22,987 | 9,863 | 26.99 |
| Maite Barrios | 14,234 | 7,600 | 9,439 | 43.23 |
| Wolfgang Glänzel | 10,572 | 20,145 | 9,439 | 30.73 |
| Félix Moya Anegón | 18,691 | 23,583 | 8,625 | 20.40 |
| Cassidy Sugimoto | 13,079 | 8,081 | 8,458 | 39.97 |
| Ronald Rousseau | 8,066 | 19,118 | 6,934 | 25.51 |

The same thing can be said about the "profile views" indicator: the counts obtained back in September are always higher than the ones available two months later on November the 9th (Table 11). To the best of our knowledge, there has not been an announcement regarding any changes in the profile views indicator.

*Table 11. Top 10 authors with the highest profile view counts on ResearchGate (9th of November, 2015), compared to the same indicator on the 10th of September, 2015.*

| AUTHOR NAME | SEPTEMBER 10th (2015) | NOVEMBER 9th (2015) | MISMATCH (%) |
|---|---|---|---|
| | PROFILE VIEWS | PROFILE VIEW | |
| Nader Ale Ebrahim | 19,821 | 13,281 | 67.00 |
| Chaomei Chen | 7,760 | 3,937 | 50.73 |
| Loet Leydesdorff | 4,227 | 1,758 | 41.59 |
| Bakthavachalam Elango | 2,883 | 1,756 | 60.91 |
| Zaida Chinchilla | 5,840 | 1,569 | 26.87 |
| Mike Thelwall | 4,297 | 1,568 | 36.49 |
| Lutz Bornmann | 3,129 | 1,439 | 45.99 |
| Wolfgang Glänzel | 3,012 | 1,301 | 43.19 |
| Kevin Boyack | 3,256 | 1,135 | 34.86 |
| Peter Ingwersen | 2,335 | 1,025 | 43.90 |

In any case, a high Pearson correlation between the sum of *Downloads* and *Views*, and the new *Reads* indicator ($r = 0.93$, $n = 499$; $\alpha = 0.95$; p-value < 2.2e-16) is observed; and also between the Profile View counts collected in September and the ones collected in November ($r = 0.93$; $n = 535$; $\alpha = 0.95$; p-value < 2.2e-16).

### 3.4.5. General strengths and shortcomings of academic profiles

Lastly, Table 12 summarizes the main strengths and weaknesses of each of the platforms analyzed in this study.

*Table 12. Advantages and disadvantages of academic profiles provided by social platforms*

| ADVANTAGES | DISADVANTAGES |
|---|---|
| **GOOGLE SCHOLAR CITATIONS** | |
| ▪ Widest coverage (all languages, sources and disciplines)<br>▪ User-friendly<br>▪ High growth rate<br>▪ Automatic updates<br>▪ Alerts (new citations to your work, or publications from other authors) | ▪ Scarce quality control<br>▪ Open to manipulation<br>▪ Inherits mistakes from Google Scholar |
| **RESEARCHERID** | |
| ▪ Offers advanced bibliometric indicators | ▪ No automatic updates<br>▪ Not very user-friendly<br>▪ Inherits mistakes from WoS<br>▪ Not used by many authors<br>▪ Only WoS CC publications count towards citation metrics |
| **RESEARCHGATE** | |
| ▪ Increasingly used by the scientific community: very high growth rate<br>▪ Offers usage data (views and downloads)<br>▪ User-friendly<br>▪ Correlates with citation data<br>▪ Social functions to contact other authors | ▪ No automatic updates (one co-author must upload the document)<br>▪ Lack of transparency in its indicators<br>▪ Still not used by many authors<br>▪ Sends too many e-mails (by default) |
| **MENDELEY PROFILES** | |
| ▪ Increasingly used by community<br>▪ Offers usage data (reads)<br>▪ Correlates with citation data<br>▪ Allows discipline analysis<br>▪ Social functions (follow other authors) | ▪ No automatic updates<br>▪ Quality of metadata depends on user input |

It is clear that none of the platforms considered and analyzed in this Working Paper is without its problems and limitations. At the same time, all of them offer new insights for measuring scientific impact.

*Google Scholar* offers the widest coverage, situated on approximately 160 million hits on May 2014 (Orduna-Malea et al, 2014). Its indexing criteria (all academic documents openly stored in the academic web space) makes this database the only place where every academic document is indexed regardless of its typology (not only journal articles but also books, book chapters, reports, thesis dissertations, conference proceedings, etc.), its language, or its discipline. Thanks to this wide variety of sources, *Google Scholar* is able to measure not only scientific but also educational and professional impact in the broadest sense of the term. At the same time, as regards strict scientific impact, there is a high correlation (r = 0.8) between the number of citations of these documents in GS and their citations in WoS (Martin-Martin et al, 2014).

*Google Scholar Citations* includes citation scores for authors, areas of interest, and institutional information. Additionally, in this platform, the owner of the profile can improve the bibliographic information provided by Google Scholar, and merge duplicates Google Scholar hasn't been able to detect. This impressive collection of data, together with the development of functionalities (such as detecting and merging duplicates), makes *Google Scholar* the best tool for the bibliometric analysis of some disciplines, especially those within the areas of the Humanities, Social Sciences, and Engineering.

Unfortunately, Google Scholar is not without its problems. The possibility to edit records in the profiles does not solve its parsing problems, for which there doesn't seem to be a clear explanation sometimes. We must point out however that the system is improving year by year. Moreover, in an academic big data environment, these errors (which we deem affect less than 10% of the records in the database) are of no great consequence, and do not affect the core system performance significantly.

On the other hand, the philosophy of the product (oriented to the user, lacking any bibliographic control) makes the tool rather open to confusing data, mistakes (described in section 3.4.1), and to manipulation, a really serious problem in the academia at the moment. Scientific misconduct should not be disregarded as mere spam.

Moreover, *Google Scholar* is user-friendly but not bibliometrician-friendly. Google Scholar's agreements with big publishers to collect data from their servers and present them in the search engine come at a price: among other things, the impossibility of offering an API which would no doubt be highly welcomed by the scientific community. An API would allow us to keep working on our understanding the production, dissemination, and consumption of scientific information worldwide.

*ResearchGate* is the second most-used platform among the tools analyzed in this work. The high number of users that this platform is currently attracting reinforces the validity of the metrics it provides (essentially because of the great amount of documents that have been already uploaded to the system). This is reflected in the extraordinary correlation that RG Score achieves with the h-index and total citations from *Google Scholar*. Moreover, there is no better platform to calculate number of downloads per document.

We believe this is a logic result, because the RG Score is basically made up of the number of publications an author has published, the citations to these publications, and the JCR Impact Factor of the journals where these articles are published. Usage indicators may also have some weight, but not much yet.

Nonetheless, the lack of transparency in the calculation of the different metrics (especially the RG Score) prevents it from being useful, since they cannot be replicated.

This the reason why the following questions still arise: what was *ResearchGate* really measuring before the changes in the View and Download indicators took place? What is it really measuring now? Why isn't *ResearchGate* more open about the way it computes the indicators they display?

Moreover, the introduction of subjective values (such as the participation in question & answers in the platform) may introduce some bias (high participation in the social platform does not have anything to do with academic impact, though it serves to incentive the use of the platform). In any case, the weight of this parameter doesn't seem to be significant.

Likewise, changes in the company policies, such as the elimination of some services (the complete list of documents ranked according to number of reads is no longer available), makes this platform unpredictable and unreliable at the moment. Other specific limitations are related to the quantity of documents indexed in the platform; references not properly identified, or incorrectly attributed citations.

Regarding *Mendeley*, we should acknowledge the validity of the Readers indicator, which strongly correlates to both the Downloads indicator provided by *ResearchGate* (different sides of usage) and to citation-based metrics from *Google Scholar*. However, we found some limitations in this platform while studying the Bibliometric community (which may be extrapolated to other academic communities).

First, calling the number of users that have saved a bibliographic record in their personal collection "readers" is absolutely incorrect (Delgado López-Cózar and Martín-Martín, 2015). The term should be changed to one that more accurately represents the nature of the indicator, because the current one can lead to misunderstandings and misinterpretations.[11]

Second, the fact that there are no automatic profile updates makes the system completely dependent on user activity. A total of 149 out of the 336 profiles analyzed (44.3%) didn't include a single document (Figure 34), and only 23% of the researchers have an effective presence in the platform. This fact strongly limits the use of *Mendeley* for the purpose of evaluating authors.

***Figure 34. Example of empty academic profile in Mendeley***

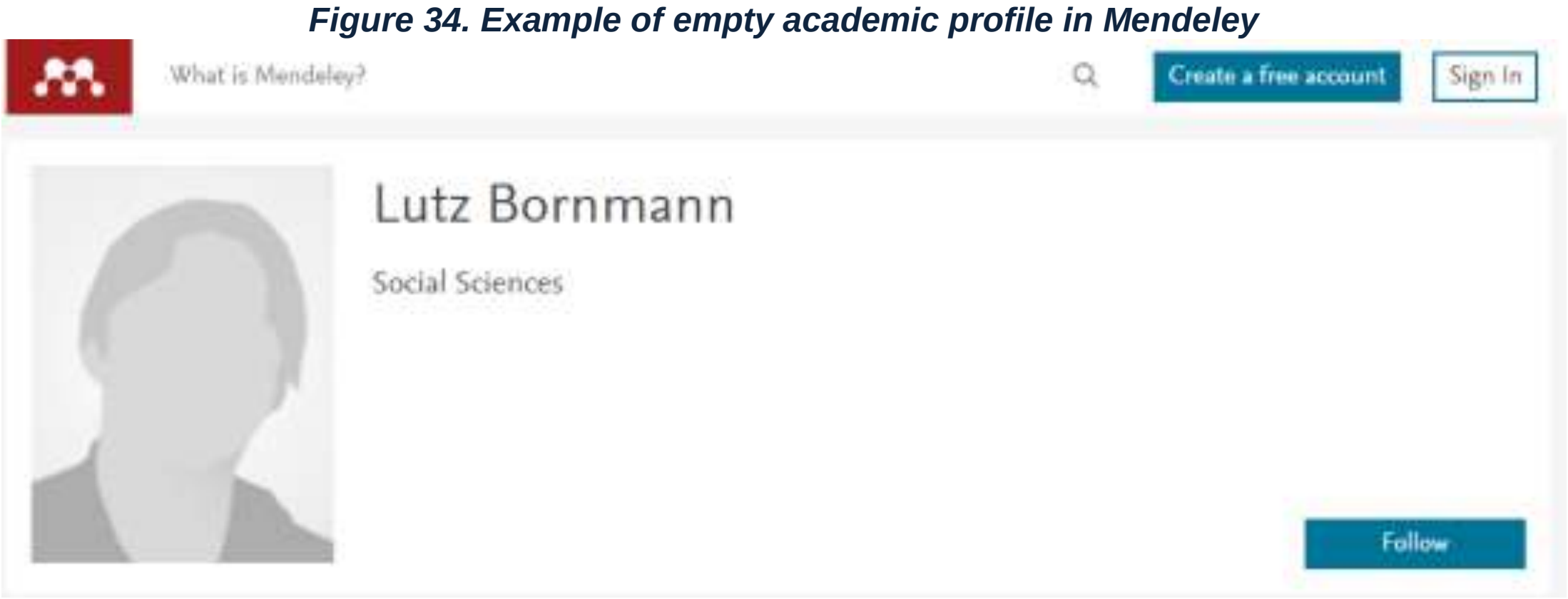

The last academic profiling service we analyzed was *ResearcherID*. There is no automatic profile updates in this platform, and a great percentage of user profiles (34.4%) have no public publications displayed (Figure 35), that is, the profile only contains basic information about the subject interests of the author and its affiliation. Only 26% of the authors in our sample had a ResearcherID profile with at least one document, and most of these profiles were out of date.

[11] The new "Reads" metric provided by *ResearchGate* suffers from the same problem, as it is combining online accesses to the document and downloads, which are not the same even though they claim they are.

*Figure 35. Example of an empty academic profile in ResearcherID*

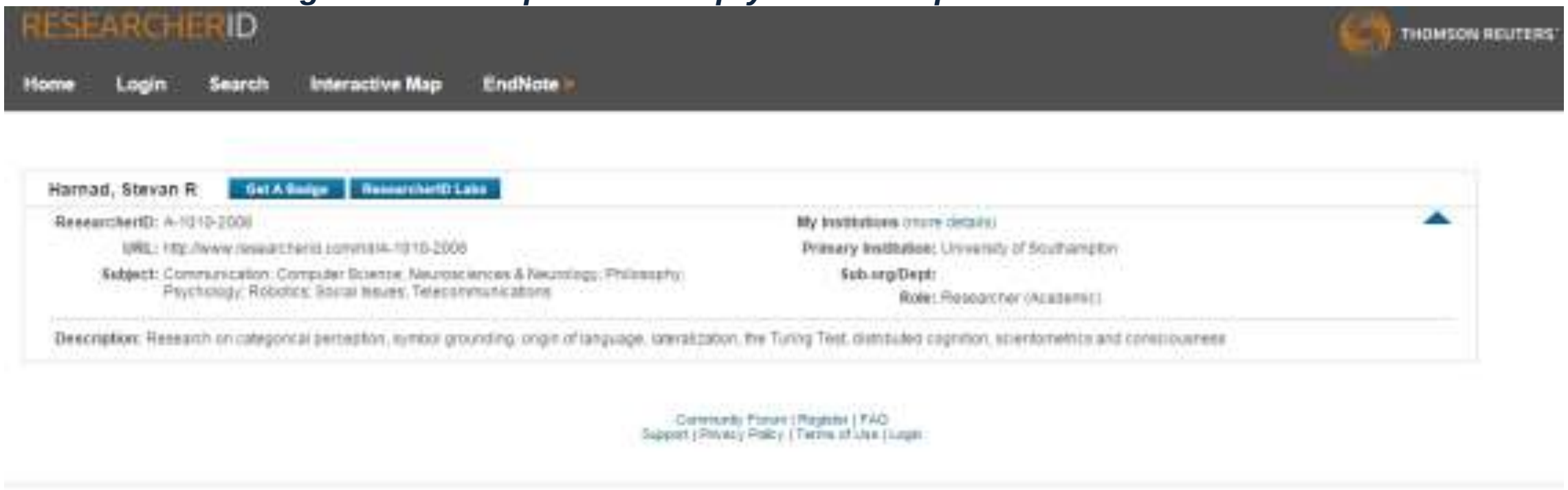

Apart from this lack of real use, we found several errors which had been inherited from the citation scores available in the Web of Science. That is, WoS is not error-free in the attribution of citation scores. For all these reasons, we do not consider *ResearcherID* a valuable tool for bibliometric purposes.

# 4. CONCLUSIONS

Although this work is focused on the analysis of a specific academic community (Bibliometrics), the results obtained allowed us to obtain a number of important findings, summarized below.

Firstly, *Google Scholar* (with its associated platform for academic profiles *Google Scholar Citations*), provides a very precise and accurate picture of the bibliometric community. The data collected, not only at the author-level but also at the document-level and source-level (journal and books), clearly responds to our mental image of the field. That is, *Google Scholar* helped identify the most influential authors (core and related) and sources (journals and publishers) in the discipline.

The level of use of other social platforms is quite far from the one found for *Google Scholar Citations*, not only in the number of user profiles created, but also in the regularity with which they are updated. *ResearchGate*'s growth rate is impressive and currently stands as the second most used profile platform by the bibliometric community. Its usage indicators (*Downloads* and *Views*) and its social network features (communication and information sharing among users) provide a perspective that *Google Scholar Citations* lacks.

The social tools analyzed here have a number of significant limitations, which clearly get in the way of generating academic mirrors complementary to those based merely on citations. In the case of *ResearchGate* these limitations are caused by the opacity of the indicators and unexpected changes in the policies of the company, whereas in *Mendeley* and *ResearcherID* the problems arise from the existence outdated profiles. This issue has a negative effect on the accuracy of the information provided by these platforms, as seen in Figure 5.

*Twitter,* on the other hand, presents a completely different picture. Its author-level indicators do not correlate with citation-based indicators (from *Google Scholar*) nor with usage indicators (provided by *ResearchGate* and *Mendeley*), but they do correlate with other network indicators (which measure an author's participation in the community as well as his/her ability to connect with other users). This lack of correlation should however not be understood negatively. Instead, we should interpret it as a sign that it these indicators measure a different dimension of the author's impact on the Web.

Two different kinds of indicators were found in these platforms: first, all metrics related to academic performance. This first group can further be divided into usage metrics (views and downloads) and citation metrics. Second, all metrics related to connectivity and popularity (followers). *ResearchGate* provides examples for these two sides of academic performance, since *Google Scholar Citations* profiles do not offer data about downloads or reads.

In the process of conducting this analysis, we identified a series of errors that allowed us to outline the main limitations of each product. May this serve as a sign that this study hasn't been made with an intention to exalt a particular database over the others. On the contrary, the intention was to thoroughly, comprehensively, conscientiously, and neutrally test the possibilities of *Google Scholar* as a tool for scientific evaluation.

In this sense, the empirical results indicate that *Google Scholar* should be the preferred source for relational and comparative analyses in which the emphasis is put on author clusters. Individual data should be taken with some caution as it may be subject to some errors. Despite these errors (as well as the lack of more advanced filtering features), *Google Scholar* has been able to measure the academic community dedicated to measuring; and has done it successfully: detecting "those who count" (bibliometricians).

Lastly, the results should be understood within the context of the bibliometric community. They may be different in other academic communities where the greater or lesser use of technologies can clearly influence the data. Furthermore, there is a certain positive bias in the use of these platforms because within the bibliometric community, these platforms are part of the object of study of the discipline, as is the case of this work.

# REFERENCES

Barjak, F., Li, X., & Thelwall, M. (2007). “Which factors explain the web impact of scientists’ personal homepages?” *Journal of the American Society for Information Science and Technology, 58*(2), 200–211.

Becher, T., & Trowler, P. (2001). *Academic tribes and territories: Intellectual enquiry and the culture of disciplines*. McGraw-Hill Education (UK).

Bensman, S. J. (2007). “Garfield and the impact factor”. *Annual Review of Information Science and Technology*, *41*(1), 93-155.

Bonitz, M. (1982). “Scientometrie, Bibliometrie, Informetrie”. *Zentralblatt für Bibliothekswesen*, *96* (2):19–24.

Borgman, C. L. & Furner, J. (2002). “Scholarly Communication and Bibliometrics”. *Annual Review of Information Science and Technology*, *36*, 3-72.

Braun, T. (1994). “Little scientometrics, big scientometrics… and beyond?” *Scientometrics*, *30*, 373–537.

Broadus, R.N. (1987a). “Early approaches to bibliometrics”. *Journal of the American Society for Information Science*, *38*, 127–129.

Broadus, R. N. (1987b). “Toward a definition of ‘bibliometrics’”. *Scientometrics*, 12, 373–379.

Brookes, B. C. (1988). “Comments on the scope of bibliometrics”. In: L. Egghe, R. Rousseau (Eds). *Informetrics 87/88. Select Proceedings of the First International Conference on Bibliometrics and Theoretical Aspects of Information Retrieval*. Amsterdam, Elsevier Science, 29–41.

Brookes, B. C. (1990). “Biblio-, Sciento-, Infor-metrics??? What are we talking about?”. In: L. Egghe, R. Rousseau (Eds). *Informetrics 89/90. Selection of Papers Submitted for the Second International Conference on Bibliometrics, Scientometrics and Informetrics*. Amsterdam, Netherlands, Elsevier, 31–43.

Castells, M. (2002). *La galaxia internet*. Barcelona: Plaza & Janés.

Cronin, B. (2001). “Bibliometrics and beyond: some thoughts on web-based citation analysis”. *Journal of Information science*, *27*(1), 1-7.

De Bellis, N. (2009). *Bibliometrics and citation analysis: from the science citation index to cybermetrics*. Maryland: Scarecrow Press.

Delgado López-Cózar, E. & Martín-Martín, A. (2015). “Thomson Reuters coquetea con las altmetrics: usage counts para los artículos indizados en la Web of Science”. *EC3 Working Papers*, 20.

Delgado López-Cózar, E., Robinson-García, N. & Torres-Salinas, D. (2014). “The Google Scholar Experiment: how to index false papers and manipulate bibliometric indicators”. *Journal of the Association for Information Science and Technology*, *65*(3), 446-454.

Franceschini, F., Maisano, D. & Mastrogiacomo, L. (2015). “Research quality evaluation: comparing citation counts considering bibliometric database errors”. *Quality & Quantity*, *49*(1), 155-165.

García-Pérez, M. A. (2010). “Accuracy and completeness of publication and citation records in the Web of Science, PsycINFO, and Google Scholar: A case study for the computation of h indices in Psychology”. *Journal of the American Society for Information Science and Technology*, *61*(10), 2070-2085.

Garfield, E. (1983). “Idiosyncrasies and errors, or the terrible things journals do to us”. *Current Contents*, *2*, 5-11

Garfield, E. (1990). “Journal editors awaken to the impact of citation errors-how we control them at ISI”. *Current Contents*, *41*, 5-13

Glänzel, W. & Schoepflin, U. (1994). “Little scientometrics, big scientometrics … and beyond?” *Scientometrics*, *30*, 375–384.

Godin, B. (2006). “On the origins of bibliometrics”. *Scientometrics*, *68*(1), 109-133.

González-Díaz, C.; Iglesias-García, M.; Codina, L. (2015). “Presencia de las universidades españolas en las redes sociales digitales científicas: caso de los estudios de comunicación”. *El profesional de la información*, *24*(5), 640-647.

Gorbea Portal, S. (1994). “Principios teóricos y metodológicos de los estudios métricos de la información. *Investigación Bibliotecológica*, *8*, 23-32.

Haustein, S., Peters, I., Bar-Ilan, J., Priem, J., Shema, H. & Terliesner, J. (2014). “Coverage and adoption of altmetrics sources in the bibliometric community”. *Scientometrics*, *101*(2), 1145-1163.

Hertzel, D.H. (1987). “History of the development of ideas in bibliometrics”. In: A. Kent, (Ed.). *Encyclopedia of library and information sciences*, Vol. 42 (Supplement 7), Marcel Dekker, New York, 144–219

Hood, W. & Wilson, C. (2001). “The literature of bibliometrics, scientometrics, and informetrics”. *Scientometrics*, *52*(2), 291-314.

Jacsó, P. (2005). “Google Scholar: the pros and the cons”. *Online information review*, *29*(2), 208-214.

Jacso, P. (2006a). “Deflated, inflated and phantom citation counts”. *Online information review*, *30*(3), 297-309.

Jacsó, P. (2006b). “Dubious hit counts and cuckoo's eggs”. *Online Information Review*, *30*(2), 188-193.

Jacsó, P. (2008). “Google scholar revisited”. *Online information review*, *32*(1), 102-114.

Jacsó, P. (2010). “Metadata mega mess in Google Scholar”. *Online Information Review*, *34*(1), 175-191.

Jamali, H. R., Nicholas, D. & Herman, E. (2015). “Scholarly reputation in the digital age and the role of emerging platforms and mechanisms”. *Research Evaluation*, rvv032.

Kramer, Bianca; Bosman, Jeroen (2015): 101 Innovations in Scholarly Communication - the Changing Research Workflow. Available at: https://101innovations.wordpress.com/

Larivière, V. (2012). “The decade of metrics? Examining the evolution of metrics within and outside LIS”. *Bulletin of the American Society for Information Science and Technology*, *38*(6), 12-17.

Larivière, V., Sugimoto, C. & Cronin, B. (2012). “A bibliometric chronicling of library and information science’s first hundred years”. *Journal of the American Society for Information Science and Technology*, *63*(5), 997-1016.

Lawani, S. M. (1981), “Bibliometrics: its theoretical foundations, methods and applications”. *Libri*, *31*, 294–3

Martín-Martín, A., Orduña-Malea, E., Ayllón, J.M. & Delgado López-Cózar, E. (2014). “Does Google Scholar contain all highly cited documents (1950-2013)?”. *EC3 Working Papers*, 19.

Más-Bleda, A. & Aguillo, I. F. (2013). “Can a personal website be useful as an information source to assess individual scientists? The case of European highly cited researchers”. *Scientometrics*, *96*(1), 51-67.

Mas-Bleda, A., Thelwall, M., Kousha, K. & Aguillo, I. F. (2014). “Do highly cited researchers successfully use the social web?”. *Scientometrics*, *101*(1), 337-356.

McCain, K. W. (2010). “The view from Garfield’s shoulders: Tri-citation mapping of Eugene Garfield’s citation image over three successive decades”. *Annals of Library and Information Studies*, *57*, 261-270.

Mikki, S., Zygmuntowska, M., Gjesdal, Ø. L. & Al Ruwehy, H. A. (2015). “Digital Presence of Norwegian Scholars on Academic Network Sites—Where and Who Are They?”. *PloS one*, *10*(11), e0142709.

Moed, H. F. & Vriens, M. (1989). “Possible inaccuracies occurring in citation analysis”. *Journal of Information Science*, *15*(2), 95-107.

Narin, F. & Moll, J.K. (1977). "Bibliometrics". *Annual Review of Information Science and Technology*, *12*, 35-58.

Nicolaisen, J. & Frandsen, T. F. (2015). "Bibliometric evolution: Is the journal of the association for information science and technology transforming into a specialty Journal?". *Journal of the Association for Information Science and Technology*, *66*(5), 1082-1085.

Orduna-Malea, E., Ayllón, J. M., Martín-Martín, A. & López-Cózar, E. D. (2015). “Methods for estimating the size of Google Scholar”. *Scientometrics*,*104*(3), 931-949.

Peritz, B.C. (1984). “On the careers of terminologies; the case of bibliometrics”, Libri, *34*: 233–242

Poyer, R. K. (1979). “Inaccurate references in significant journals of science”. *Bulletin of the Medical Library Association*, *67*(4), 396.

Sengupta, I.N. (1992). “Bibliometrics, informetrics, scientometrics and librametrics: an overview”, *Libri*, *42*, 75–98.

Shapiro, Fred R. (1992). “Origins of Bibliometrics, Citation Indexing, and Citation Analysis: The Neglected Legal Literature”. *Journal of the American Society for Information Science*, *43*(5), 337–39.

Sher, I. H., Garfield, E., & Elias, A. W. (1966). “Control and Elimination of Errors in ISI Services”. *Journal of Chemical Documentation*, *6*(3), 132-135.

Thelwall, M. (2008). “Bibliometrics to webometrics”. *Journal of Information Science*, *34*(4), 605-621.

Thelwall, M., & Kousha, K. (2015). “ResearchGate: Disseminating, communicating, and measuring Scholarship?”. *Journal of the Association for Information Science and Technology*, *66*(5), 876-889.

Van Raan, A. (1997). “Scientometrics: State-of-the-art”. *Scientometrics*, *38*(1), 205-218.

Van-Noorden, R. (2014). “Online collaboration: Scientists and the social network”. *Nature news, 512*(7513), 126-129.

White, H. D. & McCain, K. W. (1998). “Visualizing a Discipline: An Author Co-Citation Analysis of Information Science, 1972–1995”. *Journal of the American Society for Information Science*, *49*(4), 327-355.

White, H.D. & McCain, K.W. (1989). “Bibliometrics”. *Annual review of information science and technology*, *24*, 119-186.

Whitley, R. (1984). *The intellectual and social organization of the sciences*. UK: Oxford University Press.

Wilson, C.S. (1999). “Informetrics”. *Annual Review of Information Science and Technology*, *34*, 107-247.